\documentclass[lettersize,journal]{IEEEtran}

\usepackage{tabularx}
\usepackage{threeparttable}
\usepackage{mathrsfs}
\usepackage{color}
\usepackage{amsfonts,amssymb}
\usepackage{amsfonts}
\usepackage{subfigure}
\usepackage{booktabs}

\usepackage{amsmath}
\usepackage{enumerate}
\usepackage{algorithm}
\usepackage{algorithmic}
\usepackage{bm}
\usepackage{makecell}
\usepackage{multirow}
\usepackage{mathtools}
\usepackage{bbm}
\usepackage{dsfont}
\usepackage{pifont}
\usepackage{amsthm}
\usepackage{cite}
\usepackage{balance}
\usepackage{enumerate}

\def\BibTeX{{\rm B\kern-.05em{\sc i\kern-.025em b}\kern-.08em
    T\kern-.1667em\lower.7ex\hbox{E}\kern-.125emX}}

\newtheorem{remark}{Remark}
\newtheorem{theorem}{Theorem}

\newtheorem{corollary}{Corollary}
\newtheorem{assumption}{Assumption}

\newtheorem{definition}{Definition}

\let \sss=\scriptscriptstyle

\begin{document}

\title{
I Can Read Your Mind: \\ Control Mechanism Secrecy of Networked Dynamical Systems under Inference Attacks
}
\author{Jianping He$^{\dag}$,~\IEEEmembership{Senior Member IEEE}, Yushan Li$^{\dag}$,~\IEEEmembership{Student Member IEEE}, Lin Cai$^{\ddag}$,~\IEEEmembership{Fellow IEEE},\\ and Xinping Guan$^{\dag}$,~\IEEEmembership{Fellow IEEE}
\thanks{ 
    ${\dag}$: The Dept. of Automation, Shanghai Jiao Tong University, Key Laboratory of System Control and Information Processing, Ministry of Education of China, and Shanghai Engineering Research Center of Intelligent Control and Management, Shanghai, China. E-mail address: \{jphe, yushan\_li,xpguan\}@sjtu.edu.cn. 

    ${\ddag}$: The Dept. of Electrical and Computer Engineering, University of Victoria, BC, Canada. Email address: cai@ece.uvic.ca.
}%
}

\maketitle

\begin{abstract}
Recent years have witnessed the fast advance of security research for networked dynamical system (NDS). 
Considering the latest \textit{inference attacks} that enable stealthy and precise attacks into NDSs with observation-based learning, 
this article focuses on a new security aspect, i.e., how to protect \textit{control mechanism secrets} from inference attacks, including state information, interaction structure and control laws. 
We call this security property as control mechanism \textit{secrecy}, which provides protection of the vulnerabilities in the control process and fills the defense gap that traditional cyber security cannot handle. 
Since the knowledge of control mechanism defines the capabilities to implement attacks, ensuring control mechanism secrecy needs to go beyond the conventional data privacy to cover both transmissible data and intrinsic models in NDSs. 
The prime goal of this article is to summarize recent results of both inference attacks on control mechanism secrets and countermeasures. 
We first introduce the basic inference attack methods on the state and structure of NDSs, respectively, along with their inference performance bounds. 
Then, the corresponding countermeasures and performance metrics are given to illustrate how to preserve the control mechanism secrecy. 
Necessary conditions are derived to guide the secrecy design. 
Finally, thorough discussions on the control laws and open issues are presented, beckoning future investigation on reliable countermeasure design and tradeoffs between the secrecy and control performance.

\end{abstract}

\section{Introduction}\label{sec:introduction}
In the last decades, traditional control systems are becoming increasingly diverse, networked, and integrated with numerous physical and cyber components. 
Networked dynamical systems (NDSs) are characterized by the locality of information exchange between individual nodes (subsystems) and the coordinated capability to implement control and safety-critical tasks with high-reliability requirements \cite{olfati2007consensus}, such as multi-robot systems \cite{oh2015survey,choi2018detecting}, vehicle and traffic networks \cite{hubaux2004security,ni2019toward}. 

While NDSs have many promising applications, the networked working nature and physical openness make NDSs vulnerable to a wide range of security risks. 
It arises as an urgent and critical problem to secure NDSs under various cyber/physical attacks \cite{cardenas2008secure}. 
Specifically, from the perspective of external observations, the state, structure, and control laws of NDSs are critical elements to evaluate and analyze the operating performance (we call the three elements as control mechanism). 
This article reveals that, with the rapid development of artificial intelligence, external attackers/adversaries can infer the control mechanism based on a small number of observations, which is equivalent to read the mind of the NDS systems. 
Furthermore, the attacker can leverage the inferred knowledge to achieve more stealthy and intelligent attacks. 
Different from the mainstream security research in the literature, we focus on the mentioned inference attacks, and introduce the notion of control mechanism secrecy to investigate the security performance under inference attacks. 

\begin{figure}[t]
\centering
\subfigure[Multi-component type NDS, such as manufacturing industries, oil refineries, and smart homes. ]{\label{fig:component}
\includegraphics[width=0.4\textwidth]{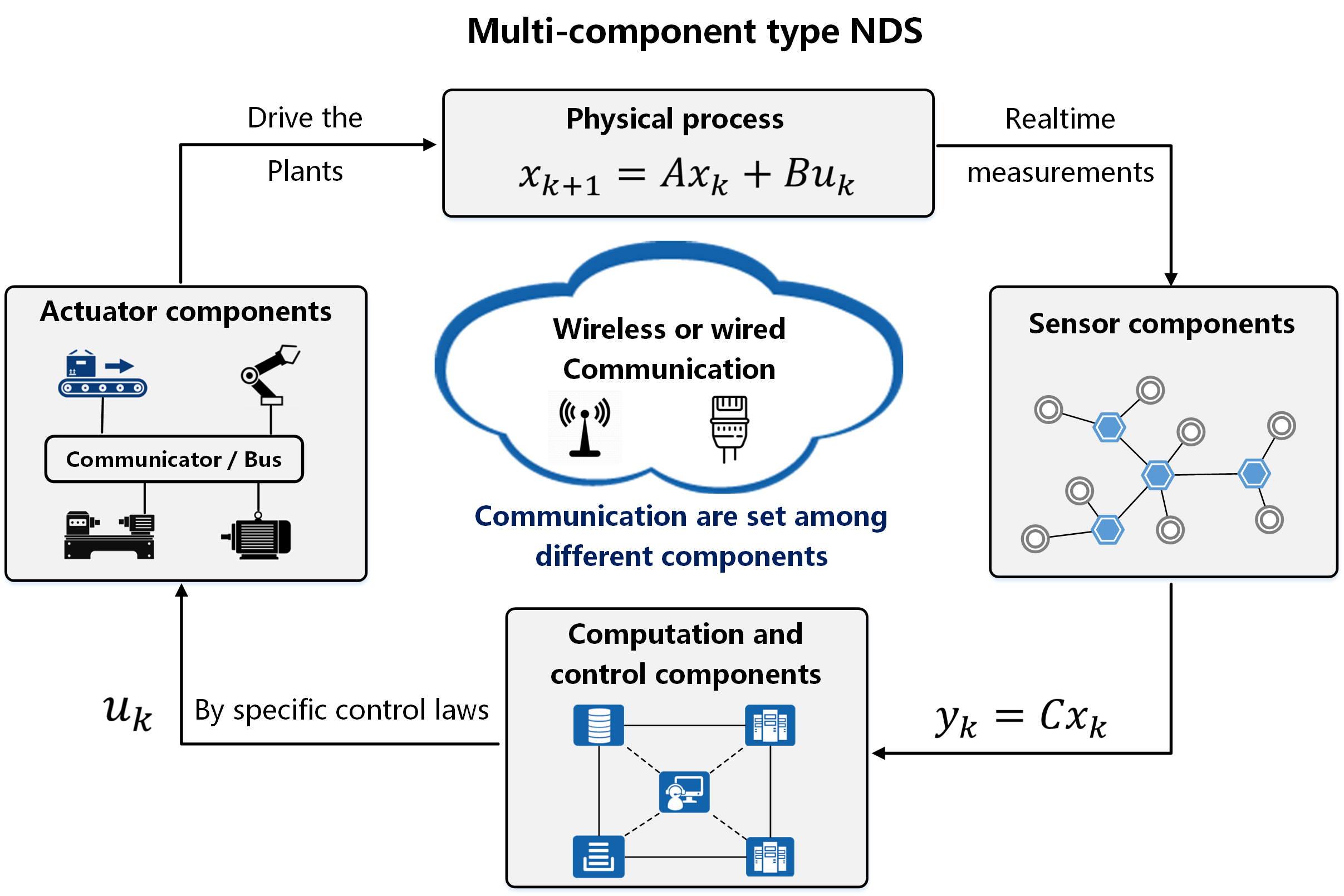}}
\subfigure[Multi-agent type NDS, such as smart grid, robot formation, automobiles, VANETs and etc.]{\label{fig:agent}
\includegraphics[width=0.45\textwidth]{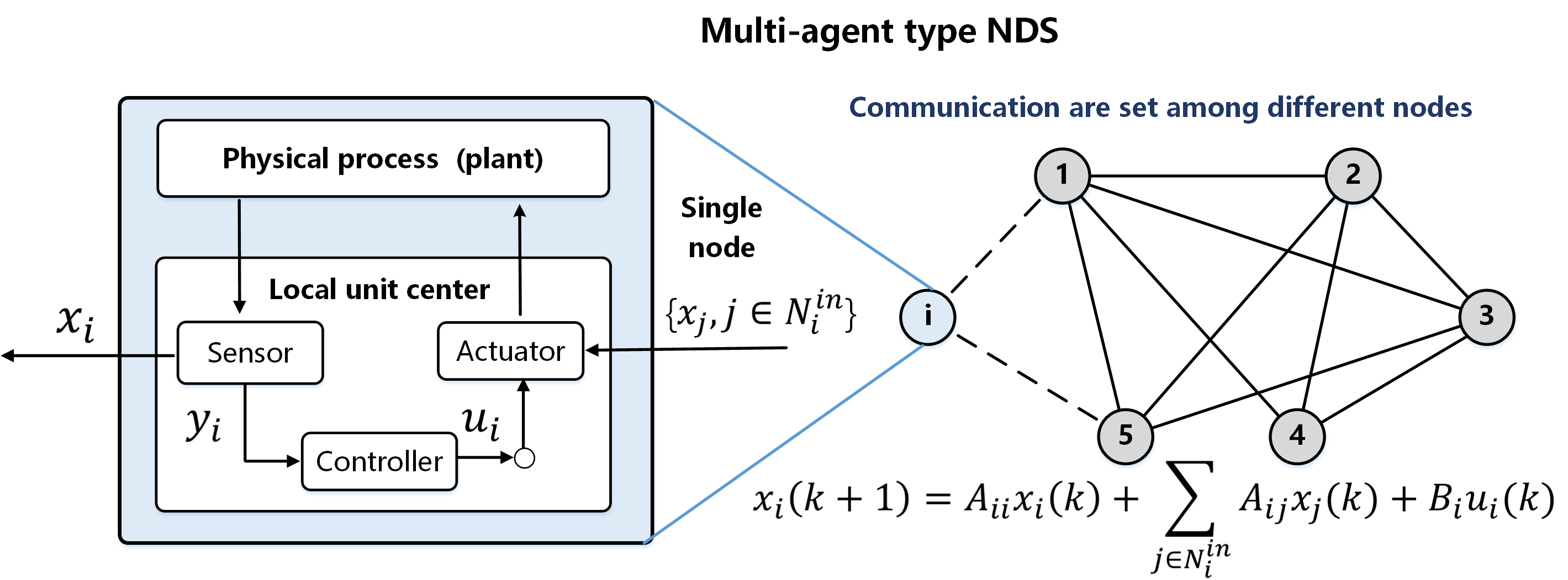}}
\caption{Two common types of NDSs.}
\label{fig:NDS_type}
\end{figure}

\begin{figure*}[t]
\centering
\includegraphics[width=0.82\textwidth]{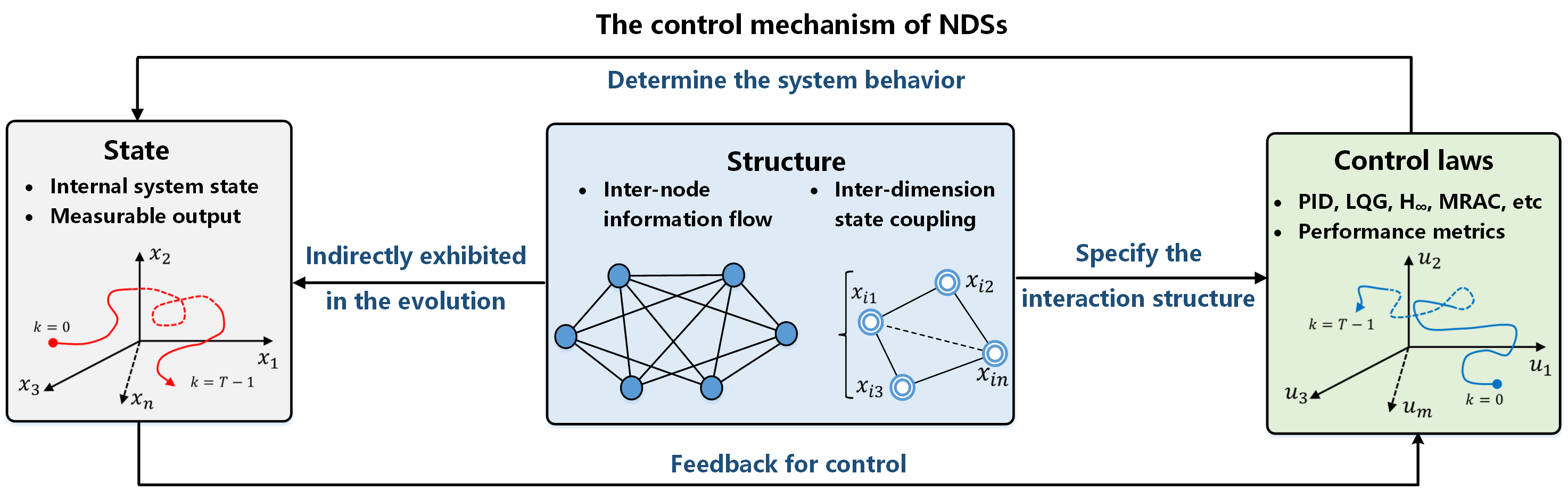}
\caption{The elements of the control mechanism. 
The state, structure, and control laws are deeply coupled with each other: 
the state evolution is a direct result of the control laws, while the control laws need to use the state and structure to control the system behavior. 
}
\label{fig:frame1}
\vspace{-10pt}
\end{figure*}

\subsection{What is Control Mechanism Secrecy}

To interpret the contents of mechanism secrecy, the control mechanism needs to be specified first. 
NDSs can be categorized into two types: multi-component type and multi-agent type, as shown in Fig.~\ref{fig:NDS_type}. 
For both types, they are all composed of sensors for state measurements, communication/transmission devices for information exchange, and controlled actuators for driving plants. 
Given the common feature, the system state, internal interaction structure, and control laws are the fundamental three elements that characterize the dynamics of NDSs. 
The detailed meanings are given as follows. 
\begin{itemize}
\item The states represent a group of specific features that characterize the operating process, e.g., the displacement in robotic systems. 
In some situations, the states cannot be obtained directly due to environmental constraints, but are measured by deployed/embedded sensors. 
These indirectly obtained values related to the states are called the outputs of the system. 

\item NDSs have two types of internal structure: i) the coupling structure that measures the mutual influence of different states in a single node (subsystem), and ii) the interconnection structure that specifies the information interaction flows between different nodes. 

\item The control laws of NDSs deal with two critical aspects: i) how the system inputs are selected and used to accomplish the control objective, ii) how the system reacts to the injected inputs and evolves. 
\end{itemize}

An illustration of the control mechanism is shown in Fig.~\ref{fig:frame1}. 
Accordingly, the control mechanism secrecy of NDSs has three aspects: 
first, to protect the historical states and futuristic states from being estimated and predicted, respectively; 
second, to protect the intra-state structure and the interconnection structure among nodes from being inferred; 
third, to protect the input design and model parameters from being regressed. 
In summary, the control mechanism secrecy aims to conceal the dynamical properties against external inference attacks without compromising the operating performance. 

\subsection{Why Control Mechanism Secrecy}

According to famous Kerckhoffs’s principle \cite{van2014encyclopedia} and Shannon’s maxim ``\textit{the enemy knows the system}''\cite{shannon1949communication}, 
it is dangerous to assume that the attacker lacks the knowledge about the system. 
The two principles have guided security research in control systems, with rich achievements to guarantee the system performance under various attacks. 
On the other hand, a growing amount of works focus on the defense design under the assumption that the attacker has perfect knowledge about the system. 
This assumption is for worst-case attacks, and generally leads to a passive defense manner. 
The worst-case assumption can be counter-productive, which leads to overly conservative defense strategies with compromised system performance. 
This dilemma motivates us to investigate the unsolved issues inspired by Shannon’s maxim, i.e., the fundamental control mechanism secrecy. 
We need to answer two important questions:  
\textit{i) under what conditions and to what extent can the attacker infer the control mechanism}, and \textit{ii) in what way to enhance the control mechanism secrecy}. 

Here we use an example of multi-robot systems to illustrate how to ``read your mind''. 
Suppose a multi-robot system is deployed in an unknown environment for reconnaissance, while a malicious attacker (like an UAV or other robots equipped with sensors) can observe the trajectories of the robots. 
From the observed information, the attacker can infer the real historical states which may contain sensitive location information of the robot base \cite{shokri2011quantifying}, or predict the future trends of the trajectory evolution \cite{li2020evolvegraph}. 
Also, if the robots exhibit a process of reaching certain regular pattern, then the attacker can infer the internal interaction structure among the robots \cite{vasquez2018network,li2021topology}, and find the critical robot that has dominant impacts on the system \cite{zhang2014analysis}. 
Even worse, the control laws about how the robots are driven are also likely to be learned if sufficient data of the control process are collected \cite{deisenroth2015gaussian,tolstaya2020learning}. 
Once these elements of control mechanism are disclosed to the attacker, the robot system will face more severe security risks. 
For instance, the attacker can use the knowledge to launch one-shot strike against the critical robot precisely to disrupt the whole system \cite{quinonez2020savior}, or choose to be a spy robot that sneaks into the system and stealthily misguides the system \cite{li2019learning,licitra2019single}.

Based on the above analysis, if we can protect the control mechanism secret, the capabilities of the attackers will be substantially constrained, thereby largely reducing the defense burdens. 
Therefore, it is critical to investigate the control mechanism secrecy, as another layer of proactive defense for the system. 
Moreover, it provides much needed insights to characterize the capabilities of the state-of-the-art inference attacks, and is an under-explored research issue beckoning further investigation.



\begin{figure*}[t]
\centering
\includegraphics[width=0.80\textwidth]{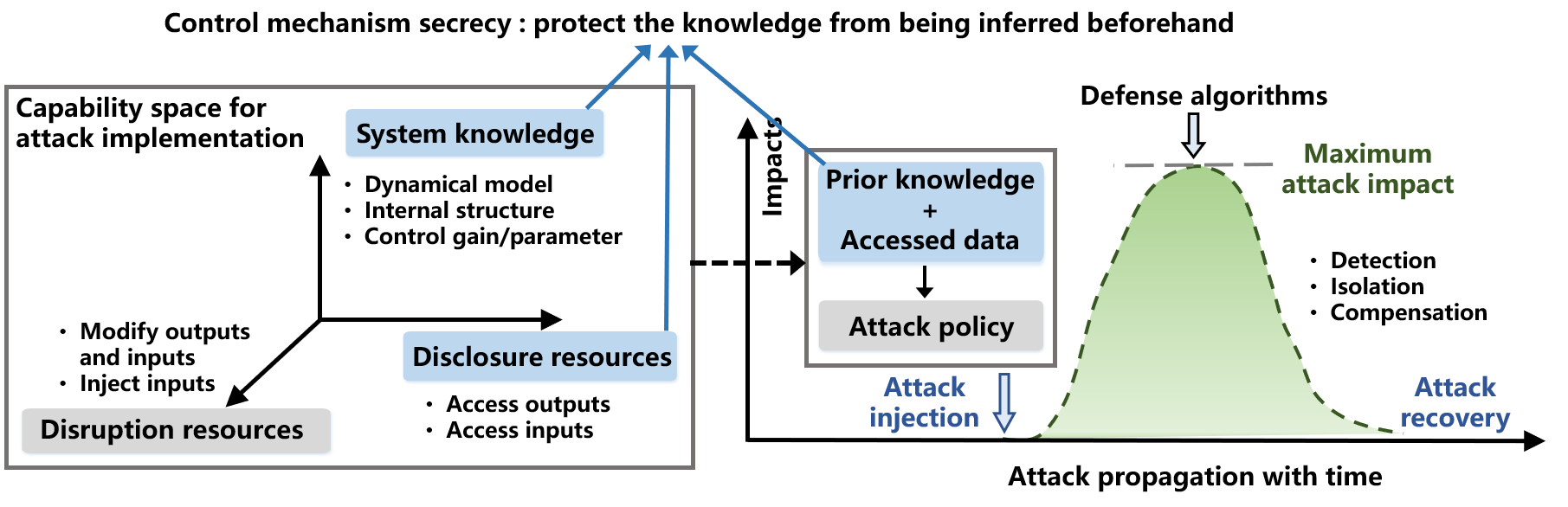}
\caption{Location illustration of the mechanism secrecy in the research fields. 
The control mechanism secrecy can be cast into the system knowledge and disclosure resource dimensions in the attack space. 
}
\label{fig:attack_location}
\vspace{-10pt}
\end{figure*}

\subsection{Related Work}

Control mechanism secrecy is not a brand new notion, as it is inspired by numerous pioneering works. 
In this part, we delineate how the control mechanism secrecy is associated with and also different from the existing security research. 

\textit{Perspective of the attack-defense process}. 
In NDSs, the interaction between the attack and defense can take three stages: 
i) attack design, ii) attack injection and effects spreading, and iii) attack detection and mitigation. 
Numerous attack models and defense methods have been developed to address different security aspects (see \cite{giraldo2017security,sandberg2022secure} for a detailed review). 
The previous works mainly addressed three objectives: 
a) attack modeling and analysis (e.g., \cite{teixeira2015secure,zhang2015optimal,conti2016survey,anguluri2019probabilistic,sanchez2019bibliographical}), 
i.e., analyzing the attack capabilities with certain system knowledge and resources, and illustrating how to maximize the attack impacts;   
b) attack detection and identification (e.g., \cite{pasqualetti2013attack,bai2017data,guan2017distributed,giraldo2018survey,anguluri2020centralized}), 
i.e., designing detectors based on outputs to judge whether the system is under attacks (like classic $\mathcal{X}^2$ detector), and investigating the limits of detection performance when the attacks are massive. 
c) attack mitigation and countermeasures (e.g., \cite{fawzi2014secure,mo2015physical,pajic2016attack,barboni2020towards,ding2020secure}), 
i.e., developing appropriate remedial strategies to alleviate the attacks injected into NDSs, for instance, reconstructing the underlying states from corrupted sensor observations or designing attack-resilient control algorithms. 

The above works on different aspects of the system security achieved prominent results to effectively defend various types of attacks. 
Nevertheless, they rely on deterministic attack model supported by known system knowledge, and barely consider what exactly an attacker could know about the system and how they can infer it. 
Specifically, the attack space can be characterized as the model knowledge, disclosure, and disruption resources \cite{teixeira2015securea}, where the former two are indispensable to achieve stealthy and powerful attack impacts. 
The control mechanism secrecy is complementary to consider how an external attacker can obtain such information to support the attack design and what kinds of smart attack can be launched with the learnt information, and how the defender can protect the information from being inferred. 
In a word, the control mechanism secrecy focuses on the stage before injecting attacks, as shown in Fig.~\ref{fig:attack_location}.

\textit{Privacy v.s. mechanism secrecy}. 
The value of a system element is private if an adversary cannot distinguish the real value from its candidate value set based on the publicly available information \cite{han2018privacy}. 
For control mechanism secrecy, its state secrecy has a close relationship with the data privacy. 
They both try to characterize the security of the system, especially when the user focuses on the sensitive data in the system. 
For example, many recent works have made progress in preserving the state privacy of consensus-based NDSs \cite{mo2016privacy,he2018privacy,wang2019privacy,ruan2019secure}, which belong to the state secrecy of the control mechanism secrecy. 
On the other hand, data privacy mainly focuses on the leakage risk of transmissible and computable data, while the mechanism secrecy also needs to deal with the intrinsic model elements like the interaction topology and control laws, i.e., the secrecy of the structure and control law. 
In addition, privacy is mainly concerned with whether the sensitive data in the system can be revealed to outsiders (e.g., the state is privacy-preserving), which is more like a binary form. 
The control mechanism secrecy investigates the relationship between the inference performance and system dynamics, including how accurate the inference methods can achieve by specific methods and available observations, how the inference error will affect the system dynamics, and how to design efficient countermeasures to degrade the inference performance, et al. 
Most of these issues still remain unsolved and need further investigation.

\textit{Security objective}. 
The premise of analyzing the security lies that we need to characterize the security into specific objectives/metrics first. 
In the literature, the famous confidentiality-integrity-availability (CIA) metrics are commonly used to describe the system security, which represents the unauthorized information release, modification and use, respectively \cite{dibaji2019systems}. 
The security issues concerned with the CIA metrics include information monitoring, data corruption, and communication blocking/delay, which mainly focus on the cyber aspect of the system. 
Control mechanism secrecy is proposed to characterize the \textit{deducibility} of the system information (especially through physical observations), which can be regarded as a new security dimension along with the CIA properties in NDSs. 

\textit{Methodology of the mechanism secrecy}. 
Mechanism secrecy is associated with many theories and techniques in control, optimization and learning fields. 
Considering different settings of the inference attack, different methods can be utilized and tailored for problem-solving. 
For instance, if the model form of the system is explicitly specified and known by the external observer, inferring the system state and structure can be regarded as a problem of parameter identification. 
If the model form of the system is totally unknown, then the problem may fall into the realm of machine learning.

\subsection{Main Points and Organization}
In this article, we point out that utilizing advanced inference and learning techniques to explore the internal properties of NDSs is a new and critical research trend. 
These techniques can be leveraged by adversaries to support highly intelligent and stealthy attack behaviors, thus incurring severe risks for NDSs. 
To characterize such security risks and fill the gap that traditional cyber security cannot handle, we propose a new security property, namely control mechanism secrecy, and construct a comprehensive analysis framework by surveying prior related results. 
Specifically, we interpret the control mechanism as three critical elements: state information, interaction structure and control laws. 
Then, we demonstrate how to infer the control mechanism and how to protect its secrecy from the three aspects, respectively. 

The contents are organized as follows. 
Section \ref{sec:formulation} introduces the modeling of NDSs and the formulation of the mechanism secrecy. 
Section \ref{sec:inference_methods} outlines the basic methods of inferring the control mechanism. 
The counterpart defense strategies are studied in Section \ref{sec:secure_methods}. 
Detailed discussions on more complicated scenarios and some meaningful open issues are provided in Section \ref{sec:discussions}, inspiring more efforts on future investigation. 
Finally, we present illustrative numerical examples in Section \ref{sec:simulations} and summarize this article in Section \ref{sec:conclusions}.

\section{Formulation of Mechanism Secrecy}\label{sec:formulation}
Let $\mathcal{G}=(\mathcal{V},\mathcal{E})$ be a directed graph that models the networked system, where $\mathcal{V}=\{1, \cdots, N\}$ is the finite set of nodes and $\mathcal{E}\subseteq \mathcal{V}\times \mathcal{V}$ is the set of interaction edges.
An edge $(i,j)\in \mathcal{E}$ indicates that $i$ will use information from $j$.
The adjacency matrix $A^{\sigma}=[a^{\sigma}_{ij}]_{N \times N}$ of $\mathcal{G}$ is defined such that ${a}^{\sigma}_{ij}\!>\!0$ if $(i,j)$ exists, and ${a}^{\sigma}_{ij}\!=\!0$ otherwise.
Denote ${\mathcal{N}_i}=\{j\in \mathcal{V}:a^{\sigma}_{ij}>0\}$ as the in-neighbor set of $i$, and $d_i=\left| {\mathcal{N}_i} \right |$ as its in-degree. 
Define $L=\text{diag}\{A^{\sigma} \bm{1}\}-A^{\sigma}$ as the Laplacian matrix of $\mathcal{G}$, where $\bm{1}$ is a vector of all ones. 
Then we have $L\bm{1}=0$. 
A directed path is a sequence of nodes $\left \{r_1, r_2, \cdots, r_j\right\}$ such that $(r_{i+1},r_i)\!\in\!\mathcal{E}$, $i=1, 2, \cdots, j-1$.
A directed graph has a (directed) spanning tree if there exists at least a node having a directed path to all other nodes.
$\mathcal{G}$ must have a spanning tree to guarantee that at least one node's information can reach all other nodes.

\subsection{System Model}

Consider the nodes in a NDS are represented by $\mathcal{V}$. 
For $i\in\mathcal{V}$, its dynamics is described by 
\begin{equation}\label{eq:general_model}
\begin{aligned}
x_i(k+1) &= f(x_i(k),u_i(k),\mathcal{X}_i(k), \omega_i(k) ), \\
y_i(k) &= g(x_i(k),v_i(k) ),
\end{aligned}
\end{equation}
where $\mathcal{X}_i(k)=\{x_j(k),j\in\mathcal{N}_i^{in}\}$ represents the state set of node $i$'s all neighbors,  
$x_{i} \in \mathbb{R}^{n_i}$, $y_{i} \in \mathbb{R}^{m_i}$ and $u_i\in \mathbb{R}^{q_i}$ are the state, output and input vectors of node $i$, respectively. 
Besides, $\omega_i(k)$ and $v_i(k)$ are mutually independent process and measurement noises, which has both zero means and variances of $\Sigma_i^{\omega}$ and $\Sigma_i^{v}$, respectively. 
For the noises $\omega_i(k)$ and $v_i(k)$, the following assumption is commonly used.
\begin{assumption}[Gaussian noise assumption]\label{assu:Gaussian}
The process noise $\omega_i(k) \sim \mathcal{N}(0,\Sigma_i^{\omega})$ and the observation noise $v_i(k) \sim \mathcal{N}(0,\Sigma_i^{v})$ are i.i.d. zero-mean Gaussian noises. 
\end{assumption}

Note that (\ref{eq:general_model}) is an abstract formulation that can represent most system models in real applications. 
In the literature, the linear time-invariant systems are most widely investigated, and the nodal dynamics are given by 
\begin{equation}\label{eq:nodal_model}
\!\begin{aligned}
x_{i}(k+1) &=A_{i i} x_{i}(k) \!+\! \!\!\!\sum_{j \in \mathcal{N}_i^{in}}^{N} \!\!A_{i j} x_{j}(k) \! + \! B_{i} u_{i}(k) \!+ \! w_{i}(k),\! \\
y_i(k) &= {C_i} {x_i}(k)+v_i(k),
\end{aligned}
\end{equation}
where $A_{i j} \in \mathbb{R}^{n_i \times n_j}$, $B_{i} \in \mathbb{R}^{n_i \times q_i}$ and ${C_i}\in \mathbb{R}^{m _i \times n_i}$. 
Note that $A_{i j} x_{j}(k)$ represents the effect of node $j$ on node $i$, and $B_{i} u_{i}(k)$ represents the internal input to make node $i$ accomplish specific control objectives. 
In a global form, the dynamics of a NDS is given by 
\begin{equation}\label{eq:global_model}
\begin{aligned}
x(k+1) &=A x(k) + B u(k)+w(k),\\
y(k) &= {C} {x}(k)+v(k),
\end{aligned}
\end{equation}
where $x=[x_1^{\mathsf{T}}, \cdots, x_N^{\mathsf{T}}]^{\mathsf{T}}$, 
$y=[y_1^{\mathsf{T}}, \cdots, y_N^{\mathsf{T}}]^{\mathsf{T}}$, 
$\omega=[\omega_1^{\mathsf{T}},  \cdots, \omega_N^{\mathsf{T}}]^{\mathsf{T}}$, 
and $v=[v_1^{\mathsf{T}}, \cdots, v_N^{\mathsf{T}}]^{\mathsf{T}}$. 
In addition, we have 
\begin{align}\nonumber
A=\left[\begin{array}{ccc}
A_{11} & \cdots & A_{1 N} \\
\vdots & \ddots & \vdots \\
A_{N 1} & \cdots & A_{N N}
\end{array}\right], ~
B=\left[\begin{array}{ccc}
B_{1} & \cdots & 0 \\
\vdots & \ddots & \vdots \\
0 & \cdots & B_{N}
\end{array}\right],
\end{align}
$B=\operatorname{blkdiag}(B_{1},\cdots, B_{N})$, and $C=\operatorname{blkdiag}(C_{1},\cdots, C_{N})$. 
Correspondingly, the canonical controllability and observability matrices of the system are given by 
\begin{align}
Q_c&=\left[B,AB,\cdots, A^{n-1}B \right] \in \mathbb{R}^{n \times nq},~ \\
Q_o&=\left[C^\mathsf{T},(CA)^\mathsf{T},\cdots, (C A^{n-1})^\mathsf{T} \right]^\mathsf{T} \in \mathbb{R}^{nm \times n} ,
\end{align}
where $n=\sum_{i=1}^{N} n_i$, $q=\sum_{i=1}^{N} q_i$ and $m=\sum_{i=1}^{N} m_i$. 
The system \eqref{eq:global_model} is controllable if $\operatorname{rank}(Q_c)=n$, and observable if $\operatorname{rank}(Q_o)=n$. 
Given time horizon $T$, we define the corresponding controllability and observability matrices as 
\begin{align}
M_c&=\left[B,AB,\cdots, A^{T-1}B \right] \in \mathbb{R}^{n \times Tq},~ \\
M_o&=\left[C^\mathsf{T},(CA)^\mathsf{T},\cdots, (C A^{T-1})^\mathsf{T} \right]^\mathsf{T} \in \mathbb{R}^{Tm \times n}.
\end{align}

In most existing literature concerning system security, the parameters $(A,B,C)$ and input $u$ are known to the system (or at least the local information is known). 
For an external attacker, the knowledge is unknown initially. 
Unless otherwise specified, we mainly focus on linear system models (\ref{eq:nodal_model}) and (\ref{eq:global_model}) in this article. 

Next, we present asymptotically stable matrix class $\mathcal{S}_a$ and the (strictly) marginally stable matrix $\mathcal{S}_m$ as follows:
\begin{equation}
\begin{aligned}
\mathcal{S}_a=&\{Z\in \mathbb{R}^{n \times n}, \rho_{\max}(Z)<1\}, \\
\mathcal{S}_m=&\{Z\in \mathbb{R}^{n \times n}, \rho_{\max}(Z)=1~\text{~and the geometric} \\
&\text{multiplicity of eigenvalue one is $1$} \} .
\end{aligned}
\end{equation}

\begin{table*}[t]
\centering
\caption{\label{tab:systems} Representative Networked Dynamical Systems} 
\begin{tabular}{ccccc}
\toprule
System type  &  Model conditions &   Representative literature & Examples  \\
\midrule
Integrated CPS    &  \makecell[c]{ The system is controllable \\ and observable }  & \makecell[c]{ \cite{teixeira2015securea,fang2020stealthy,sui2021vulnerability,zhou2021unified} }  & \makecell[c]{ Industrial process control systems, \\ SCADA, transportation systems } \\
\midrule
Multi-sensor network  & $(A,B)$ is controllable  &  \cite{chong2015observability,guan2018distributed,mao2022computational,ren2020secure}  & \makecell[c]{ Environment monitoring,\\ distributed state tracking }\\
\midrule
Interconnected subsystems  &  $\mathcal{G}$ is connected  &  \cite{weerakkody2017graphtheoretic,lu2020privacy,anguluri2020centralized,katewa2021security}  &  \makecell[c]{ Smart grid, water network, \\ chemical reactor network, robot formation } \\
\midrule
Consensus-based network  &   \makecell[c]{ At least $\mathcal{G}$ has a spanning \\ tree and $A$ is row-stochastic}  & \cite{mo2017privacy,he2019consensus,ruan2019secure,yemini2021characterizing} & \makecell[c]{ Time synchronization, \\  opinion agreement, flocking} \\
\bottomrule
\addlinespace[0.5ex]
\setlength\tabcolsep{0.5ex}
\vspace{-15pt}
\end{tabular}
\end{table*}

\subsection{Common Models Variants}
Note that in the literature, the results concerning the security of NDSs may not always be based on a unified model like \eqref{eq:nodal_model}, 
and the variant models can be used in different scenarios. 
We summarize four commonly used model variants (for simplicity we ignore the process and observation noise terms), which are given as follows. 
\begin{itemize}
 \item Integrated CPS:
\end{itemize} 
\begin{equation} \label{eq:int_cps}
\begin{aligned}
x(k+1) &=A x(k) + B u(k), \\
y(k) &= Cx(k).
\end{aligned}
\end{equation}
This model is considered when the system of interest is presented in a whole and the defender needs to analyze the global performance under attack, which means that a central unit with global knowledge about the system is required. 

\begin{itemize}
 \item Multi-sensor network:
\end{itemize} 
\begin{equation} \label{eq:multi_sensor}
\begin{aligned}
x(k+1) &=A x(k) + B u(k), \\
y_i(k) &= {C_i} {x_i}(k).
\end{aligned}
\end{equation}
The multi-sensor model can be seen as an extended version of \eqref{eq:int_cps}, where multiple sensors with different capabilities are used to measure the same dynamic process. 

\begin{itemize}
 \item Interconnected subsystems:
\end{itemize} 
\begin{equation} \label{eq:inter_sub}
\begin{aligned}
x_{i}(k+1) &=\sum_{j=1}^{N} A_{i j} x_{j}(k) + B_{i} u_{i}(k), \\
y_i(k) &= {C_i} {x_i}(k).
\end{aligned}
\end{equation}
This model is used where one needs to consider the self-dynamics and information locality of each subsystem. 

\begin{itemize}
 \item Consensus-based network:
\end{itemize} 
\begin{equation} \label{eq:cons_model}
\begin{aligned}
x_i(k+1) &=  \sum_{j =1}^{N} a_{ij}{x_j}(k), \\
y_i(k) &= {x_i}(k).
\end{aligned}
\end{equation}
The consensus-based model is most common in the literature, which mainly describes the agreement-achieving process in distributed computing and control fields. 

It is worth noting that, the former two models \eqref{eq:int_cps} and \eqref{eq:multi_sensor} can be both illustrated as the multi-component type NDSs (as shown in Fig.~\ref{fig:component}). 
The networked characteristic of them reflects in that the components are connected in wire/wireless manners, and their security risks are usually presented in the way where some dimensional values of the input/outputs (or the complete output of a single one in multi-sensor cases) can be maliciously tampered. 
The latter two models \eqref{eq:inter_sub} and \eqref{eq:cons_model} can be both illustrated as the multi-agent type NDSs (as shown in Fig.~\ref{fig:agent}). 
The common security risks of such systems are presented in inter-agent forms, e.g., some agents in the network are corrupted and transmit false information, or the connections between agents are disturbed. 
More detailed characteristics and examples are given in Table \ref{tab:systems}. 

\begin{figure*}[t]
\centering
\includegraphics[width=0.75\textwidth]{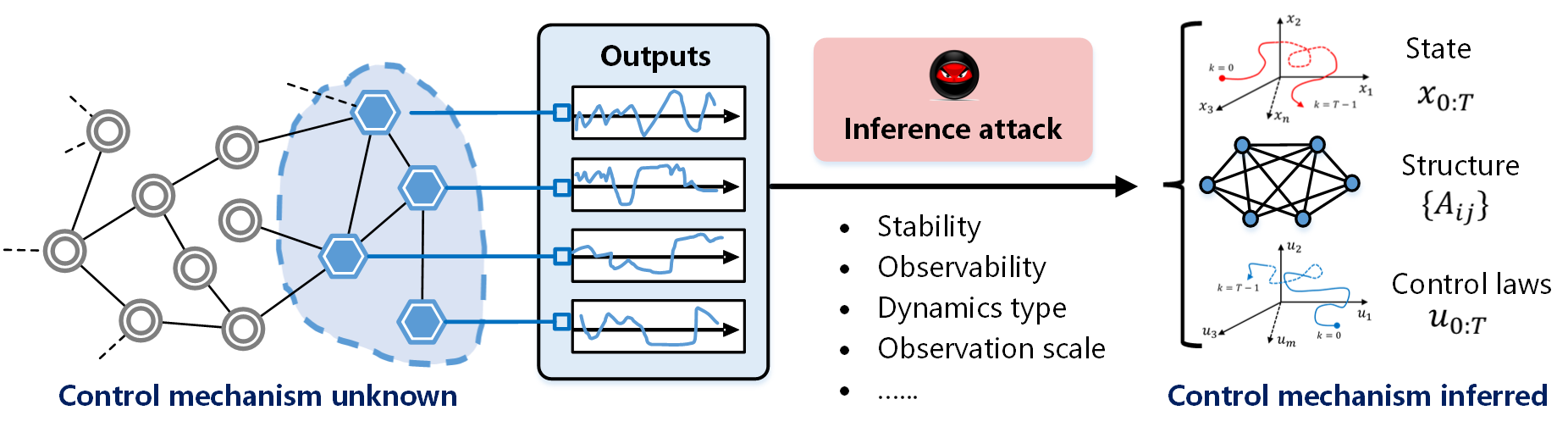}
\vspace{-5pt}
\caption{Illustration of the inference attack against the control mechanism of NDSs. 
}
\label{fig:inference_attack}
\vspace{-10pt}
\end{figure*}

\subsection{Security Analysis}
In this part, we present a detailed analysis of the control mechanism. 
Generally, both the internal state $x$ and the output $y$ can be regarded as the state of the system. 
The output is the indirect reflection of the internal state, and is available or observable in most situations. 
Next, we analyze how the control mechanism can be susceptible to security breaches when there is an external observer.  
\begin{itemize}
\item First, the state directly reflects the system behavior in the running process. 
When the system is observable, the state of the system can be easily obtained from the outputs if the system matrices $A$ and $C$ are also known. 
If the values of $A$ and $C$ are unknown but their dimensions are known, then the problem turns to the identification of these matrices, where the identified matrices are used to restore the states. 
Even if the system is not observable, it is possible to infer the underlying states with growing observations, e.g., by regression or Bayesian methods. 

\item Second, the structure of NDSs specifies the information flow between different nodes (e.g., $A_{ij}$ in model \eqref{eq:nodal_model}), and the coupling relationships between different state dimensions of a single node (e.g., $B_{i}$ in model \eqref{eq:nodal_model}). 
Even though the matrix values may not be estimated accurately by the malicious observer, their corresponding binary attributes are much easier to infer, which also leaks critical system information and causes great risks. 
This security vulnerability will be worse when the system reacts to external inputs, or the states are fully measurable. 

\item Third, the control laws are the cornerstone for the running process of the system. 
In practice, the design of the control laws is determined by multiple factors, including the stability, practical model, and control objectives (or performance metrics) of the system. 
Although the control laws of the system are generally unavailable for an external observer, it is highly possible that the rules can be approximated based on observed system interactions, e.g., by state-of-the-art data-driven techniques. 
Consequently, the approximated rules can be leveraged to predict the system behavior for malicious purposes. 
\end{itemize}

In a summary, the three elements determine the evolution process and convergence performance of the system, which are of vital importance for system implementation. 
Unfortunately, the state-of-art information techniques have enabled an external adversary without any prior knowledge to infer the control mechanism. 
When the sensitive information is leaked and mastered by the adversary, 
it increases the capabilities of precise attacks against the system and reduces the trial-and-error costs of attack implementation. 
Traditional cyber-security countermeasures hardly take the leakage risks into account, 
potentially increasing the defense costs and undermining the efficiency of alleviating attack impacts.  
Therefore, it is essential for NDSs to develop methods that guarantee the control mechanism secrecy.

\subsection{Inference Attack against NDSs}
We now present the setting for the external attacker, who aims to infer the control mechanism of NDSs from external observations. 
Since the control mechanism of NDSs includes the state, structure and control laws, attacks can be launched against these three aspects. 
In the following, we define the inference attack, which is also illustrated in Fig.~\ref{fig:inference_attack}. 

\begin{definition}[Inference Attack]\label{def:infer_attack}
An external attacker can access the observations $\{y_i: i\in\mathcal{V}\}$ in the dynamics (\ref{eq:general_model}) and excites the system nodes with designed inputs $\{u_i^{e}: i\in\mathcal{V}\}$. 
Then, the attacker infers the control mechanism by finding the following groups of mapping relationships 
\begin{align}
&\phi_{X}:\{y_i: i\in\mathcal{V}\}\to \{x_i: i\in\mathcal{V}\}, \label{eq:infer_a}\\
&\phi_{A}:\{y_i, y_j: i,j\in\mathcal{V}\}\to \{A_{ij}: i,j\in\mathcal{V}\}, \label{eq:infer_b}\\
&\phi_{U}:\{y_i, y_j, j\in\mathcal{N}_{i}: i\in\mathcal{V}\}\to \{u_i : i\in\mathcal{V} \}. \label{eq:infer_c}
\end{align}
\end{definition}

It is remarkable that the inference attack is oriented from observations of the NDS and requires no prior knowledge about the control mechanism. 
Therefore, this kind of attack is pervasive in the attack-defense context for an external attacker to master the control mechanism to a great extent. 
The inferred knowledge further facilitate the attack capability, like sneaking into the system stealthily or striking the vulnerable parts precisely. 
In addition, the three inference processes \eqref{eq:infer_a}-\eqref{eq:infer_c} about control mechanism are not necessarily independent of each other but can be deeply coupled. 

\subsection{Example: Coordination of Mobile Robots}
Finally, for better understanding, we take the coordination problem of mobile robotic network as a typical example to illustrate the mechanism secrecy of NDSs. 
In the following example, a mobile robotic network of $N$ agents aims to achieve motion coordination. 
Let $x_i=[x_{i,1},x_{i,2}]^{\mathsf{T}}$ be the state of robot $i$, where  $x_{i,1}$ and $x_{i,2}$ represent the position and velocity, respectively. 
Based on Newton’s motion law, the dynamics of robot $i$ is described by 
\begin{equation} \label{eq:example1}
\begin{aligned}
   ~& x_{i,1}(k+1)=x_{i,1}(k)+T_0 x_{i,2}(k)+({T_0^2}/{2}) {u}_i^f(k), \\
   ~& x_{i,2}(k+1)=x_{i,2}(k)+T_0 {u}_i^f(k), 
\end{aligned}
\end{equation}
where $T_0>0$ is the control period, and the feedback input ${u}_i^f(k)$ is given by \cite{cao2010sampled}
\begin{equation} \label{eq:example2}
    {u}_i^f(k)\!=\!\!\sum_{j\in\mathcal{N}_i^{in}}\!\! {a_{ij}[x_{j,1}(k)-x_{i,1}(k)+\alpha(x_{j,2}(k)-x_{i,2}(k))]}. 
\end{equation}
Then, substituting \eqref{eq:example2} into \eqref{eq:example1} and one has 
\begin{equation}\label{eq:robot_model}
\begin{bmatrix}
x_{i,1}(k+1) \\
x_{i,2}(k+1)
\end{bmatrix}
=\tilde{A}_{ii}\begin{bmatrix}
x_{i,1}(k) \\
x_{i,2}(k)
\end{bmatrix} + \sum_{j \in \mathcal{N}_i^{in}} {\tilde{A}_{ij} \begin{bmatrix}
x_{j,1}(k) \\
x_{j,2}(k)
\end{bmatrix} } , 
\end{equation}
where 
\begin{small}
$\tilde{A}_{ii}=\begin{bmatrix}
{ 1-\frac{T_0^2}{2}\sum_{j \in \mathcal{N}_i^{in}}a_{ij} }  & { T_0-\frac{\alpha  T_0^2}{2}\sum_{j \in \mathcal{N}_i^{in}}a_{ij} } \\
{ -T_0 \sum_{j \in \mathcal{N}_i^{in}}a_{ij} } & { 1-\alpha T_0 \sum_{j \in \mathcal{N}_i^{in}}a_{ij} }
\end{bmatrix}$
\end{small}, and 
\begin{small}$\tilde{A}_{ij}=\begin{bmatrix}
{ \frac{T_0^2 a_{ij}}{2} }  & { \frac{\alpha a_{ij}   T_0^2}{2} } \\
{ T_0 a_{ij} } & { \alpha T_0 a_{ij} }
\end{bmatrix}.$
\end{small}

Note that \eqref{eq:robot_model} shows an autonomous control process within the system. 
If the control objective is associated with other tasks (e.g., go to a specific region), then an extra driving control law $\tilde{u}_i$ is required.  
From \eqref{eq:robot_model}, the state-correlation structure $\tilde{A}_{ii}$ indicates that the two states of a single robot is correlated. 
The node-connectivity structure $A_{ij}$ specifies how a robot interacts with its neighbors, where $a_{ij}$ is critical. 
For an external attacker, the position and velocity of a robot are generally measurable in practice, and thus the state of the robotic network is also observable. 
Based on the measured trajectory information, the attacker can use many state-of-the-art methods to infer the real states $x$, the structure matrix $A$, and even extra control laws $\tilde{u}_i$ if any. 
Therefore, there exist severe risks that the mechanism of the robotic network is disclosed to an attacker, and effective countermeasures are necessary to be developed. 


\section{Inferring the Control Mechanism} \label{sec:inference_methods}

\begin{table*}[t]
\centering
\caption{\label{tab:state_solution} state secrecy under different prior knowledge of attackers and system properties} 
\begin{tabular}{ccccc}
\toprule
\multicolumn{2}{c}{\multirow{2}{*}{\begin{tabular}[c]{@{}c@{}}\textbf{Capability of}\\ \textbf{ the Inference Attack}\end{tabular}} }
& \multicolumn{3}{c}{ \textbf{ Properties of the System}}  \\
\cmidrule{3-5}
& & Measurable ($C$ is invertible)     & $(A,C)$ observable  & $(A,C)$ unobservable \\ 
\midrule
\multirow{3}{*}{ 
\makecell[c]{ \textbf{Prior knowledge}\\ \textbf{of the Attacker} \\ \textbf{over the System} } }
& \makecell[{}{p{2.5cm}}]{The values of $(A,C)$ are exactly known}
& \makecell[c{m{3.5cm}}]{ $x(k)$ can be directly obtained given $y(k)$} 
& \makecell[c{m{3.5cm}}]{ at least $n$ groups of observations are needed}
& \makecell[c{m{3.5cm}}]{ $T$ groups of observations such that $M_o$ is invertible, or only obtain sparse solution} \\ 
\cmidrule{2-5}
&\makecell[c{m{2.5cm}}]{ Only dimensions of $(A,C)$ are known } & \multicolumn{2}{c}{ the state is possible to be estimated by system identification techniques }  & hard to estimate \\
\cmidrule{2-5}
& \makecell[c{m{2.5cm}}]{ $(A,C)$ are totally unknown } & \multicolumn{3}{c}{ the state cannot be estimated based on the system observability } \\
\bottomrule
\addlinespace[0.5ex]
\setlength\tabcolsep{0.5ex}
\vspace{-15pt}
\end{tabular}
\end{table*}

In this section, we first elaborate that the inference performance of the malicious observer is closely related to the prior information and system properties (e.g., observability and controllability). 
Then, we introduce the basic methods to infer the state and topology structure of NDSs. 
Both the global and local inference cases are considered. 

\subsection{Prior Knowledge and Inference Conditions}
Considering the output of the system, classic observability is commonly used in the state estimation problem. 
To better describe a major class of NDSs, we further define the measurability of states as follows. 
\begin{definition}[Measurability of states]
The state of subsystem/node $i$ is fully measurable if the observation matrix is full-ranked, i.e., 
\begin{equation}
\operatorname{rank}(C_i)=n_i.
\end{equation}
\end{definition}
Note that if the states of each node are fully measurable, then the global system is also measurable, i.e., $\operatorname{rank}(C)=n$. 
Consequently, if the state of the global system is measurable, the observability of the system is also guaranteed, i.e., $\operatorname{rank}(M_o)=\operatorname{rank}(C)=n$ holds. 
Since $\operatorname{rank}( C^{\mathsf{T}} C )=n$ for systems with fully measurable states, we directly assume $C=I$ for these systems in the following sections.

Given system model \eqref{eq:global_model}, we characterize the prior knowledge of the observer into three categories: i) the values of $(A,C)$ are known, ii) only the dimensions of $(A,C)$ are known, and iii) $(A,C)$ are unknown. 
The first situation describes the most powerful observer, while the last one presents a totally ignorant observer. 
Since the available outputs are the most critical elements for estimating the states and the measurability and observability of the system will determine the inference cost to obtain the final estimate. 
To better illustrate this effect, we summarize the conclusion in Table \ref{tab:state_solution}. 
In this section, we mainly focus on the case where $(A,C)$ is known and the system satisfies the state measurability condition, and discuss the other cases in the following sections. 

\subsection{Inferring the State}

Inferring the state of NDSs can be regarded as conventional state estimation problems, which aim to reconstruct the system state from the obtained measurements. 
Considering different observation ranges on NDSs, the problems are divided into two types: global and local state estimation.


\subsubsection{Global State Estimation}
For simplicity, we first assume the state transition matrix, $A$, and the observation matrix, $C$, are known. 
Given a time horizon $T$, one aims to use an estimator $\phi_X(y(k),k=0,\cdots,T-1) : \mathbb{R}^{ mT} \to \mathbb{R}^{n}$ to reconstruct the initial system state $x(0)$. 
For simple expressions, the observations and noises in $T$ steps are organized as 
\begin{align}
\bm{y}_{T} &= [y^{\mathsf{T}}(0), \cdots, y^{\mathsf{T}}(T-1)]^{\mathsf{T}}, \nonumber\\
\bm{v}_{T} &=[v^{\mathsf{T}}(0), \cdots, v^{\mathsf{T}}(T-1)]^{\mathsf{T}}. \nonumber
\end{align}
When there is no input $u$ and process noise $w$, one has 
\begin{equation}\label{eq:observation_eq}
\bm{y}_{T}=M_o x(0)+\bm{v}_{T}.
\end{equation}
Based on \eqref{eq:observation_eq}, we show how the initial state is estimated in the following theorem.

\begin{theorem}
Consider the observation model \eqref{eq:observation_eq} and Assumption \ref{assu:Gaussian} holds. 
If $\operatorname{rank}(M_o)=n$ holds, then the unbiased estimator for $x(0)$ is given by 
\begin{equation}\label{eq:OLS_estimator}
\hat{x}(0;T)= (M_o^{\mathsf{T}}M_o)^{-1}M_o^{\mathsf{T}}\bm{y}_{T},
\end{equation}
whose estimation error satisfies
\begin{equation}\label{eq:OLS_error_limit}
\mathop {\lim }\limits_{T \to \infty } \Pr\{ \| \hat{x}(0;T)-x(0) \| = 0 \}=1. 
\end{equation}
\end{theorem}

The estimator $\hat{x}(0;T)$ easily follows from \eqref{eq:observation_eq} as long as the observability matrix $\operatorname{rank}(M_o)=n$ is guaranteed. 
Apparently, if there are no noises involved in the observations, then $x(0)$ can be accurately estimated. 
For other states $x(k),0<k<T$, one only needs to take $x(k)$ as the new initial state of interest, and use the data in the slot $[k,T-1]$ to calculate $\hat{x}(k)$\footnote{For example, let $\bm{y}_{k:T}=[y^{\mathsf{T}}(k), y^{\mathsf{T}}(k+1),\cdots, y^{\mathsf{T}}(T-1)]^{\mathsf{T}}$ and $M_o(T-k)=\left[C^\mathsf{T},(CA)^\mathsf{T},\cdots, (C A^{T-k-1})^\mathsf{T} \right]^\mathsf{T}$. 
Then, $x(k)$ can be estimated by $\hat{x}(k;T)= \left(M_o^{\mathsf{T}}(T-k)M_o(T-k)\right)^{-1}M_o^{\mathsf{T}}(T-k)\bm{y}_{k:T}$ when $\operatorname{rank}(M_o(T-k))=n$ holds.}. 
The unbiased asymptotic estimation error \eqref{eq:OLS_error_limit} is explained by the following remark. 

\begin{remark}
Under Gaussian noises, the estimator by \eqref{eq:OLS_estimator} is called the maximum likelihood estimate (MLE) of $x(0)$, whose estimation performance can be represented by the covariance of the error vector (e.g., see \cite{mo2017privacy}). 
Here, the estimation error is given by
\begin{equation}\label{eq:estimation_error}
\| \hat{x}(0;T)- x(0) \|= \| (M_o^{\mathsf{T}}M_o)^{-1}M_o^{\mathsf{T}} \bm{v}_{T} \|. 
\end{equation}
Then, the limit of \eqref{eq:OLS_error_limit} is derived by adopting the concentration inequality of Gaussian matrix (see Chapter 2 in \cite{wainwright2019high}), and the convergence rate of the estimation error is mainly determined by the stability of $A$. 
Note that if the noises are not in Gaussian form, \eqref{eq:OLS_estimator} cannot be interpreted as MLE but the optimal estimator with least square errors. 
\end{remark}

It is worth mentioning that if there is other available prior knowledge about the system, the condition $\operatorname{rank}(M_o)=n$ can be relaxed. 
For instance, in scenarios when $mT<n$, it is common to assume that $x(0)$ is a sparse vector. 
Then, the state vector can be obtained by solving the compressed sensing problem, given by  
\begin{align}\label{eq:sparse_estimator}
\min _{x(0)}\|x(0)\|_{0}~~\text{s.t.},~\bm{y}_{T}=M_o x(0),
\end{align}
whose solution is characterized by the following result
\begin{theorem}[see \cite{candes2006stable}] \label{th:sparse_solution}
Let $\hat{x}(0;T)$ be the solution of problem \eqref{eq:sparse_estimator} and $q = \| \hat{x}(0;T)\|_{1}$ ($1\le q\le n$). 
If $mT>2q$ and all subsets of $2q$ columns of $M_o$ are full rank, then $\hat{x}(0;T)$ is unique. 
\end{theorem}

Theorem \ref{th:sparse_solution} reveals the possibility of obtaining a unique estimate of $x(0)$ when the observations are insufficient. 
In the literature, this type of problem is called sparse identification and has received increasing attention recently \cite{hayden2016sparse,dobbe2019blind,mao2022computational}.

\subsubsection{Local State Estimation}
Here we consider a more general situation where only local outputs of the system are available for the observer. 
Suppose the observer has access to the outputs of node $j$ and its in-neighbors $\mathcal{N}_j$, and it aims to estimate the initial state $x_i(0), i\in\mathcal{N}_j$. 
When the system is noise-free, the output of node $j$ is given by 
\begin{equation}\label{eq:local_output}
y_j(k)=C_j x_j(k)=C_j A_{[j,\tilde{\mathcal{N}}_j]} x_{\tilde{\mathcal{N}}_j}(k-1),
\end{equation}
where $\tilde{\mathcal{N}}_j=\{j\}\cup \mathcal{N}_j$, $A_{[j,\tilde{\mathcal{N}}_j]}=[A_{jl_1},\cdots,A_{jl_{|\tilde{\mathcal{N}}_j|}}]$ and $x_{[\tilde{\mathcal{N}}_j]}\!=\![x_{l_1},\cdots,x_{l_{|\tilde{\mathcal{N}}_j|}}]$. 
Recursively, it follows from \eqref{eq:local_output} that 
\begin{equation}\label{eq:local_output2}
y_j(k)=C_j [ A^k x(0) ]_{j}. 
\end{equation}
Note that the observer only knows the local information $C_i$, $A_{[j,\tilde{\mathcal{N}}_j]}$ and $x_{\tilde{\mathcal{N}}_j}$, the term $[ A^k x(0) ]_{j}$ in \eqref{eq:local_output2} when $k\ge2$ cannot be computed due to 
\begin{equation}
[ A^k x(0) ]_{j} \neq [A^k]_{[j,\tilde{\mathcal{N}}_j]} x_{\tilde{\mathcal{N}}_j}(0) \neq (A_{[j,\tilde{\mathcal{N}}_j]})^{*k} x_{\tilde{\mathcal{N}}_j}(0),
\end{equation}
where the power exponent symbol $(\cdot)^{*k}$ represents an element-wise power operation in the matrix. 
Since estimating $x_{i}(0)$ from the local information of node essentially requires global knowledge in general sense, it cannot be cast as a simple state observability problem, but can be treated as a joint problem combined with other techniques (e.g., system identification and Bayesian methods). 
More formally, the solvability of this problem heavily depends on the topology of the system.

\begin{theorem}[Necessity of local state estimate \cite{mo2017privacy,rezazadeh2018privacya,altafini2020systemtheoretic}] 
Consider the observation model \eqref{eq:local_output}. 
To accurately estimate $x_i(0)$ ($i\in\mathcal{N}_j$) by the information of node $j$, the following neighboring condition must be satisfied, given by 
\begin{equation}\label{eq:local_topology_condition}
\tilde{\mathcal{N}}_i \subseteq \tilde{\mathcal{N}}_j. 
\end{equation}
\end{theorem}

Note that the condition \eqref{eq:local_topology_condition} guarantees that node $j$ has complete information of its neighbor nodes. 
Under this situation, $x_i(0)$ can be estimated by node $j$ with high confidence. 
Similar to the global estimation case, the accuracy of $\hat{x}_i(0)$ can be directly represented by $\| \hat{x}_i(0)-x_i(0) \|$. 
Since the focus of this article is to characterize the mechanism secrecy under inference attack and discuss possible countermeasures, more detailed state estimation methods are omitted here.

In the literature, the security issues concerning the state estimation mainly focus on how to estimate the state from observations in the presence of sensor attacks, and further design secure controllers to alleviate the influence of the attacks. 
It is remarkable that the state estimation procedure in these works is conducted by the system itself, which has partial or complete knowledge about the nominal system parameters, e.g., matrices $A$ and $C$. 
In this article, we reveal that it is possible that these estimation methods can be leveraged by external adversaries to infer the sensitive state information of the system, causing severe threats to the system.

\subsection{Inferring the Topology Structure}\label{subsec:infer_struc}

The structure of NDSs contains two aspects: the coupling structure between different states of a single node, and the interaction topology between different nodes. 
For simplicity, we begin with this one-dimensional node state case and focus on the topology structure, by considering the following input-free global model
\begin{equation}\label{eq:model_example_1}
\begin{aligned}
x(k)&=Ax({k-1})+\omega({k-1}), \\
y(k)&=x(k)+v(k),
\end{aligned}
\end{equation}
where $\omega(k)$ and $v(k)$ satisfy Assumption \ref{assu:Gaussian}. 

\textit{Methods Review}. 
Inferring the interaction topology from observations over the NDSs emerges in various applications in last decades, including multi-robot formation \cite{liu2019dynamic}, social networks \cite{ahmed2009recovering}, and brain connectivity patterns \cite{monti2014estimating}. 
With the topology obtained, one can trace the information flow over a social network, or identity the critical node with maximum influence in a communication network. 
Generally, the feasibility of topology inference lies in that the states of the target systems are fully measurable, so as to establish an explicit expression between adjacent observations. 
Otherwise, one can hardly extract the underlying topology from direct observations.

\begin{table}[t]
\small
\caption{\label{tab:topology_methods} Representative methods of topology inference}
\begin{tabular}{lll}
\toprule[1pt]
\multicolumn{1}{c}{Methods}     & \multicolumn{1}{c}{Principles}    \\ \hline
\makecell[c]{ Granger estimator \\ e.g., \cite{granger1969investigating,brovelli2004beta,santos2019local}  } & \makecell[c{p{5cm}}]{utilize the node causality exhibited in the consecutive two states in expected sense. Multiple observation rounds over the dynamic process are needed, and the topology is symmetric}  \\ \hline
\makecell[c]{Spectral decomposition \\ e.g., \cite{segarra2017network,schaub2019spectral,zhu2020network} } & \makecell[c{p{5cm}}]{utilize the diagonalization of the sample matrices and reconstruct the symmetric topology by finding suitable eigenvalue/eigenvector pairs} \\ \hline 
\makecell[c]{Kernel method \\ e.g., \cite{karanikolas2016multi,karanikolas2017multi,wang2018inferring} } & \makecell[c{p{5cm}}]{suitable for nonlinear dynamic topology, key idea: select appropriate kernel basis functions to approximate the nonlinear dynamics}\\ \hline
\makecell[c]{ Sparse identification \\ e.g., \cite{hayden2016sparse,shahid2016fast,onuki2016graph} } & \makecell[c{p{5cm}}]{ consider the connections between nodes are sparse, and take the sparsity as a priori known} \\
\bottomrule[1pt]
\end{tabular}
\end{table}

Mathematically, topology inference can be regarded as an inverse problem. 
In the literature, a large body of research has been developed to tackle the problem due to their massive employments \cite{giannakis2018topology,brugere2018network}. 
Generally, the following four types of methods are commonly used: Granger estimator, spectral decomposition, kernel-based methods, and sparse identification. 
The basic principles and characteristics of these methods are summarized in Table \ref{tab:topology_methods}. 
These works mainly focus on symmetric topology and asymptotic inference performance.

For example, if multiple observation round are available over the system  (\ref{eq:global_model}), then the node directionality can be described by the Granger causality \cite{granger1969investigating,santos2019local}, given by 
\begin{equation}\label{eq:Granger}
{R_1^x}(t)  = WR_0^x(t\!-\!1), 
\end{equation}
where $R_0={\mathbb{E}}\left[ {{x_{t}} x_{t}^{\mathsf{T}}} \right]$ and $R_1={\mathbb{E}}\left[ {{x_{t}} x_{t-1}^{\mathsf{T}}} \right]$ are the autocorrelation and one-lag autocorrelation matrices. 
Accordingly, when $t\to\infty$, one can infer the topology by 
\begin{equation}
W={R_1^x}(\infty){(R_1^x)^{-1}}(\infty). 
\end{equation}
Note that this result is based on observations over multiple process rounds, and the observation noises are often ignored. 
In this part, we use the ordinary least square method to illustrate how to infer a directed topology from observations in a single round, and present their non-asymptotic performance.


\subsubsection{Global Topology Inference}

Since the topology to be inferred is represented by a matrix variable and to differentiate with the notations in the last subsection, 
we organize the state/observation/noise vectors of $T$ steps as matrices
\begin{equation}
\begin{aligned}
X_T^- &=[x_{0},x_{2},\cdots,x_{T-1}] ,~X_T^+ =[x_{1},x_{2},\cdots,x_{T}] , \\
Y_T^- &=[y_{0},y_{2},\cdots,y_{T-1}] ,~Y_T^+ =[y_{1},y_{2},\cdots,y_{T}] , \\
\Omega_T &=[\omega_0,\omega_1\cdots,~\omega_{T-1}],~\Upsilon_T =[\upsilon_1,\upsilon_2\cdots,\upsilon_{T}] .
\end{aligned}
\end{equation} 
Then, the whole dynamic process is compactly written as
\begin{equation}
\begin{aligned}
X_T^+ = A X_T^- +\Omega_T,~Y_T^+ = A X_T^+  +  V_T.
\end{aligned}
\end{equation}
For every two adjacent observations, it follows that
\begin{align}\label{eq:two-observation}
y_t&=Wx_{t-1}+\omega_{t-1}+v_t \nonumber \\
&=Ay_{t-1}-Av_{t-1}+\omega_{t-1}+v_t .
\end{align}
Note that (\ref{eq:two-observation}) only represents the quantitative relationship between adjacent observations, not a causal dynamical process. 
Then, the popular OLS estimator is derived solving the following problem
\begin{equation}
\begin{aligned}
\textbf{P}_\textbf{1}:~~
\mathop {\min }\limits_{A}\sum\limits_{t = 1}^{T} \| y_t-A y_{t-1} \|^2 \Rightarrow \mathop {\min }\limits_{A} \| Y_T^{+} - A Y_T^- \|_F^2
\end{aligned}
\end{equation}
Then, by finding the derivative, one obtains the optimal solution as 
\begin{equation} \label{OLS_estimator}
\hat A_o \!= \!Y_T^+ (Y_T^-)^\mathsf{T} (Y_T^- (Y_T^-)^\mathsf{T})^{-1}. 
\end{equation}
It should be noted that, the OLS estimator essentially treats all the terms in $y_t-Ay_{t-1}=-A\upsilon_{t-1}+\omega_{t-1}+v_t $ as interference noises. 
In fact, the observation noise will influence the inference performance of \eqref{OLS_estimator}. 

For simplicity, we define the following sample covariance matrix and its one-lag version as
\begin{equation}\label{eq:sample11}
\begin{aligned}
\Sigma_0(T)=\frac{1}{T}(Y_T^-) (Y_T^-)^\mathsf{T},~\Sigma_1(T)=\frac{1}{T}(Y_T^+) (Y_T^-)^\mathsf{T}.
\end{aligned}
\end{equation}
Then, a revised version of is given by the following theorem. 
\begin{theorem}[Causality in single observation round; see \cite{lys2021cdc}]\label{th:causality_estimator}
Considering the systems \eqref{eq:model_example_1} and given observations $\{ y_{t}\}_{t=1}^{T}$, if $A\in \mathcal{S}_a$, we have
\begin{equation}\label{eq:conclusion2}
\Sigma_1(\infty)=A (\Sigma_0(\infty) -\sigma_{\upsilon}^2I),
\end{equation}
where $\Sigma_1(\infty)=\mathop {\lim }\limits_{T \to \infty }\Sigma_1(T)$ and $\Sigma_0(\infty)=\mathop {\lim }\limits_{T \to \infty }\Sigma_0(T)$.
\end{theorem}

Different from the Granger causality in \eqref{eq:Granger}, Theorem \ref{th:causality_estimator} relaxes the dependence on multiple observation rounds, and presents the observation causality for a single round, while taking the observation noises into account. 
Then, given a finite horizon $T$, we propose the causality-based estimator as  
\begin{equation}\label{causality_estimator}
\hat A_c\!=\!\Sigma_1(T) (\Sigma_0(T) -\sigma_{\upsilon}^2I)^{-1}.
\end{equation}
\begin{remark}
We demonstrate that although the estimator $\hat A_c$ is derived from Theorem \ref{th:causality_estimator} where $A \in \mathcal{S}_a$ holds, it is also applicable when $A \in \mathcal{S}_m$. 
In fact, Theorem \ref{th:causality_estimator} is directly based on the Chebyshev inequality, where the bounded state constraint precludes us from proving the convergence and accuracy of $\hat A_c$ when $A \in \mathcal{S}_m$. 
To tackle this issue, we can resort to the concentration measure in Gaussian space. 
\end{remark}

Next, we explicitly characterize the convergence and accuracy of the two estimators. 
\begin{theorem}[Convergence speed and accuracy of $\hat{A}_o$ and $\hat{A}_c$; see \cite{li2021topoJourna}]\label{th:converge-speed}
Considering the systems \eqref{eq:model_example_1}, with probability at least $1-\delta$, the non-asymptotic bound of the OLS estimator $\hat{A}_o$ satisfies 
\begin{equation} \label{eq:wo_bound}
\| \hat{A}_o-A \| \sim\left \{
\begin{aligned}
&\mathcal{O}(\sqrt{ \frac{ \log{T} }{T} })+\mathcal{O}(\sigma_{\upsilon}^2),~&& \text{if}~A \in \mathcal{S}_m, \\
&\mathcal{O}(\frac{1}{\sqrt{T}})+\mathcal{O}(\sigma_{\upsilon}^2),~&&\text{if}~A \in \mathcal{S}_a. 
\end{aligned}\right.
\end{equation}
and the non-asymptotic bound of the causality-based estimator $ \hat{A}_c $ satisfies 
\begin{equation}
\| \hat{A}_c-A \| \sim\left \{
\begin{aligned}
&\mathcal{O}(\sqrt{ \frac{ \log{T} }{T} }),~&& \text{if}~A \in \mathcal{S}_m, \\
&\mathcal{O}(\frac{1}{\sqrt{T}}),~&&\text{if}~A \in \mathcal{S}_a. 
\end{aligned}\right.
\end{equation}
\end{theorem}
In terms of sample scale $T$, Theorem \ref{th:converge-speed} demonstrates the convergence speed of the inference error bound by using $\hat{A}_o$ and $\hat{A}_c$. 
We conclude that the extra cost for the estimators when $W\in \mathcal{S}_m$ is longer error converging time (or larger observation sample scale), requiring $\mathcal{O}(\sqrt{  \log{T} })$ times than that when $W\in \mathcal{S}_a$. 
In terms of accuracy, when $T\to \infty$, the inference error will converge to a constant by $\hat{A}_o$, while that of $\hat{A}_c$ will converge to zero, which shows the latter one has better inference accuracy.

Finally, we demonstrate the relationship of the causality-based estimator $\hat{A}_c$ with the Granger estimator $\hat{A}_g$ and OLS estimator $\hat{A}_o$, by revealing the equivalence condition of their observation matrices in single and multiple observation rounds, respectively. 
Taking $\|W\|_{F}^2$ as the regularizator, $\textbf{P}_\textbf{1}$ is transformed to
\begin{equation}
\begin{aligned}
\textbf{P}_\textbf{2}:~
\mathop {\min }\limits_{A}\sum\limits_{k = 1}^{T} \| y_k- A y_{k-1} \|^2+\beta\| A \|_{F}^2,
\end{aligned}
\end{equation}
where $\beta$ is a regularization parameter.
Then, we present the following theorem. 
\begin{theorem}[see \cite{lys2021cdc}]\label{th:equivalent1}
Considering the systems \eqref{eq:model_example_1}, for $\hat{A}_c$ and $\hat{A}_g$, if $A\in \mathcal{S}_a$, when $T\to\infty$, we have
\begin{equation}\label{eq:equivalent}
\Sigma_0(\infty)=R_0^x(\infty)+\sigma_{\upsilon}^2 I,~\Sigma_1(\infty)=R_1^x(\infty)+\sigma_{\upsilon}^2 A.
\end{equation}
For $\hat{A}_c$ and $\hat{A}_o$, $\hat A_c$ is equivalent to solving $\textbf{P}_\textbf{2}$ with $\beta=-\sigma_{\upsilon}^2$, which is a de-regularization form of $\hat A_o$. 
\end{theorem}
This theorem is significant in two aspects. 
First, it reveals that the expected state covariance matrix of $T\to \infty$ is identical with the sample covariance matrix along the single time horizon, which is an interesting result that describes the relationship between multiple and single observation rounds. 
Second, it provides a new interpretation for using LS methods to infer the interaction topology from the perspective of node causality.
Also, it shows the idea of how to set a reasonable regularization term and parameters for the LS problem modeling when both the input and output data are corrupted.

\subsubsection{Local Topology Inference}

Denote by $\mathcal{V}_{\sss F}$ the observable and $\mathcal{V}_{\sss F'}=\mathcal{V}\backslash\mathcal{V}_{\sss F}$ the unobservable subsets in $\mathcal{V}$, respectively. 
Given the system dynamics described by (\ref{eq:model_example_1}), the inference robot $r_a$ has only observations of $\mathcal{V}_{\sss F}\subseteq \mathcal{V}$. 
Based on this division, the state evolution in (\ref{eq:model_example_1}) is divided into 
\begin{equation}\label{eq:devide_state}
\left[ {\begin{aligned}
{x}_{k+1}^{\sss F}\\
{x}_{k+1}^{\sss F'}
\end{aligned}} \right] \!=\! \left[ {\begin{aligned}
A_{\sss FF}~A_{\sss FF'}\\
A_{\sss F'F}~A_{\sss F'F'}
\end{aligned}} \right]\left[ {\begin{aligned}
{x}_{k}^{\sss F}\\
{x}_{k}^{\sss F'}
\end{aligned}} \right] \!+ \!\left[ {\begin{aligned}
\omega_{k}^{\sss F}\\
\omega_{k}^{\sss F'}
\end{aligned}} \right],
\end{equation}
where ${x}_{k+1}^{\sss F'}$ is the state of the unobservable part $\mathcal{V}_{\sss F'}=\mathcal{V}\backslash\mathcal{V}_{\sss F}$. 
Note that the available information $\{ y_{k}^{\sss F}\}$ for the external observer satisfies  
\begin{equation}\label{eq:local_observation}
y_{k+1}^{\sss F}= {A_{\sss FF}} y_{k}^{\sss F} + \omega_{k}^{\sss F}  + {A_{\sss FF'}}x_{k}^{\sss F'} + v_{k+1}^{\sss F}-{A_{\sss FF}}v_{k}^{\sss F}, 
\end{equation}
which only represents the explicit relationship of every two consecutive observations, not a real process. 
Then, the local topology inference problem is to obtain the structure matrix $A_{\sss FF}$ from the observations $\{ y_{k}^{\sss F}\}$. 
It is intuitive that one may adopt a truncated version of \eqref{OLS_estimator} to calculate $A_{\sss FF}$, given by 
\begin{equation} \label{eq:truncated-estimator}
\hat A_{\sss FF} = Y_{T,F}^+ (Y_{T,F}^-)^\mathsf{T} (Y_{T,F}^- (Y_{T,F}^-)^\mathsf{T})^{-1}. 
\end{equation}
Unfortunately, \eqref{eq:truncated-estimator} is far from the strict least square solution by basic linear algebra, i.e., 
\begin{equation}
\hat A_{\sss FF} \neq [ Y_T^+ (Y_T^-)^\mathsf{T} (Y_T^- (Y_T^-)^\mathsf{T})^{-1} ]_{\sss FF}. 
\end{equation}
From \eqref{eq:local_observation}, the unequal effect is incurred by the non-negligible terms $\{  {A_{\sss FF'}}y_{k}^{\sss F'} \}$ even when $T\to \infty$. 
Hence, it is extremely hard to infer the local topology $A_{\sss FF}$ with high accuracy from noisy $\{ y_{k}^{\sss F}\}$. 
\begin{remark}
Under specific conditions, it is possible to obtain an (asymptotically) accurate estimate of the local topology by estimator (\ref{eq:truncated-estimator}). 
Recent works \cite{matta2018consistent,santos2019local,cirillo2021learninga} have made some progress in deriving the conditions, for example, i) the topology is in symmetric Erd\H{o}s-R{\'e}nyi random graph form with vanishing connection probability, 
and ii) the ratio of the observable nodes to all nodes converges to constant as the network goes to infinity. 
These methods cannot infer directed topology where the specified node conditions are not available. 
\end{remark}

To obtain a reliable local topology estimate, an alternative method is to shrink the inference range \cite{li2021topology}. 
Let $\mathcal{V}_{\sss H}$ be a subset of $\mathcal{V}_{\sss F}$ such that
\begin{equation}
\mathcal{N}_i \subseteq \mathcal{V}_{\sss F},~\forall i \in \mathcal{V}_{\sss H}. 
\end{equation}
Denote $\mathcal{V}_{\sss H'}=\mathcal{V}_{\sss F}\backslash{\mathcal{V}_{\sss H}}$ and $A_{\sss HF}=[A_{\sss HH}~A_{\sss HH'}]$. 
Then, the optimal estimation of ${W}_{\sss HF}$ in the sense of least square can be calculated by 
\begin{equation}\label{solution1}
\hat A_{\sss HF}= Y_{T,H}^+ (Y_{T,F}^-)^\mathsf{T} (Y_{T,F}^- (Y_{T,F}^-)^\mathsf{T})^{-1}, 
\end{equation}
which is free of the influence from the unobservable part $\mathcal{V}_{\sss F'}$. 
Note that a drawback of this method is that one needs to determine the subset $\mathcal{V}_{\sss H}$ first. 
In many situations where no more prior knowledge is available, one can select a small $\mathcal{V}_{\sss H}$ (e.g., single node) to obtain a relatively conservative estimation.

\subsection{Inference by Known Excitation: System Identification}
In this part, we point that if i) the system is driven by inputs which are known to the attacker, or ii) the attacker is able to inject self-designed excitation inputs into the system, then it is possible to reconstruct the state and structure from the input-output data, which can be regarded as a system identification (SI) problem.

In the context of SI, the attacker access the inputs and outputs of NDSs (e.g., the velocities and positions of mobile robot as the inputs and outputs, respectively), while the state $x$, the process noise $\omega$ and the measurement noise $v$ are unavailable. 
Conventionally, the following assumption is usually made in the literature for consistent and convergent identification results \cite{6213241}. 
\begin{assumption}\label{assum:system_identi}
The system \eqref{eq:global_model} is controllable, observable, and minimal realization with the system order known. 
The excitation input $u$ is persistently exciting, and the initial state is a zero vector. 
\end{assumption}

To describe the analytical relationship between the input and output, the Markov parameter matrix $G(T)$ is commonly used, which is defined as  
\begin{equation}\label{eq:Markov_Matrix}
G(T) = \left[ CB, CAB,\cdots, CA^{T-2}B, CA^{T-1}B \right].
\end{equation}
Upon defining the upper-triangular Toeplitz matrices corresponding to $\{u(k)\}$,
\begin{small}$$U_{T}=\begin{bmatrix}
u(0)& u(1) & u(2) & \cdots & u(T-1) \\
0 & u(0)& u(1) & \cdots & u(T-2) \\
0 & 0 & u(0)& \cdots & u(T-3)\\
\vdots & \vdots & \vdots & \ddots & \vdots \\
0 & 0 & 0 & \cdots & u(0) 
\end{bmatrix}.$$
\end{small}
\!\!Note that under Assumptions \ref{assu:Gaussian} and \ref{assum:system_identi}, the process and measurement noises are zero-mean and $x(0)=\bm{0}$. 
With these in mind, one can form the following least-squares problem,
\begin{equation}
\hat{G}=\mathop {\arg\min}\limits_{G\in\mathbb{R}^{m \times Tq}} \| {Y_T^-} - G {U_T}\|^2_{F}.
\end{equation}
To solve the above problem, many popular methods can be adopted, such as the subspace method. 
Ideally, one wishes the estimation error $\|\hat{G}-G\|$ to be as small as possible. 
Many recent works have achieved fruitful advances \cite{oymak2019non,simchowitz2019learning,zheng2020non}. 

With the solution $\hat{G}$ obtained, one can further use the celebrated Ho-Kalman Algorithm \cite{ho1966effective} and singular value decomposition techniques to estimate all the system parameters, including the matrices $A$, $B$ and $C$. 
It should be noted that this derived system model is a similarity transformation of the original system \eqref{eq:global_model} but shares the same inputs and outputs. 
Mathematically, it can be represented as 
\begin{equation}
\begin{array}{ll}
\left\{\begin{array}{l}\label{eq:sysmodel0}
\tilde{x}(k+1)=\tilde{A} \tilde{x}(k)+\tilde{B}u(k)+{\omega}(k), \\
y(k)=\tilde{C} \tilde{x}(k)+v(k),
\end{array}\right.
\end{array}
\end{equation}
where $\tilde{A} = P^{-1}AP$, $\tilde{B} = P^{-1}B$ and $\tilde{C} = CP$ (with $P \in \mathbb{R}^{n \times n}$ is a nonsingular matrix). 
Consequently, under the model \eqref{eq:sysmodel0}, the system state $x(0)$ is easy to estimate in the same manner as \eqref{eq:OLS_estimator} by using $\tilde{C}$ and $\tilde{A}$. 
For states $x(k),0<k<T$, the corresponding estimates can be easily obtained from the data in the slots $[k,T-1]$, by taking $x(k)$ as the new initial state of interest. 

Compared with directly inferring the state or structure of NDSs, the SI methods utilize the excitation input and enjoy the merits of not relying on much prior knowledge (like $A$ and $C$ need to be known in observability based methods). 
Besides, both the state and structure can be simultaneously inferred by first estimating the Markov matrix $G(T)$. 
We observe that the key of the SI methods is to approximate the real Markov matrix, not the real matrices $A$, $B$, $C$. 
In other words, the value accuracy of the obtained $\tilde{A}$, $\tilde{B}$, $\tilde{C}$ along with the corresponding states cannot be guaranteed, as they are essentially approximating the similarity transformations of the real ones. 
Therefore, directly using SI methods may be inappropriate if one aims to infer the real value of the system states and matrices. 
However, if one only cares for the system properties in matrix space (e.g., the supports and eigenvalues of $A$), then the SI methods is an effective alternative.

\begin{table*}[t]
\centering
\caption{\label{tab:state_results} A summary of topology inference results } 
\begin{tabular}{ccccc}
\toprule
Inference goal  &   Basic setting &  Estimator / Method  &   Performance / Conditions  \\
\midrule
Global state    &    \makecell[c]{   $(A,C)$ is known and observable }  &  $\hat{x}(0;T)= (M_o^{\mathsf{T}}M_o)^{-1}M_o^{\mathsf{T}}Y$  & \makecell[c]{ $\mathop {\lim }\limits_{T \to \infty } \Pr\{ \| \hat{x}(0;T)-x(0) \| = 0 \}=1$ } \\
\midrule
Local state     &  \makecell[c]{ $ \{ C_i, A\}$ is observable, consensus model } &  Maximum likelihood estimator & \makecell[c]{ $\tilde{\mathcal{N}}_i \subseteq \tilde{\mathcal{N}}_j$ is necessary }\\
\midrule
Global structure  & \makecell[c]{  Consensus-based model \eqref{eq:model_example_1} \\ only observations $\{y(k)\}$ are available }  &   $\hat A_c\!=\!\Sigma_1(T) (\Sigma_0(T) -\sigma_{\upsilon}^2I)^{-1}$    &   \makecell[c]{ $\mathop {\lim }\limits_{T \to \infty } \Pr\{ \| \hat A_c-A \| = 0 \}=1 $ }\\
\midrule
Local structure &  \makecell[c]{  $\exists \mathcal{V}_{\sss H} \subseteq \mathcal{V}_{\sss F} $, $\forall i \in \mathcal{V}_{\sss H},~\mathcal{N}_i \subseteq \mathcal{V}_{\sss F}$  }  & $\hat A_{\sss HF}= Y_{T,H}^+ (Y_{T,F}^-)^\mathsf{T} (Y_{T,F}^- (Y_{T,F}^-)^\mathsf{T})^{-1} $    & \makecell[c]{ $\mathop {\lim }\limits_{T \to \infty } \Pr\{ \| \hat{A}_{\sss HF}- A_{\sss HF} \| = 0 \}=1 $ }\\
\midrule
\makecell[c]{ Global joint \\ inference } &  \makecell[c]{ The system is controllable and observable \\ excitation input is available and known  }  &  \makecell[c]{ System identification procedure \\ infer Markov matrix $G$ }   & \makecell[c]{ The similar transformation of $(x,A,B,C)$ \\ are derived by Ho-Kalman Algorithm} \\
\bottomrule
\addlinespace[0.5ex]
\setlength\tabcolsep{0.5ex}
\vspace{-15pt}
\end{tabular}
\end{table*}

\section{Preserving The Control Mechanism Secrecy}\label{sec:secure_methods}

Since the control mechanism is highly vulnerable to inference attacks, it is very necessary to develop effective methods to preserve the control mechanism secrecy. 
In this section, we first present the main ideas of introducing secrecy and related method review. 
Then, we demonstrate how to measure the secrecy degree of the state and topology, and summarize possible techniques to preserve the control mechanism secrecy. 

\subsection{Methods Review}

In the literature, the encryption and noise-adding schemes are two mainstream approaches to secure the system information from being accurately inferred. 

\textit{Encryption}. 
The encryption-based mechanism is commonly used in many communication network systems, where each node sends cryptographic information and decodes the received using private/public keys (see \cite{acar2019survey} for a review). 
The encryption mechanism (especially homomorphic encryption) is a powerful tool to protect data privacy during data exchange and sharing, 
making the data computable by others while unrevealed to them \cite{kogiso2015cybersecurity,farokhi2017secure,schulzedarup2019encrypted,lu2018privacy,zhang2019admm}. 
As encryption by cryptographic algorithms is adopted during communication and computation (e.g., sensor measurements and controllable signals), it is effective against cyber attacks but not inference attacks. 
Since many NDSs which are physically open in the environment, even if the encrypted communication and computation scheme does not reveal the data, external observations of the system operation can still reveal the states/outputs, 
e.g., the displacements and velocities of a mobile robotic system can be directly observed and measured. 
Therefore, the encryption-based mechanism is not effective for the control mechanism secrecy of physically open NDSs.

\textit{Noise-adding}. 
The noise-adding scheme protects the system information by adding admissible random signals to the local dynamics. 
Due to its simple implementation, the noise-adding scheme has received considerable attention in recent years and is widely used for the security of NDSs \cite{mo2016privacy,geng2015optimal,han2018privacy,he2018privacy,he2019consensus}. 
Among these works, the notion of differential privacy has been widely applied to design the noise-adding method and analyze the performance \cite{dwork2006differential,cortes2016differential,he2020differential}. 
Differential privacy means that the presence or absence of any individual record in the database will not affect the statistics significantly. 
The formal definition is given as follows. 
\begin{definition}[$(\epsilon,\delta)$-differential privacy]
A randomized mechanism $\mathcal{A}$ with domain $\Omega$ is $(\epsilon,\delta)$-differentially private if, for any pair $x_1$ and $x_2$ $(x_1, x_2\in\Omega\subseteq \mathbb{R}^n)$ of $\sigma$-adjacent state vector and any set $\mathcal{O} \subseteq \operatorname{Ra}(\mathcal{A})$, where $\operatorname{Ra}(\mathcal{A})$ is the domain of the output under mechanism $\mathcal{A}$,
\begin{equation}
\Pr\{\mathcal{A}(x_1) \in \mathcal{O}\} \leq e^{\epsilon} \Pr\{\mathcal{A}(x_2) \in \mathcal{O}\}+\delta. 
\end{equation}
If $\delta=0$, we say $\mathcal{A}$ is $\epsilon$-differentially private. 
\end{definition}
However, due to the distributed and dynamical nature of NDSs, the additive noises on each node will adversely traverse in the system, cause perturbation on the dynamic process, and possibly make the resulting behavior deviate from the desired one. 
Therefore, the differential privacy may not be the perfect support to design the noise, especially when there is a tradeoff between the operational performance and security requirements \cite{katewa2018privacy}. 
For example, \cite{he2020differential} has proved that the commonly used average consensus in NDSs and $\epsilon$-differential privacy guarantees are impossible to be realized simultaneously. 
Based on the above analysis, an admissible noise-additive scheme should be capable of addressing the tradeoff between the desired dynamics and mechanism secrecy of NDSs, which is still an open issue (more details will be discussed in Section \ref{subsec:tradeoff}). 
In this article, we mainly focus on investigating the secrecy design using noise-adding methods and analyzing the performance.

\subsection{Preserving Initial State Secrecy by Adding Noises}\label{subsec:secure_state}

In this subsection, we consider the scenario where node $j$ wishes to estimate the initial state of node $i$, i.e., $x_i(0)$. 
To begin with, we focus on the consensus model where $u_i$ is absent and $x_i$ is one-dimensional. 
The results can be extended to general NDSs. 
Let $\theta_i(k)$ be the added random noise on node $i$ at iteration $k$, and $x_i^+(k)$ be the state sent out by node $i$ in iteration $k$, given by
\begin{align}\label{nod}
x_i^+(k)=x_i(k)+\theta_i(k). 
\end{align}
Note that $\theta_i(k)$ may be dependent of the noises before iteration $k$. 
When node $i$ receives the information from its neighbor nodes, its state is updated by
\begin{align}\label{eq:distri_noise}
x_i(k+1)=h_i(x_i^+(k), \mathcal{X}_i^+(k)),
\end{align}
where $\mathcal{X}_i^+(k)=\{x_j^+(k): j\in \mathcal{N}_i\}$ is the received neighbor information set, and $h_i$ represents the state-transition function of node $i$. 
The equation \eqref{eq:distri_noise} defines a distributed iteration algorithm that protects the state secrecy, since mixed random noises are used for state update in each iteration.

\subsubsection{Performance Metric For State Secrecy}
First, we present an uniform formulation for state secrecy. 
Define the output sequences of node $i$ in the running process until iteration $k$ by
\begin{equation}
\mathcal{I}_{i}^{out}(k)=\{x_i^+(0), ...,  x_i^+(k)\},
\end{equation}
Note that during the running process of any neighbor node $j\in N_i$, it can receive the information from not only node $i$, but also its own neighbors $\mathcal{N}_j^{in}$, which may contain common neighbors of $\mathcal{N}_i^{in}$ that help to estimate $x_i(0)$. 
Therefore, we define the information sequence of node $j$ to estimate node $i$ until iteration $k$ by
\begin{align}
I_j^i(k)&=\{x_i^+(0), x_\ell^+(0), ...., x_i^+(k), x_\ell^+(k) ~|~ \nonumber \\&~~~~~\ell=j~\text{or}~\ell\in N_i\cap N_j\},
\end{align}
Apparently, when $k=0$, node $j$ only has one-step information of node $i$, i.e., $I_j^i(0)=I_{i}^{out}(0)=\{x_i^+(0)\}$.

Next, considering $I_j^i(k)$ is the only available local knowledge to estimate node $i$, we define the $\epsilon$-optimal distributed estimation of $x_i(0)$ as follows ($\epsilon\geq 0$ is a small constant).

\begin{definition} [$\epsilon$-optimal distributed estimation]\label{opestde}
Let $\mathcal{I}_{\nu}^{out}(k)$ be the possible output given $x_i(0)=\nu$ at iteration $k$. Considering $\epsilon$-accurate estimate, under $\mathcal{I}_j^i(k)$,
\begin{align*}
\hat{x}_i^*(k)=\arg\max_{\hat{x}_i \in \mathcal{X}_i} \Pr\left\{\mathcal{I}_{\nu}^{out}(k)=\mathcal{I}_{i}^{out}(k)\mid  \forall |\nu-\hat{x}_i|\leq \epsilon  \right\},
\end{align*}
is named the $\epsilon$-optimal distributed estimation of $x_i(0)$ at iteration $k$. 
Then,  $\hat{x}_i^*=\lim_{k\rightarrow \infty} \hat{x}_i^*(k)$ is named the $\epsilon$-optimal distributed estimation of $x_i(0)$.
\end{definition}

Note that for any estimate $\hat{x}_i$ satisfying $|x_i-\hat{x}_i|\leq\epsilon$, we call it as a $\epsilon$-accurate estimate of $x_i(0)$. 
For simplicity, in the following parts, we drop the $\epsilon$ and directly call $\hat{x}_i^*$ as the optimal distributed estimation. 
Then, to further quantify the security degree of the noise-adding algorithm \eqref{eq:distri_noise} and derive the relationship between estimation accuracy and state secrecy, we introduce the following  $(\epsilon, \delta)$-data-secrecy definition.
\begin{definition}[$(\epsilon, \delta)$-state secrecy]
A distributed randomized algorithm is $(\epsilon, \delta)$-state-secret, iff
 \begin{equation}\label{deprivacy}
\delta=\Pr\{|\hat{x}_i^*-x_i(0)|\leq \epsilon\},
\end{equation}
where $\delta$ is the disclosure probability that the initial state $x_i(0)$  can be successfully estimated by others using the optimal distributed estimation in a given interval $[x_i(0)-\epsilon, x_i(0)+\epsilon]$.
\end{definition}

\begin{remark}
By calling $\hat{x}_i^*$ as the optimal estimation of $x_i(0)$, we mean that it is the best estimation of $x_i(0)$ that other nodes could ever obtain with the available information (e.g., the network topology and updating rule are known), regardless of the specific methods they use. 
Therefore, the $(\epsilon, \delta)$-state secrecy in fact depicts the largest probability that $x_i(0)$ is disclosed with tolerable error $\epsilon$. 
\end{remark}

\subsubsection{Secrecy Performance under Noise-adding Method}
As the initial states of the nodes are important and confidential variables for the system, we next focus on the performance analysis of inferring the initial states. 

\begin{theorem}[see \cite{he2018preservingTIT}] \label{theorem3ik}
Considering the noise-adding algorithm (\ref{eq:distri_noise}),  under $I_{j}^{i}(k)$, the optimal distributed estimation of $x_i(0)$ satisfies
\begin{align}\label{opex0sdik}
\hat{x}_i^*(k) &=x_i^+(0)- e_{\theta_i(0)|I_j^i(k)}(x_i^+(0)),
\end{align}
where
\begin{align}\label{ethetaik}
&e_{\theta_i(0)|I_j^i(k)}(x_i^+(0)) \nonumber \\
=&\arg\max_{y \in \{x_i^+(0)-\mathcal{X}_i\}}\int_{y-\epsilon}^{y+\epsilon}f_{\theta_i(1), ..., \theta_i(k)} (\tilde{\theta}_i(1), ..., \tilde{\theta}_i(k))\nonumber \\& f_{\theta_i(0)|\theta_i(k)=\tilde{\theta}_i(k), ..., \theta_i(1)=\tilde{\theta}_i(1)}(z) \text{d} z,
\end{align}
where $\tilde{\theta}_i(k)\!=\!x_i^+(k)\!-\!h_i(x_i^+(k\!-\!1), x_j^+(k\!-\!1)\!:\!j\!\in\! N_i)$, $f_{\theta_i(1), ..., \theta_i(k)} (\cdot)$ is the joint probability density function (PDF) of random variables $\{\theta_i(1), ..., \theta_i(k)\} $, 
and $f_{\theta_i(0)|\theta_i(k)=\tilde{\theta}_i(k), ..., \theta_i(1)=\tilde{\theta}_i(1)}(\cdot)$ is the conditional PDF of $\theta_i(0)$ when $\{\theta_i(k)=\tilde{\theta}_i(k), ..., \theta_i(1)=\tilde{\theta}_i(1)\}$.

\end{theorem}

Note that \eqref{ethetaik} is a general expression of the noise estimation, whose detailed form is closely related to the noise dependence and the connectivity relationships between different nodes.  
Next, we present the explicit expression of optimal $e_{\theta_i(0)|I_j^i(k)}(x_i^+(0))$ under $I_{j}^{i}(k)$ and three scenario conditions, respectively, as follows.

\begin{corollary}\label{th:error_estimation}
Considering the noise-adding algorithm \eqref{eq:distri_noise} and given $I_{j}^{i}(k)$, we have the following results of $e_{\theta_i(0)|I_j^i(k)}(x_i^+(0))$. 
\begin{enumerate}
\item If noises $\theta_i(0), ..., \theta_i(k)$ are independent of each other, 
\begin{align}\label{opex0sdck}
e_{\theta_i(0)|I_j^i(k)}(x_i^+(0))=e_{\theta_i(0)}(x_i^+(0)).
\end{align}

\item If $N_i\nsubseteq N_j$ for $\forall ~j\in N_i$ or the other nodes do not know all the information used for the updating by node $i$,
\begin{align}\label{gethetakkk}
&{e}_{\theta_i(0)|I_j^i(k)}(x_i^+(0))\nonumber \\
=&\arg\max_{y \in \{x_i^+(0)-\mathcal{X}_i\}}\int_{y-\epsilon}^{y+\epsilon}\oint_{\Theta_{\tilde{\theta}_i(1)|I_j^i(1)}}\cdot\cdot\cdot\oint_{\Theta_{\tilde{\theta}_i(k)|I_j^i(k)}} \nonumber \\&f_{\theta_i(1), ..., \theta_i(k)} (z_k, ...,z_1) f_{\theta_i(0)|\theta_i(k)=z_k, ..., \theta_i(1)=z_1}(z_0) \nonumber \\& \text{d} z_k \cdot\cdot\cdot \text{d} z_1 \text{d} z_0,
\end{align}
where $\Theta_{\tilde{\theta}_i(k)|I_j^i(k)}$ is the set of all possible values of $\tilde{\theta}_i(0)$. 

\item If $N_i\subseteq N_j$ and $N_i$ is known to node $j$,
\begin{align}\label{thetaek}
&e_{\theta_i(0)|I_j^i(k)}(x_i^+(0))= \arg\max_{y \in \{x_i^+(0)-\mathcal{X}_i\}}\int_{y-\epsilon}^{y+\epsilon}\nonumber
\\ &~~~~f_{\theta_i(0) | \theta_i(1)=\tilde{\theta}_i(1), ..., \theta_i(k)=\tilde{\theta}_i(k)}(z) \text{d} z.
\end{align}
\end{enumerate}
\end{corollary}

\begin{figure}[t]
\centering
\includegraphics[width=0.44\textwidth]{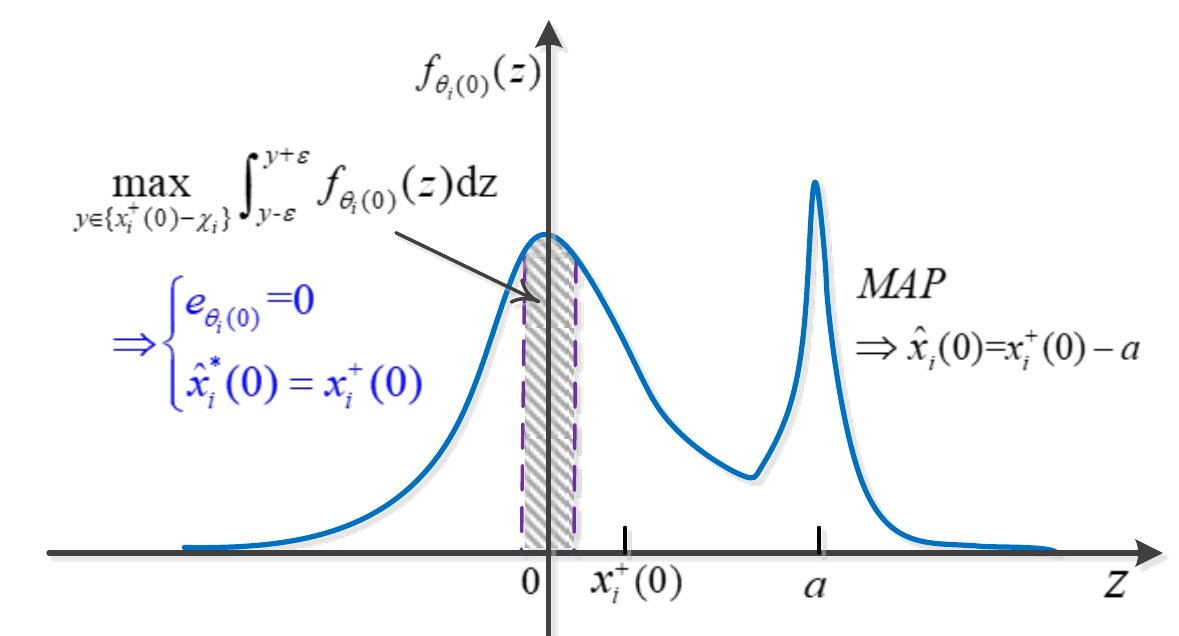}
\caption{The optimal distributed estimation of $e_{\theta_i(0)}(x_i^+(0))$ and the optimal estimation of $\theta_i(0)$ with MAP (source: \cite{he2018preservingTIT}). 
Note that under uniform prior distribution over $x_i(0)$, the estimate by MAP is the maximum value point of $f_{\theta_i(0)}(z)$, while the optimal estimate in the sense of $(\epsilon, \delta)$-state secrecy is the point that has the largest $\epsilon$-shaded area over $f_{\theta_i(0)}(z)$. } 
\label{fig:theta_example}
\end{figure}

From corollary \ref{th:error_estimation}, the first point is that if the added noises are mutually independent, the growing knowledge of data at following iterations will not help to estimate $x_i(0)$ better. 
Second, when $N_i\nsubseteq N_j$ for $\forall ~j\in N_i$ holds, node $j$ cannot obtain all the information of from $\mathcal{N}_i$ for node $i$'s state updating. 
Therefore, the exact value of $\tilde{\theta}_i(k)$ in \eqref{gethetakkk} cannot be obtained, i.e., $\tilde{\theta}_i(k)$ is not a deterministic value but in a possible value set $\Theta_{\tilde{\theta}_i(k)|I_j^i(k)}$. 
Third, if all information (including the global topology used during the iterative process) is known, the estimation accuracy of $x_i(0)$ may
be increased when the added noises are correlated with each other, i.e., the relationships of the noises can be used to decrease the uncertainty of $e_{\theta_i(0)|I_j^i(k)}(x_i^+(0))$. 
To better illustrate the meaning of $e_{\theta_i(0)|I_j^i(k)}(x_i^+(0))$, we present a simple example under $I_j^i(0)$. 
As we can see from Fig.~\ref{fig:theta_example}, supposing the prior distribution over $x_i(0)$ is uniform, $e_{\theta_i(0)}(x_i^+(0))$ locates the point $y$ which has the largest $\epsilon$-shaded area over $f_{\theta_i(0)}(z)$. 
Therefore, the estimation $e_{\theta_i(0)|I_j^i(k)}(x_i^+(0))$ is different from the classic maximum a posteriori estimation (MAP) problem that ignores the estimation accuracy of $|x_i-\hat{x}_i|$.

Next, define the set of available sequence $I_j^i(k)$ that can ensure $\epsilon$-accurate estimation, given by 
\begin{align}
\mathcal{S}_i(k)=&\{I_j^i(k)  \mid   |e_{\theta_i(0)|I_j^i(k)}(x_i^+(0)) -\theta_i(0)| \leq \epsilon\}.
\end{align}
Then, define $\mathcal{S}_i^1(k)$ be the set of all the first element in the sequence set $\mathcal{S}_i(k)$ (i.e., all  possible $x_i^+(0)$ included in $\mathcal{S}_i(k)$), and correspondingly define 
\begin{align}
\mathcal{S}_i^0(k)=&\{\theta_i(0) \mid  x_i^+(0) \in \mathcal{S}_i^1(k)\}.
\end{align}
Clearly, one has $\mathcal{S}_i^1(k)=x_i(0)+\mathcal{S}_i^0(k)$. 
The following theorem provides properties of the disclosure probability under $I_j^i(k)$, which is denoted by $\delta(k)$. 

\begin{theorem}[see \cite{he2018preservingTIT}]\label{theorem2ki}
Considering the noise-adding algorithm (\ref{eq:distri_noise}), the disclosure probability $\delta$ at iteration $k$ satisfies
\begin{align}\label{eq:prob_delta}
\delta (k) \leq  \oint_{\mathcal{S}_i^0(k)} f_{\theta_i(0)}(z) \text{d} z.
\end{align}
Specifically, if any of the following conditions holds,
\begin{enumerate}
\item the added noises are independent of each other;
\item $\Theta_{\tilde{\theta}_i(\ell)|I_j^i(\ell)}\supseteq \Theta_i$ or $\Theta_{\tilde{\theta}_i(\ell)|I_j^i(\ell)}=\mathcal{R}$ holds for $\ell=1,..., k$ and $\forall k\geq1$;
\end{enumerate}
then the equality in \eqref{eq:prob_delta} holds for $\forall k\geq 0$, i.e.,  
\begin{align}\label{priacygeneralcase}
\delta(k)=\delta=\oint_{\mathcal{S}_i(0)} f_{\theta_i(0)}(z) \text{d} z.
\end{align}
\end{theorem}

Theorem \ref{theorem2ki} provides the upper bound of disclosure probability under $I_j^i(k)$, and the conditions to reach the upper bound. 
Finally, we demonstrate how to design the added noises to secure the $(\epsilon, \delta)$-state secrecy. 
If node $j$ cannot know all the information used in the iterative process of node $i$ ($\delta=1$ if all the information is known to node $i$), 
then protecting the security of $x_i(0)$ can be transformed to solve an unconstrained minimization problem as follows,
\begin{equation}
 \begin{split}\label{problem:p2}
\min_{f_{\theta_i(0)}(y)} ~~ & \delta=\oint_{\mathcal{S}_i(0)} f_{\theta_i(0)}(y) \text{d} y, 
\end{split}
\end{equation}
which means that one needs to find an appropriate distribution form for $\theta$. 
Based on the conclusion in Theorem \ref{theorem2ki}, the value of $\delta$ is determined by set $\mathcal{S}_i(0)$, where $\mathcal{S}_i(0)$ is bounded by $\epsilon$. 
Therefore, it is very intuitive for one to design a $f_{\theta_i(0)}(y)$ with a large variance, such that $\delta$ is smaller than any given small value. 
For instance, considering $\mathcal{X}_i=\mathcal{R}$, one has $\mathcal{S}_i(0) =[e_{\theta_i(0)} - \epsilon, e_{\theta_i(0)} + \epsilon]$. Then, a uniform distribution with $ f_{\theta_i(0)}(y)\leq {1\over L}$ ($L$ is a constant) can ensure that
\begin{align}\label{eq:uni_case}
\delta=\oint_{\mathcal{S}_i(0)} f_{\theta_i(0)}(y) \text{d} y \leq {2\epsilon \over L}.
\end{align}
This example means that for any given $\delta$, one can always find a large $L$ that makes \eqref{eq:uni_case} hold to achieve $(\epsilon, \delta)$-state-security. 
As for the convergence of this kind of noise-adding algorithm, the readers are referred to \cite{mo2017privacy}.

\subsection{Making Future States Unpredictable}\label{subsec:unpredictability}
The historical states of NDSs contain the regularity of the state trajectory, which can be exploited by external observers for predicting future states. 
The future states reflect the trend about how the system will evolve, and need to be secret to outside. 
In this subsection, we investigate how to make the future state of a node unpredictable. 

Similar to the last subsection, we continue to consider the distributed algorithm \eqref{eq:distri_noise}, with a modification that an extra input $\eta_i(k)$ is introduced to enhance the unpredictability. 
Then, the modified algorithm is given by 
\begin{align}\label{eq:noise_input}
x_i(k+1)=h_i(x_i^+(k), \mathcal{X}_i^+(k)) + \eta_i(k),
\end{align}
where $x_i^+(k)$ is calculated by \eqref{nod}. 
Note that if $\eta_i(k)$ is a well-defined and regular function about $k$, the regularity will make the future state easy to be predicted by many mature methods e.g., ARIMA or RNN \cite{6213241}. 
Therefore, letting $\eta_i(k)$ be associated with randomness will be an effective method to enhance the unpredictability. 
Denote by $f_{\eta_i}(z)$ the probability density function of $\eta_i$, which satisfies
\begin{equation}\label{eq:theta_expectation}
\begin{aligned}
\mathbb{E}[\eta_i]=0,~\mathbb{D}[\eta_i]\leq\sigma_{\eta}^{2}.
\end{aligned}
\end{equation}

Next, we present the prediction model of $x_i(k+1)$. 
Slightly different from the scenario in the last subsection where node $j$ wishes to infer the state of node $i$, here an external attacker is assumed to have the observations of node $i$ and all its neighbors $\mathcal{N}_i^{in}$ till iteration $k$, which is represented by 
\begin{equation}
\mathcal{I}_{a}^{i}(k) = \{ \mathcal{I}_{i}^{out}(k), \mathcal{I}_{j}^{out}(k), j \in \mathcal{N}_i^{in} \}. 
\end{equation}
It is easy to see that the available local information $\mathcal{I}_{a}^{i}(k)$ no less than $\mathcal{I}_{j}^{i}(k)$, i.e., $\mathcal{I}_{j}^{i}(k) \subseteq \mathcal{I}_{a}^{i}(k)$. 
Based on $\mathcal{I}_{a}^{i}(k)$, let $\hat{h}_i(k)$ and $\hat{\eta}_i(k)$ be the (posteriori) estimates of ${h}_i(k)$ and $ {\eta}_i(k)$, respectively. 
In this article, we assume $\hat{h}_i(k)$ and $\hat{\eta}_i(k)$ are mutually independent, and their estimation errors are denoted as $\varepsilon_{h}(k)=\hat{h}_i- h_i$ and $\varepsilon_{\eta}(k)=\hat{\eta}_i(k)- {\eta}_i(k)$. 
Then, the attacker's one-step state prediction of node $i$ at iteration $k$ is given by  
\begin{equation}\label{eq:prediction}
 \begin{aligned}
 \hat x_i(k+1| \mathcal{I}_{a}^{i}(k) )= \hat{h}_i(k)  +\hat{\eta}_i(k),
 \end{aligned}
\end{equation}
whose prediction error is given by 
\begin{equation}
\varepsilon_{x}(k+1)=\varepsilon_{h}(k)+\varepsilon_{\eta}(k). 
\end{equation}

Apparently, the prediction accuracy is determined by the estimation error $\varepsilon_{h}(k)$ and $\varepsilon_{\eta}(k)$, and $\varepsilon_{\eta}(k)$ is closely related to the design of $\eta_i$. 
To find an appropriate design of $\eta_i$ and echo with the proposed $(\epsilon, \delta)$-state secrecy, we introduce the expectation $\mathbb{E}[ \| \varepsilon_{x}(k+1) \| ]$ and the probability measure $\Pr\{\|\varepsilon_{x}(k+1) \|\leq \epsilon^{2}\}$, respectively, as the optimization objectives to find the optimal $f_{\eta_i}(y)$. 
The problems are formulated as the following two problems
\begin{equation}\label{eq:problem1}
\begin{aligned}
\textbf{P}_\textbf{3}:\qquad&\max\limits_{f_{\eta_i}(z)}\min\limits_{\hat{\eta}_i(k)} \mathbb{E}[ \| \varepsilon_{x}(k+1) \|^2 ] \\
&\text{s.t.} ~~\mathbb{E}[\eta_i]=0,~\mathbb{D}[\eta_i]\leq\sigma_{\eta}^{2},\\
\end{aligned}
\end{equation}
and
\begin{equation}\label{eq:problem2}
\begin{aligned}
\textbf{P}_\textbf{4}:\qquad&\min\limits_{f_{\eta_i}(z)}\max\limits_{ \hat{\eta}_i(k) } \Pr\{\|\varepsilon_{x}(k+1) \|^2\leq \epsilon^{2}\} \\
&\text{s.t.} ~~\mathbb{E}[\eta_i]=0,~\mathbb{D}[\eta_i]\leq\sigma_{\eta}^{2}.\\
\end{aligned}
\end{equation}
Note that $\mathbb{E}[ \| \varepsilon_{x}(k+1) \|^2 ]$ reflects the mean square error of the attacker's prediction, 
while $\Pr\{\|\varepsilon_{x}(k+1) \|^2 \leq \epsilon^{2}\}$ describes the accuracy probability that the prediction error is within a preset bound. 

\begin{remark}
Formation of $\textbf{P}_\textbf{3}$ and $\textbf{P}_\textbf{4}$ can be interpreted as follows. 
The best protection can be achieved in two steps. 
The first is to decide the best prediction that an attacker could obtain, and the second step aims to ensure the best prediction is as unreliable as possible. 
These two steps can be seen as a game process between the NDS and the attacker. 
\end{remark}

\subsubsection{Optimal Solution of $\textbf{P}_\textbf{3}$}


Since $\varepsilon_{h}(k)$ and $\varepsilon_{\eta}(k)$ are mutually independent with each other, the objective function $\mathbb{E}[ \| \varepsilon_{x}(k+1) \|^2 ]$ in $\textbf{P}_\textbf{3}$ is expanded as 
\begin{align}\label{eq-c1}
\mathbb{E}[ \| \varepsilon_{x}(k+1) \|^2 ]= &\mathbb{E}\left[ \|\hat{h}_i(k) + \hat{\eta}_i(k)- {h}_i(k)-{\eta}_i(k) \|^{2} \right] \nonumber\\
  = &\mathbb{E}\left[ \varepsilon_{h}^2(k) \right] + \mathbb{E}\left[  \varepsilon_{\eta}^2(k)   \right] \nonumber\\
  = & \mathbb{E} \left[ \varepsilon_{h}^2(k) \right] +\mathbb{D}\left[  \hat{\eta}(k) \right] 
\end{align}
Then, we have the following result. 
\begin{theorem}[see \cite{li2020unpredictable}]\label{th:optimal_p1}
Considering the noise-adding algorithm \eqref{eq:noise_input}, $f_{\eta_i}(z)$ is the optimal distribution for $\textbf{P}_\textbf{3}$ iff
\begin{equation} \label{eq:max_var}
\mathbb{D}(\eta_i)=\sigma_{\eta}^{2}.
\end{equation}
\end{theorem}
The key point of deriving Theorem \ref{th:optimal_p1} is to utilize the independence of $\varepsilon_{h}(k)$ and $\varepsilon_{\eta}(k)$. 
Although the influence of $\varepsilon_{h}(k)$ is unknown, it does not affect the design of ${\eta}(k)$, which indicates that the larger $\mathbb{D}[\eta_i]$ is, the harder an attacker can make accurate predictions. 
This property is also consistent with our intuitions. 
There are two shortcomings by adopting $\mathbb{E}[ \| \varepsilon_{x}(k+1) \|^2 ]$ as the optimization objective. 
First, $\mathbb{E}[ \| \varepsilon_{x}(k+1) \|^2 ]$ only describes the mean prediction error, while $\| \varepsilon_{x}(k+1) \|^2$ may deviate a lot from the expectation error when $\mathbb{D}(\eta_i)$ is large. 
Second, one can only determine the variance of the optimal $\eta_i(k)$, not clear what specific distribution form the PDF of $\eta_i(k)$ is. 
When solving $\textbf{P}_\textbf{4}$, by contrast, the above two issues can be addressed under certain conditions.

\subsubsection{Optimal Solution of $\textbf{P}_\textbf{4}$}
When one aims to solve $\textbf{P}_\textbf{4}$, the major difference from solving $\textbf{P}_\textbf{3}$ is that the estimation accuracy of $\hat{h}_i(k)$ needs to be considered. 
Note that the objective function in $\textbf{P}_\textbf{4}$ can be expanded as 
\begin{align}
\Pr\{\|\varepsilon_{x}(k+1) \|^2\leq \epsilon^{2}\} = & \Pr\{ \|  \varepsilon_{h}(k)+\varepsilon_{\eta}(k)\|^{2}\leq\epsilon^{2}\}\nonumber \\
=&\iint_{\Omega_{\epsilon}} f_{z_{\epsilon}}(z_1,z_2)\,\mathrm{d}{z_1}\mathrm{d}{z_2},
\end{align} 
where $f_{z_{\epsilon}}(z_1,z_2)$ is the PDF of the sum $( \varepsilon_{h}(k)+\varepsilon_{\eta}(k) )$, and $\Omega_{\epsilon}=\{(\varepsilon_{h}(k),\varepsilon_{\eta}(k))| \| \varepsilon_{h}(k)+\varepsilon_{\eta}(k) \|^2 \leq \epsilon^{2} \}$. 
As we can see, the unpredictability is not only guaranteed by the $\eta_i$, but also closely related to the accuracy of $\hat{h}_i(k)$. 
Therefore, it is extremely hard to obtain a general and analytical solution of $\textbf{P}_\textbf{4}$. 
Despite this intractability, we are still able to give some useful results under certain conditions, which is presented in the following theorem.

\begin{theorem}[see \cite{li2020unpredictable}]\label{th:prob2}
Consider the noise-adding algorithm \eqref{eq:noise_input}. 
For $\textbf{P}_\textbf{4}$, if $\hat{h}_i(k)$ is accurate, i.e., $\varepsilon_{h}(k) = 0$, then $\exists \bar{\epsilon}>0$, $\forall \epsilon\le \bar{\epsilon}$, the optimal distribution of $f_{\eta_i}(z)$ is the uniform distribution with finite maximum variances, i.e.,
\begin{equation}\label{uniform_theta}
f^{*}_{\eta_i}(z)=\left\{
\begin{aligned}
\centering
&\frac{1}{2\sqrt{3}\sigma_{\eta} },&&z\in [-\sqrt{3}\sigma_{\eta},\sqrt{3}\sigma_{\eta}]. \\
&0,&&\text{otherwise}.
\end{aligned}
\right.
\end{equation}
If $\hat{h}_i(k)$ is not accurate but deterministic, then $\exists \bar{\epsilon}>0 $, $\forall \epsilon\le \bar{\epsilon}$, $f_{\theta_i}(z)$ is optimal when the distribution of the sum $(\varepsilon_{h}(k)+\varepsilon_{\eta}(k))$ is the uniform distribution with variance $\sigma_{\eta}^{2}$. 
\end{theorem}

Theorem \ref{th:prob2} illustrates when the considered error bound $\epsilon$ is sufficiently small and $\hat{h}_i(k)$ is accurate, the uniform distribution with the maximum variance is optimal for $f_{\eta_i}(z)$. 
\begin{remark}
The solutions of $\textbf{P}_\textbf{3}$ and $\textbf{P}_\textbf{4}$ are consistent in the sense that they both require the variance to be maximum, but solving $\textbf{P}_\textbf{4}$ can further give the specific PDF form of ${\eta_i}(k)$. 
In addition, considering the parameter $\epsilon$ represents the prediction error bound of an attacker, if one can still guarantee the unpredictability under a small $\epsilon$ (i.e., $\Pr\{\|\varepsilon_{x}(k+1) \|^2\leq \epsilon^{2}\} $ is also small), then it demonstrates the noise-design of ${\eta_i}(k)$ works well. 
Therefore, it is meaningful to directly consider the situation where $\epsilon$ is sufficiently small. 
\end{remark}

It is worth noting that if $\varepsilon_{h}(k) \neq 0$, the optimal distribution of $f_{\eta_i}(z)$ relies on the characteristic of $\hat{h}_i(k)$. 
Specifically, if $\varepsilon_{h}(k)$ is also with randomness, then the solution would be much more complicated than that of a deterministic $\varepsilon_{h}(k)$ (as claimed in Theorem \ref{th:prob2}). 
Despite the coupling of $\varepsilon_{h}(k)$ and $\varepsilon_{\eta}(k)$ , we can further illustrate that the error $\varepsilon_{h}(k)$ will not affect the influence of $\varepsilon_{\eta}(k)$ on $\Pr\{\|\varepsilon_{x}(k+1) \|^2\leq \epsilon^{2} \}$. 
Denote by $\hat{h}_i^*(k) $ the optimal estimate of $\hat{h}_i(k) $, and let $\hat{\eta}_{i,1}^*(k)$ and $\hat{\eta}_{i,2}^*(k)$ be the optimal noise predictions associated with $\hat{h}_{i,1}^*(k)$ and $\hat{h}_{i}^*(k)$ ($\hat{h}_{i,1}^*(k) \neq \hat{h}_{i}^*(k) $), respectively. 
Then, we have the following corollary. 

\begin{corollary}\label{coro:degrade}
Consider the noise-adding algorithm \eqref{eq:noise_input} and $\varepsilon_{h}(k) \neq 0$. 
Given arbitrary PDF $f_{\eta_i}(z)$, $\forall \epsilon>0$, one has 
\begin{align}
 &\Pr\left\{  \|\varepsilon_{x}(k+1) \|^2\leq \epsilon^{2} |f_{\eta_i}(z), \hat{\eta}_{i,1}^*(k),\hat{h}_{i,1}(k) \right \} \nonumber \\
 \leq & \Pr\left\{\|\varepsilon_{x}(k+1) \|^2\leq \epsilon^{2} |f_{\eta_i}(z),\hat{\eta}_{i,2}^*(k), \hat{h}_i^*(k) \right \}. 
\end{align}
\end{corollary}

Intuitively, this corollary illustrates that even if there exists an estimation error in $\hat{h}_i(k)$, it will not degrade the secrecy performance brought by $\eta_i(k)$, i.e., not making the accuracy probability $\Pr\{\|\varepsilon_{x}(k+1) \|^2\leq \epsilon^{2}\} $ higher. 

Based on the above analysis, we present a brief summary of achieving the unpredictability of the states. 
First, a distribution with maximum variance for $f_{\eta_i}(z)$ is optimal in the sense of expected prediction error. 
By contrast, in the sense of $\epsilon$-accurate probability, the uniform distribution with maximum variance for $f_{\eta_i}(z)$ is optimal if $\epsilon$ is small and $\hat{h}_i(k)$ is accurate. 
Second, if i) the state of a node is high-dimensional, ii) the estimates $\hat{h}_i(k)$ and $\hat{\eta}_i(k)$ are correlated, or iii) one wishes to achieve the unpredictability of long-horizon trajectory, finding an optimal design to tackle these situations is extremely hard due to the complicated joint PDF. 
How to design better noise-adding methods and evaluation metrics for these situations still remains an open issue.

\subsection{Securing the Topology Structure of NDSs}\label{subse:sec_topo}
Considering the risk that an external attacker can infer the topology of NDSs based on observations, it is necessary to protect the topology from being accurately inferred while maintaining the normal convergence of the system dynamics. 
In Section \ref{subsec:secure_state} and \ref{subsec:unpredictability}, we have demonstrated that the noise-adding algorithm can be utilized to protect the state secrecy. 
In fact, the idea of adding noises to the states during the iteration is also promising to secure the topology, along with a more advanced design to guarantee the convergence performance meanwhile. 
Therefore, we continue to use the distributed noise-adding algorithm \eqref{eq:noise_input} for the following discussions. 

Since noises $\theta_i(k)$ and $\eta_i(k)$ in \eqref{eq:noise_input} are mutually independent, for legibility, we consider the case $\theta_i(k)=0$ and focus on the design of $\eta_i(k)$. 
Then, the global form of \eqref{eq:noise_input} with $\theta_i(k)=0$ can be explicitly rewritten as  
\begin{equation}\label{eq:global_model2}
x(k)=Ax({k-1})+\eta({k-1}), 
\end{equation}
where $\eta=[\eta_1, \cdots, \eta_N]^{\mathsf{T}}$, and the attacker has direct collections of states $\{ x(k) \}$\footnote{Here we ignore the observation noise term $v(k)$, as it is also independent with $\eta(k)$. 
By the method introduced in Section \ref{subsec:infer_struc}, its influence can be asymptoticall alleviated.}. 
Ideally, when there is no noise input injected into the system, the state will converge to a constant vector as $k\to\infty$, i.e., 
\begin{equation}
    \lim \limits_{k\rightarrow \infty } {x}(k)= x_c,
\end{equation}
where the converging state $x_c$ is determined by the setup of $A$ and the initial state $x(0)$. 
For example, if $A$ is doubly stochastic, $x_c=(\bm{1}^{\mathsf{T}}x(0) )/N $. 
Next, we present the fundamental conditions for $\eta(k)$ to guarantee the normal convergence of the system. 
\begin{theorem}[see \cite{he2019consensus}]
Consider the system \eqref{eq:global_model2}, 
the convergence of the system is guaranteed iff
\begin{equation}\label{eq:zero_convergence}
\lim_{k \to \infty}\sum_{l = 0}^{k-1} W^{k-l-1} \eta(l) = 0. 
\end{equation}
Specifically, for cases where $A$ is doubly stochastic, the convergence of \eqref{eq:global_model} is guaranteed if
\begin{equation}\label{eq:zero_convergence2}
\|\eta(k)\|_{\infty}\le \alpha \rho^k,~\text{and}~\sum\limits_{k = 0}^{\infty} \sum\limits_{i= 1}^{N} \eta_i(k)=0, 
\end{equation}
where $\alpha>0$ and $\rho\in[0,1)$. 
\end{theorem}
Note that \eqref{eq:zero_convergence} is the sufficient and necessary condition for the exact convergence to $x_c$. 
The condition \eqref{eq:zero_convergence2} in fact provides a general noise-adding method for a class of systems where $A$ is doubly stochastic. 
Following this property, one can develop an adjacent noise cancellation algorithm to approximate $\sum\limits_{k = 0}^{\infty} \sum\limits_{i= 1}^{N} \eta_i(k)=0$, while making the magnitude of $\eta_i(k)$ decay to zero over time \cite{he2018privacy}. 
In practice, an auxiliary random term $\xi$ is used, and $\eta_i(k)$ is commonly designed as 
\begin{equation}
\eta_i(k)=\xi_i(k)-\xi_i(k-1), 
\end{equation}
where $\| \xi_i(k) \| \le \frac{\alpha \rho^k}{2}$. 
Note that the above results can be easily extended to the cases when $\theta_i(k)\neq0$, which should also satisfy the zero-sum and decaying conditions. 
The general procedures are summarized as Algorithm \ref{algo:noise}, which iteratively updates the state until the stopping time $k_{\max}$.

\begin{algorithm} [t]
\caption{: Adjacent noise cancellation algorithm} \label{algo:noise}
{\small{
\begin{algorithmic}[1]
\REQUIRE{$x_i(0)$, $\{ a_{ij}, j\in\mathcal{N}_i^{in} \}$, $\alpha$, $\rho$ and $k_{\max}$. }
\ENSURE{$x_i(k_{\max})$ for each $i\in\mathcal{V}$.}
\STATE  Initialize $x_i^{+}(0)$ by \eqref{nod} and send it to $\mathcal{N}_i^{out}$. Meanwhile, receive the information $\mathcal{X}_i^+(0)=\{x_j^{+}(0),j\in \mathcal{N}_i^{in}\}$. 
\STATE Initialize $\eta_i(0)\!=\!\xi_i(0)$ such that $\xi_i(0)\!\in\![-\frac{\alpha}{2},\frac{\alpha}{2}]$, and set $k\!=\!1$. 
\WHILE {$k\le k_{\max}$}
{
    \STATE Select the auxiliary noise $\xi_i(k)$ from $[-\frac{\alpha \rho^{k} }{2},\frac{\alpha \rho^{k} }{2}]$ by the user-designed rules. 
    \STATE Set $\eta_i(k)=\xi_i(k)-\xi_i(k-1)$.
   \STATE Update $x_i(k)=h_i(x_i^+(k-1), \mathcal{X}_i^+(k-1)) + \eta_i(k)$.  
   \STATE Update $x_i^+(k)$ and send it to $\mathcal{N}_i^{out}$. 
   \STATE $k=k+1$.
}
\ENDWHILE
\end{algorithmic} } }
\end{algorithm}

\begin{remark}
Algorithm \ref{algo:noise} ensures the exact convergence when $A$ is doubly stochastic. 
This is because the row-stochasticity can ensure that each input vector will converge to a constant, and the column-stochasticity can keep the sum of the input vector unchanged with iterations. 
Meanwhile, each node does not need any global system information or the noise information of its neighbors. 
\end{remark}

Note that in Algorithm \ref{algo:noise}, the magnitude of the added noise $\eta_k$ is strictly bounded. 
From the perspective of the PDF of $\eta_k$, this condition can be relaxed to various PDF forms while ensuring the variance of $\eta_k$ exponentially decaying with $k$ \cite{mo2016privacy}. 
Under this formulation, the mean square convergence of the state is still guaranteed. 
Besides, there is another meaningful result that connects the structure and state secrecy. 
If one uses the popular differential privacy to measure the secrecy performance of the state, there exists a contradiction as summarized in the following corollary. 
\begin{corollary}[see \cite{he2020differential}]\label{th:contradiction}
Consider the system \eqref{eq:global_model2}. 
If the added noises satisfy the exact convergence condition $\lim_{k \to \infty}\sum_{l = 0}^{k-1} W^{k-l-1} \theta(l) = 0$, then the noise-adding algorithm cannot guarantee the the differential privacy of the initial state. 
\end{corollary}

Next, we illustrate how to add the noises  to protect the topology matrix $A$ from being inferred accurately. 
Considering that the attacker uses the aforementioned OLS estimator \eqref{OLS_estimator} to infer the topology, the inference error matrix is given by
\begin{equation}
\varepsilon_A = \hat{A}_o - A = \Phi_{T} (Y_T^-)^\mathsf{T} (Y_T^- (Y_T^-)^\mathsf{T})^{-1},
\end{equation}
where the added noise matrix $ \Phi_{T}\!=\![\eta(0),\eta(1),\cdots,\eta(T\!-\!1)]$. 
Then, finding the optimal noise can be formulated as the following maximization problem 
\begin{subequations}\label{eq:topo_secure}
\begin{align}
\mathop{\max }\limits_{\Phi_{T}}~~& \| \varepsilon_A\|_{F}^2  \label{eq:defense-a}\\
{\rm{s.t.}}~~& \lim \limits_{k\rightarrow \infty } {x}(k)= x_c. \label{eq:defense-b}
\end{align}
\end{subequations}
It is remarkable that the added $\eta(k)$ will continue to influence the following states after iteration $k$, and the objective function is highly nonlinear about $\Phi_{T}$. 
Therefore, directly solving \eqref{eq:topo_secure} in one-step is intractable. 
However, we can still approximate the optimal design in an iterative way, which is turned to solve the following problem at each iteration
\begin{subequations}\label{eq:iteration_secure}
\begin{align}
\textbf{P}_\textbf{5}:\quad &\mathop{\max }\limits_{\eta(k)}~~ \left\|  \left[ \Phi_{k-1},\eta_{k} \right ] (Y_k^-)^\mathsf{T} (Y_k^- (Y_k^-)^\mathsf{T})^{-1} \right \|_{F}^2  \label{eq:iteration_secure-a}\\
&{\rm{s.t.}}~~ \|\eta(k)\|_{\infty}\le \alpha \rho^k . \label{eq:iteration_secure-b}
\end{align}
\end{subequations}
Note that in the objective function of $\textbf{P}_\textbf{5}$,  $\Phi_{k-1}$ and $Y_k^-$ are deterministic and given at iteration $k$. 
Then, we have the following result. 
\begin{theorem}[see \cite{ztw2022acc}]\label{th:optimal_topo_noise}
Considering the system \eqref{eq:global_model2}, if the noise $\eta_i(k)\in [\beta_i^-(k), \beta_i^+(k)]$ where $\max\{|\beta_i^-(k)|, |\beta_i^+(k)|\}\le \alpha \rho^k$, then the optimal solution of $\textbf{P}_\textbf{5}$, ${\eta^{*}(k)}$, satisfies 
\begin{equation}\label{eq:optimal_noi}
\eta_{i}^{*}(k)\in\{\beta_i^-(k), \beta_i^+(k) \}, \forall i=1,\cdots, N. 
\end{equation}
\end{theorem}

Theorem \ref{th:optimal_topo_noise} shows that the added noise $\eta(k)$ is optimal when each node locally selects the boundary magnitude of its available noise space. 
Note that by using Algorithm \ref{algo:noise} with the noise design \eqref{eq:optimal_noi}, the exact convergence cannot be guaranteed if $A$ is not doubly stochastic. 
Under this situation, there will be a deviation between the real state $x(\infty)$ and the converging state $x_c$, and the bound of the deviation is determined by the parameter $\alpha$ and $\rho$. 
How to achieve the exact convergence for non-doubly stochastic $A$ while securing the topology still needs further investigation.

\section{Discussion and Open Issues}\label{sec:discussions}

\subsection{The Secrecy of Control Laws}
Note that the control law is the utmost critical element of the control mechanism to influence the dynamical process, which acts as the system input. 
An attacker who has inferred the the control law can easily infer the state and structure by appropriate input modeling. 
The control input of NDSs can be divided into two types: system-depend input (rely on the state and structure) and independent input. 
For the former one, if the controller has an explicit form, then the whole system can be represented in a closed-loop form, where the input $u$ is replaced by the expression of $x$ (or $y$) and $A$, e.g., the classic state-feedback controller. 
In this context, it is possible to infer the control laws by utilizing the state and structure information, and the state-of-art data-driven methods \cite{yuan2019data} are beneficial to the investigation of this direction. 
For instance, consider the Linear Quadratic Regulator (LQR) problem\cite{anderson2007optimal}. The optimal control law that solves
\begin{equation}\label{eq:LQR}
\begin{split}
    \min_{u,y} ~ & \lim_{T\to \infty} \frac{1}{T} \mathbb{E}\left[\sum_{k=0}^{T} \left(x(k)^{\top}Qx(k) + u(k)^{\top}Ru(k)\right)\right] \\
    \text{s.t.} ~ & x(k+1) = Ax(k) + Bu(k) + w(k)
\end{split}
\end{equation}
is given by
\begin{equation}\label{eq:LQR_control_law}
    u(k) = - (B^{\top}PB+R)^{-1}B^{\top}PAx(k) \triangleq -K^*x(k),
\end{equation}
where $P$ is the solution of the discrete algebraic Riccati equation that relies on $A, B, Q, R$\cite{anderson2007optimal}. Hence, the inference of \eqref{eq:LQR_control_law} is translated to the estimation of the optimal feedback matrix $K^*$ based on the observations of the state-input pairs $\{x(k),u(k)\}$. Then, various linear regression methods\cite{montgomery2021introduction} can be applied. 
As for the independent input, if the input is time-invariant, we can identify the influence of the input by using the data when the system is stable \cite{jiao2021topology}. 
However, if the independent input is time-varying, a more sophisticated input-identification procedure is required, and the latest input-output linearization \cite{zheng2021equivalence} technique may provide some feasible solutions, which still needs further investigation. 
Besides, if the external attacker has only local access to the global system, decentralized versions of the inference methods are required to obtain the local knowledge \cite{xymdec2022acc}. 

Another interesting scenario is that one can use self-designed control laws to excite the system, such that more abundant behaviors are produced to infer the system parameters. 
The latest work \cite{xyminput2022acc} has made progress in improving the inference accuracy of system models by input design, which utilizes the subspace method in system identification. 
It is promising that we can further the designed excitation input to differentiate the influence of the control laws of the system itself. 
In addition, from the defense perspective, it is also of necessity to protect the control laws from being inferred, e.g., controller gains. 
Several latest works \cite{kawano2020design,kawano2021modular,nekouei2022randomized,yazdani2022differentially} have achieved some progress in this direction, by considering a specific class of NDSs.  
As for securing the optimal control law \eqref{eq:LQR_control_law} for problem \eqref{eq:LQR}, we can similarly exploit the noise-adding mechanism by perturbing states, inputs, or both of them with uncorrelated Laplace noises or correlated adjacent-canceling noises. 
The performance can be characterized via differential privacy\cite{dwork2006differential} or $(\epsilon, \delta)$-secrecy.
However, it remains challenging to conceal the general model type and design secure controllers that could induce a much worse inference result for the attacker.

\subsection{Performance Metrics for Control Mechanism Secrecy}

Establishing appropriate metrics to evaluate the risks is critical for the control mechanism secrecy. 
In previous sections, we have used two kinds of metrics: regression error like \eqref{eq:estimation_error} and disclosure probability like \eqref{deprivacy}. 
They both reflect the difference between the estimated value and the ground truth, but also differ from each other. 
The regression error directly describes the inference error of a realization of the inference attack, which is only suitable for cases where the noise-adding method is deterministic. 
On the contrary, the disclosure probability characterizes the risk of information leakage by probability measurement, which applies to situations where the noise-adding method is with randomness. 
If the explicit distribution characteristic of the added noises is hard to describe, one can adopt the expected square error as an alternative, which is given by 
\begin{equation}
E_r(\hat{x})=\mathbb{E} [( \hat{x} -x )^\mathsf{T}(\hat{x} -x)]. 
\end{equation}
Note that $E_r(\hat{x})$ gives the estimation error in the expectation sense, which is suitable for both deterministic and random cases of the system states. 

In terms of the unpredictability of the state, current research mainly focus on the inference performance of one-step prediction \cite{he2018preservingTIT,li2020unpredictable}, which is not enough for secrecy in the long horizon. 
Suppose the attacker aims to predict $K$-step future state of the NDS, then the probability that the prediction error of each step is within $\epsilon$ can be represented as  
\begin{equation}\label{def:pred}
    \mathcal{P}_\epsilon(x_{[1:K]}) = \prod\limits_{k=1}^K \Pr\{\|\hat{x}(k|k-1)-x(k)\|_\infty\leq\epsilon\},
\end{equation}
where $\hat{x}(k|k-1)$ is the prediction of $x(k)$ at step $k-1$. 
Apparently, ensuring the state secrecy on the long horizon is much more challenging. 
On the one hand, the inferred target is no longer a single point but a trajectory composed of multiple points. 
Consequently, the uncertainty is described by multivariate probability density functions in high-dimension space, making the corresponding calculation extremely hard, let alone the possible couplings between different variables. 
On the other hand, even if the uncertainty is calculable, the result highly relies on a given system model along with its observations. 
A reliable metric for state unpredictability is desired to be independent of specific models and observations, and is able to characterize convergence trend of $\mathcal{P}_\epsilon(x_{[1:K]})$ as $K$ increases. 
How to design appropriate and analytical transformations to fill the above gap is worth further study. 
To address these challenges, an interdisciplinary approach combining model predictive control, information theory and large deviation theory may be promising. 

Last but not least, the evaluation metrics for states also provide the foundation to characterize the metrics for the secrecy of the structure and the control laws. 
Different from the state secrecy, the other two metrics will be closely related to the dynamical performance of NDSs (e.g., convergence and steady points \cite{hawkins2020differentially}). 
We discuss the details of these issues in the next subsection.

\subsection{Tradeoff Between Secrecy and Cooperation Performance}\label{subsec:tradeoff}

Although the noise-adding methods help to achieve the state secrecy, they may compromise the normal dynamical performance of NDSs. 
It is often required that the system state at certain moment should meet the preset value, apart from the secrecy requirement of state evolution. 
For example, suppose the desired state at the terminal moment $T$ is $x_{T}^*$ and we can formulate the control objective as the following linear quadratic function
\begin{align}\label{eq:desired_value}
\!\!J_c(\tilde{u}_{0:T})=&\mathbb{E}[(x_T-x_T^*)^{\mathsf{T}}H(x_T-x_T^*)] \nonumber \\ 
& + \mathbb{E}\left[\sum_{k=0}^{T-1} \left(x(k)^{\top}Qx(k) + \tilde{u}(k)^{\top}R\tilde{u}(k)\right)\right],
\end{align}
where $\tilde{u}(k)=u(k)+\eta(k)$ is the system input after adding noise $\eta(k)$, $H$ is a positive semi-definite matrix, $Q$ and $R$ are the same as that in \eqref{eq:LQR}. 
It is clearly from \eqref{eq:desired_value} that achieving the optimal trade-off between the desired terminal state and secrecy of the state evolution requires the random noises $\{\eta(k)\}$ to be well designed. 
What kind of distribution of $\eta(k)$ should be determined and how to design the input $u(k)$ coupling with $\eta(k)$ are the main challenging issues for future research. 

Note that in practice, the structure of the system is also of great importance for system convergence and stability. 
Therefore, simultaneously considering the structure disclosure risks is essential to investigate the tradeoff between secrecy and cooperation performance. 
First, the metrics considering the state aspects are not always appropriate to describe the inference capability. 
For example, in some cases, the structure matrix is not easy to obtain by using an explicit model, and one may be more interested in the eigenvalue spectrum of the matrix  \cite{segarra2017network,zhu2020network}. 
Also, the original model information of the control laws may not be priorly known, and one needs to design other evaluation models to deal with the uncertainty. 
The established metrics are supposed to be capable of describing the error influence in the system dynamics. 
Therefore, there remain many challenges in designing appropriate metrics to fully characterize the tradeoff between mechanism secrecy and cooperation performance. 

Second, due to the fundamental convergence requirement, the feasible space to design countermeasures is limited. 
Specifically, as we mentioned in Section \ref{subse:sec_topo}, when the structure matrix is not doubly stochastic, there remains a challenge to tackle with the exact convergence of the system state. 
The key point lies in eliminating the asymptotical state deviation without relying on global information \cite{ztw2022acc}. 
Furthermore, if the nodes are also associated with state-independent self-input, the coupling between the self-input and the added noise needs to be considered for the optimal noise design. 
Last but not the least, different from the state where one only needs to focus on the value itself, the system structure is composite system knowledge. 
Taking the topology matrix $A$ as an example, although the value of each element in $A$ is sensitive and important, the eigenvalue spectrum and the eigenvectors are also critical and valuable information, which reflects the convergence performance of the system dynamics. 
Therefore, simply using the error norm as in \eqref{eq:topo_secure} is not sufficient for the structure secrecy. 

\subsection{Cases of Unknown Models: Incorporating with Reinforcement Learning and Model-free Optimization}
In recent years, plenty of works have displayed increasing interests in designing optimal, stable and safe controllers from data \cite{levine2018reinforcement,de2020formulas,hewing2020learning}. 
These methods remove the design dependency on the model knowledge (e.g., specific model forms), where state-of-the-art reinforcement learning and model-free optimization techniques are used as main tools \cite{chen2019reinforcement,westenbroek2020learning,westenbroek2021combining,he2022model}. 
We point out that, when there exist interactions between the NDSs and the environment, it is possible that the online learning strategy for the system itself can also be leveraged by external attackers to infer the control mechanism. 
For example, we can formulate this process as the following model
\begin{equation}\label{eq:unknown_model}
\begin{aligned}
x(k+1) &=f(x(k),u(k),u^e(k)), \\
y(k)&= g(x(k),u(k),u^e(k)),
\end{aligned}
\end{equation}
where $u^e$ is the input describing the interaction between the attacker and the NDS. 
Here, only the observations $\{y(k)\}$ and the interaction input $\{u^e(k)\}$ are available to the attacker, and the mature tools in linear system cases can no longer be utilized. 
A promising insight of inferring the mechanism is to design appropriate inference objective function $\Phi(y,u^e)$ for the attacker, construct gradient estimates of $\Phi(y,u^e)$ to obtain a sequence of the interaction input $\{u^e(k)\}$, thus optimizing the inference objective function $\Phi(y,u^e)$ iteratively. 
Note that optimizing $\Phi(y,u^e)$ is essentially optimizing the inference error about the control dynamics. 
Therefore, the main direction of this problem research is to properly define the inference objective $\Phi(y,u^e)$ at the beginning and design the interaction strategy $u^e$ during the process.

\section{Numerical Evaluation}\label{sec:simulations}
In this section, we present some numerical tests to verify the theoretical results and evaluate the performance of some approaches considered in this article. 

\subsection{Scenario Setting}
For simplicity, we consider a full measurable NDS with 5 nodes, which subject to the following consensus-based model
$$
x_{k+1}=Ax(k),
$$
where
\begin{small}
$$
A=\begin{bmatrix}
0.9111 & 0.0444  &0       &0       &0.0444 \\
0.0444 & 0.8000  &0.0667  &0.0222  &0.0667 \\
0      & 0.0667  &0.9111  &0.0222  &0\\
0      & 0.0222  &0.0222  &0.9111  &0.0444\\
0.0444 & 0.0667  &0       &0.0444  &0.8444
\end{bmatrix},
$$
\end{small}
\!\!is a doubly stochastic matrix, and the initial state is $x(0)=[-26,-3,13,28,17]^\mathsf{T}/2$. 
One can easily obtain that $A\in\mathcal{S}_a$, and the converging state of this system $x_c=2.9$. 
For comparison of different noise-adding methods, we will selectively use Gaussian, uniform, and Laplace noises to secure the mechanism secrecy. 

\begin{figure}[t]
\centering
\includegraphics[width=0.45\textwidth]{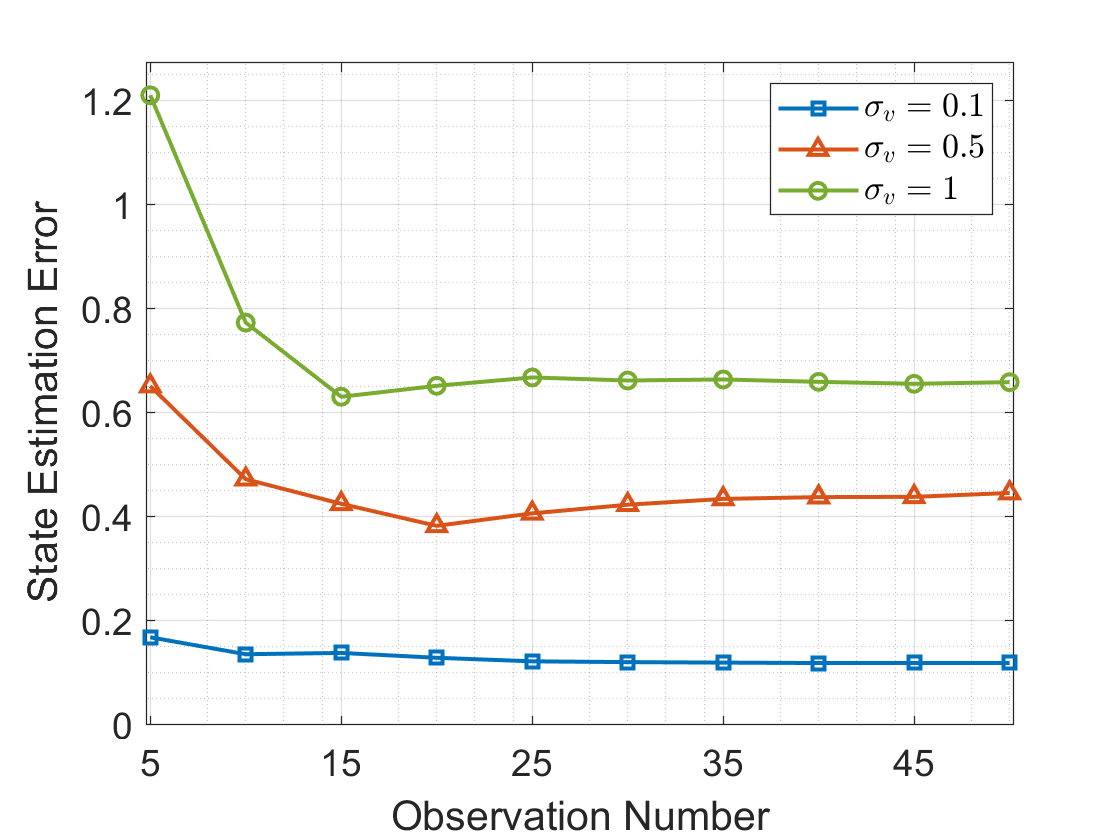}
\caption{The estimation error of $x(0)$. 
The measurement noises are Gaussian with zero mean and variances equaling to $0.1^2$, $0.5^2$ and $1$, respectively. } \label{fig:infer_state}
\end{figure}

\subsection{Verification}

\begin{figure}[t]
\centering
\includegraphics[width=0.45\textwidth]{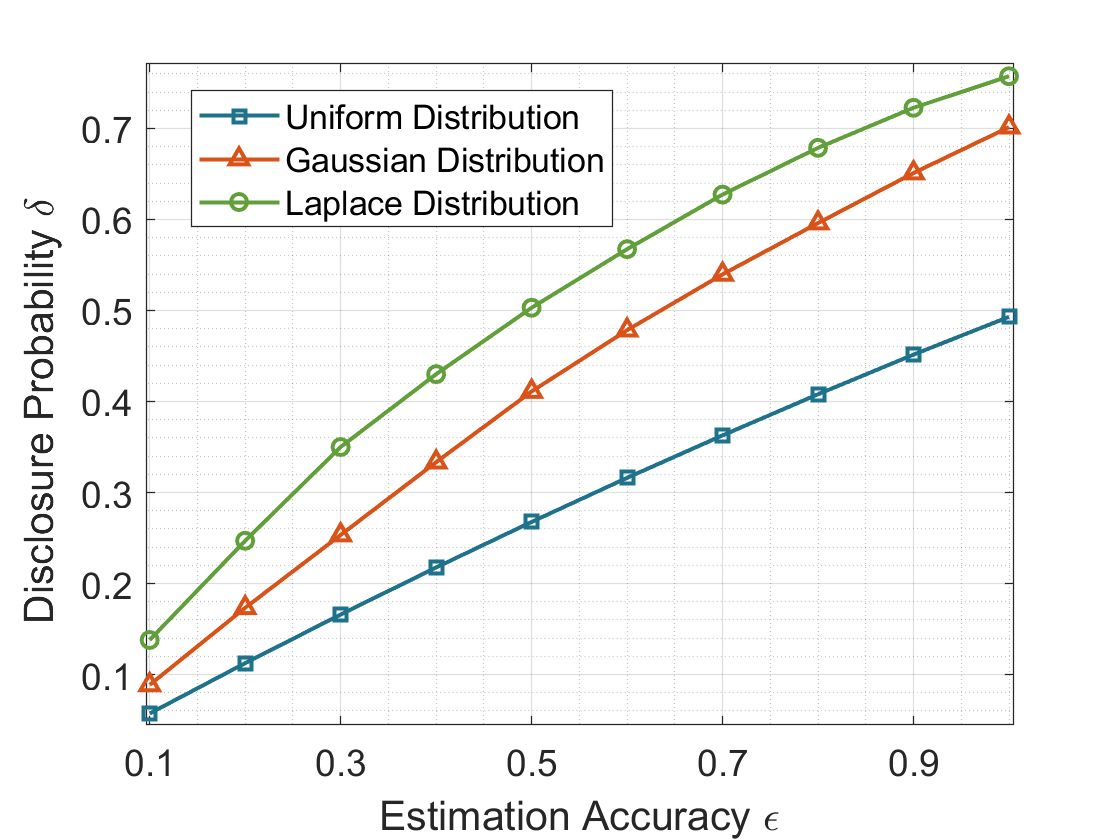}
\caption{The $(\epsilon, \delta)$-state secrecy performance, where the added noises are drawn from Gaussian, uniform and Laplace distributions, respectively. 
The process and measurement noises both satisfy zero mean and unit variance. } \label{fig:secure_state}
\end{figure}

Let us begin by examining the state secrecy. 
It is straightforward that the initial state $x(0)$ can be estimated accurately if there are no other noises involved in the system. 
Therefore, we illustrate the performance of the state estimation under different levels of noise variance $\sigma_v^2$. 
Fig.~\ref{fig:infer_state} compares the inference accuracy of the state estimator \eqref{eq:OLS_estimator}. 
Clearly, the smaller the noise variance $\sigma_v^2$ is, the smaller the estimation error is. 
Meanwhile, the inference error will decrease with the observation number and become stable. 
Fig.~\ref{fig:secure_state} further shows the $(\epsilon, \delta)$-state secrecy performance of a single node. 
Note that in this test, $3000$ simulation runs are conducted, where the Gaussian, uniform, and Laplace distributed noises are used with both zero mean and unit variance.  
In each run, one node first randomly generates a noise $\theta_i(0)$ with the given distribution, while the other node obtains $\hat{\theta}_i(0)$ by some estimation methods. 
Then, one can have the probability that $|\hat{\theta}_i(0)-\theta_i(0)|\le \epsilon$ holds. 
Clearly, one can conclude that uniform distribution is better than Gaussian and Laplace distribution in the sense of $(\epsilon, \delta)$-state secrecy. 

\begin{figure}[t]
\centering
\includegraphics[width=0.45\textwidth]{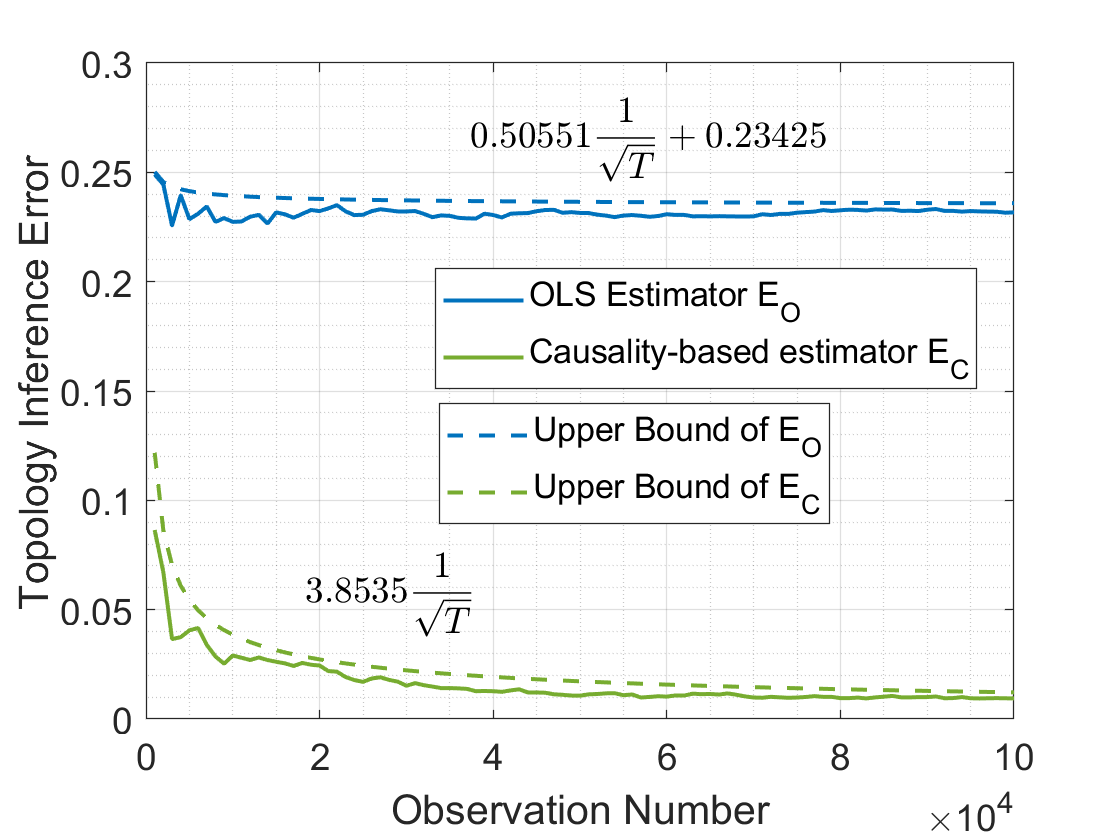}
\caption{The topology inference performance the OLS and the causality-based estimators. 
Note that the process and measurement noises both satisfy zero mean and unit variance. } 
\label{fig:infer_topo}
\end{figure}

Next, we focus on the structure secrecy performance. 
Similar to the state estimation, if there is no noise involved in the dynamic process, the attacker can accurately infer the topology matrix $A$ with more than $T\ge(N+1)$ groups of observations. 
Therefore, here we validate the inference performance of the estimator $\hat{A}_o$ in \eqref{OLS_estimator} and $\hat{A}_c$ in \eqref{causality_estimator}, by considering both process and measurement noises satisfying Assumption \ref{assu:Gaussian} with unit variances. 
As shown in Fig.~\ref{fig:infer_topo}, the inference errors of both $\hat{A}_o$ and $\hat{A}_c$ decrease as the observation number grow, which illustrates the risk that the structure of a NDS is easy to be revealed with tolerable error. 
Specifically, the inference error of $\hat{A}_c$ converges to zero while that of $\hat{A}_o$ converges to constant. 
This property echoes with the conclusions in Theorem \ref{th:converge-speed}, and can be observed from the upper bounds of the errors drawn in Fig.~\ref{fig:infer_topo}.

\begin{figure*}[t]
\centering
\subfigure[Convergence of system states.]{\label{topo_secure_a}
\includegraphics[width=0.34\textwidth]{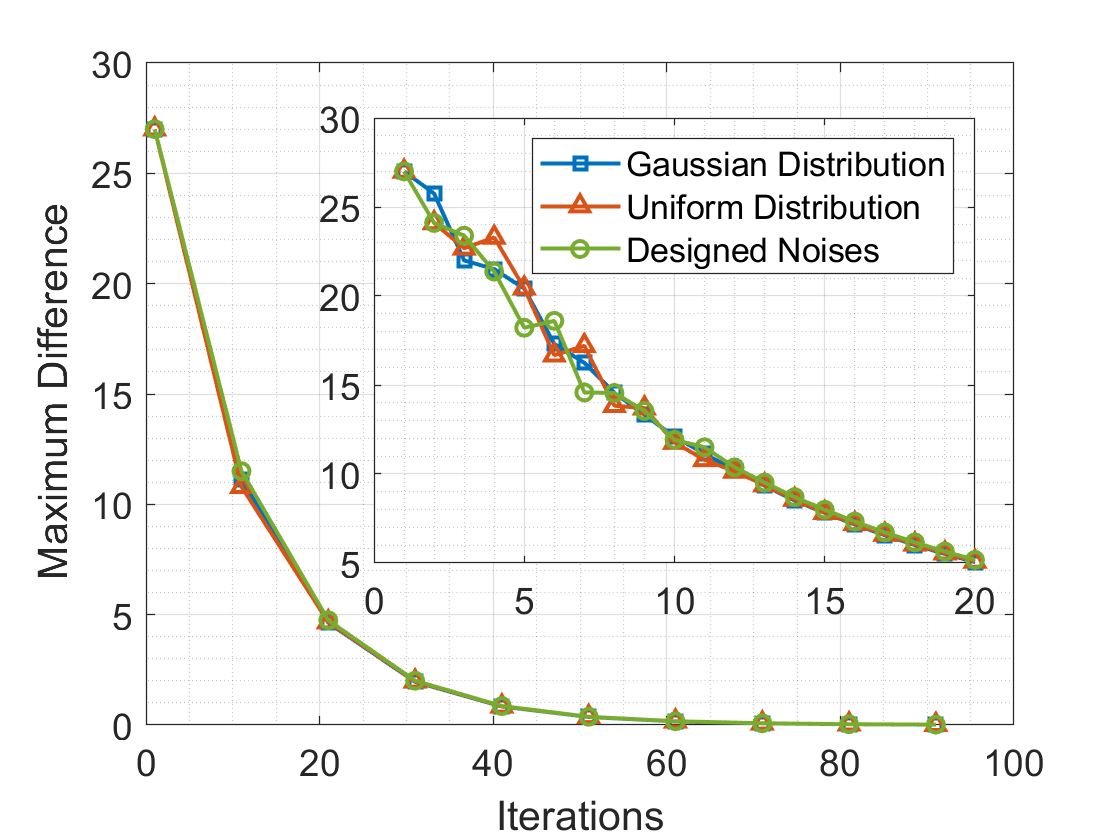}}
\hspace{-0.7cm}
\subfigure[Accuracy of the converging state.]{\label{topo_secure_b}
\includegraphics[width=0.34\textwidth]{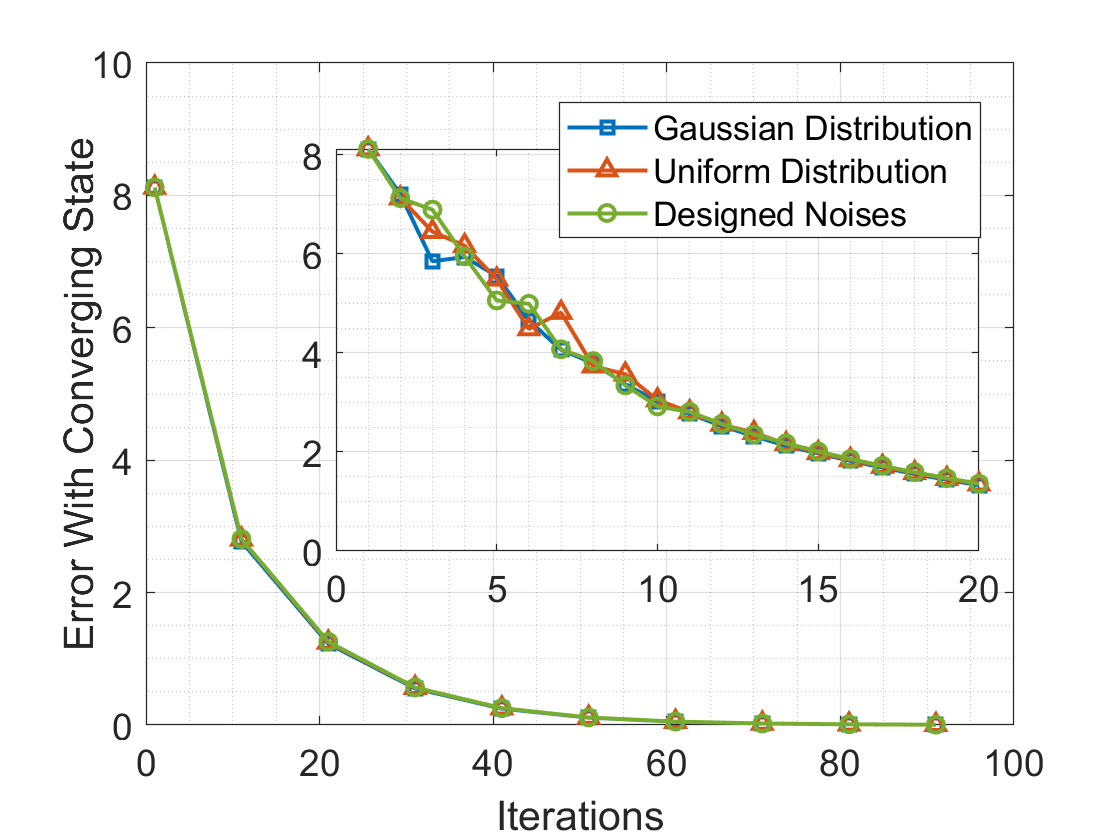}}
\hspace{-0.7cm}
\subfigure[Inference error of the topology.]{\label{topo_secure_c}
\includegraphics[width=0.34\textwidth]{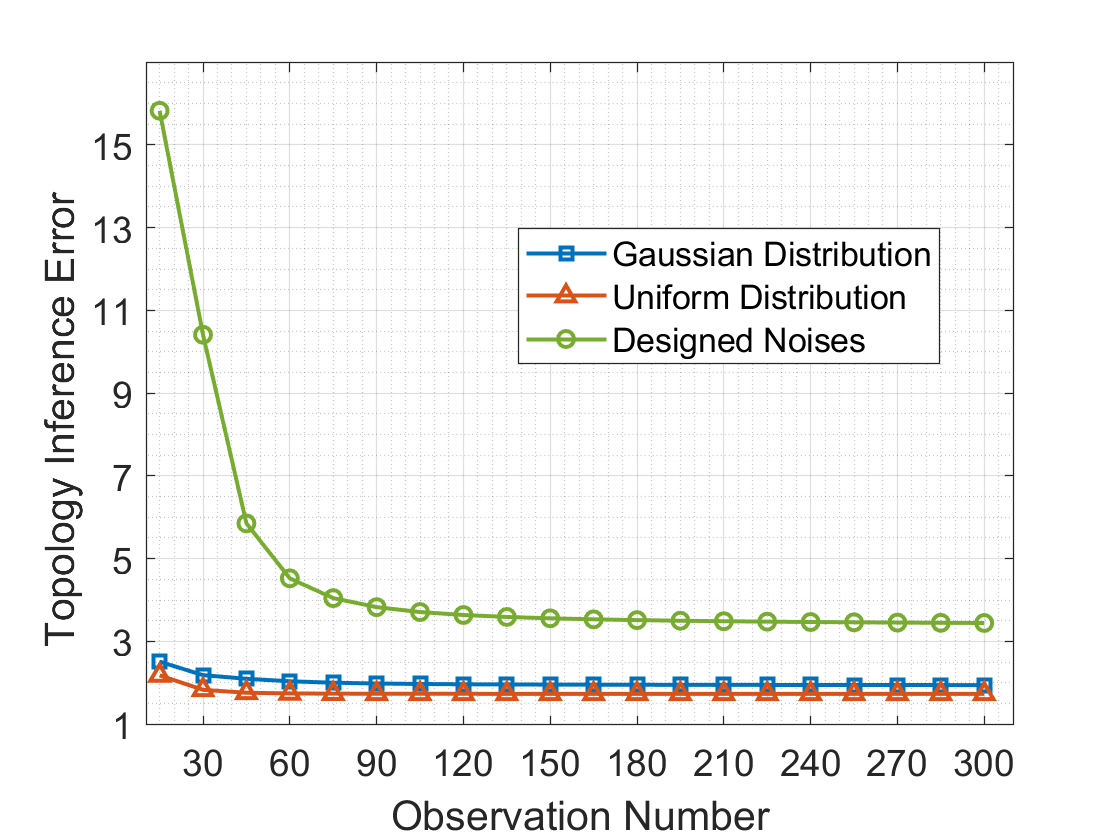}}
\caption{The system evaluation and the topology inference performance under the adjacent noise cancellation methods with three different noise distributions. 
The inset plots in (a) and (b) shows the detailed evaluation process before iteration 20. }
\label{fig:topo_secure}
\end{figure*}

Finally, we turn to validate the secrecy performance of the adjacent noise cancellation methods. 
Fig.~\ref{fig:topo_secure} the state evaluation and topology inference error when using Gaussian, uniform, and the optimal noises in $\textbf{P}_\textbf{5}$, respectively. 
The simulations here are conducted by letting the noises satisfy the exponentially decaying condition and the zero-sum condition $\sum\limits_{k = 0}^{\infty} \sum\limits_{i= 1}^{N} \eta_i(k)=0$. 
Note that according to famous $3\sigma$-rule \cite{pukelsheim1994three}, for Gaussian distribution with zero mean, the variable locates in $[-3\sigma,+3\sigma]$ with probability $0.997$, and this interval is basically regarded as the actual possible range. 
To have a fair comparison, we let the bound of the uniform and designed noises of $\textbf{P}_\textbf{5}$ be $3\sigma_{\eta}\rho^{k}$ at iteration $k$, where the decaying parameter $\rho$ is set as $0.95$. 
Clearly, one can observe from Fig.~\ref{topo_secure_a} and Fig.~\ref{topo_secure_b} that, both the convergence and accuracy of the system state are guaranteed under the adjacent noise cancellation methods. 
Remarkably, even if the convergence speed of the three kinds of noises are likewise, the topology inference errors differ a lot, as shown in Fig.~\ref{topo_secure_c}. 
Specifically, the inference error brought by the designed noise is much higher than that of the other two, which exhibits better topology-securing performance and verifies the optimality conclusions in Theorem \ref{th:optimal_topo_noise}.

\section{Concluding Remarks}\label{sec:conclusions}

This article explored the disclosure risks of the fundamental information in NDSs along with their countermeasures. 
The control mechanism of a NDS consists of confidential information including state, structure, and control laws. 
The control mechanism secrecy, in our opinion, plays a critical role in the applications of NDSs, especially in those security-related scenarios. 
We introduced the inference analysis from an attacker's view, and the protection methods from a defender's view. 
Based on the analysis, we demonstrated the inference performance of common inference methods concerning how to estimate the state and structure, and presented the results about the principles and possible methods to evaluate and secure them, such as the $(\epsilon, \delta)$-state secrecy metric and the adjacent noise cancellation algorithm. 
This collection of the latest results can be regarded as a launching pad for deeper investigation into more sound security methods. 
Furthermore, we provided in-depth discussions on the secrecy of control laws, and summarized potential ideas to infer and protect the control laws, respectively.

However, we also pointed out that even the state-of-the-art approaches in the literature cannot fully address the issues of optimal protection designs in the context of mechanism secrecy. 
Considering the situations where i) the system model is nonlinear or high-dimensional, ii) the prior knowledge about NDSs is not sufficient, 
or iii) the tradeoff between nominal system performance and the control mechanism secrecy, establishing better evaluation metrics about control mechanism and combining the excitation-based design are promising directions beckoning further research.

\section*{Acknowledgement}
The authors would like to thank Zhiyu He, Jialun Li, Xiangyu Mao and Tao Xu for their valuable comments and suggestions for this paper. 



\begin{thebibliography}{100}
\providecommand{\url}[1]{#1}
\csname url@samestyle\endcsname
\providecommand{\newblock}{\relax}
\providecommand{\bibinfo}[2]{#2}
\providecommand{\BIBentrySTDinterwordspacing}{\spaceskip=0pt\relax}
\providecommand{\BIBentryALTinterwordstretchfactor}{4}
\providecommand{\BIBentryALTinterwordspacing}{\spaceskip=\fontdimen2\font plus
\BIBentryALTinterwordstretchfactor\fontdimen3\font minus
  \fontdimen4\font\relax}
\providecommand{\BIBforeignlanguage}[2]{{%
\expandafter\ifx\csname l@#1\endcsname\relax
\typeout{** WARNING: IEEEtran.bst: No hyphenation pattern has been}%
\typeout{** loaded for the language `#1'. Using the pattern for}%
\typeout{** the default language instead.}%
\else
\language=\csname l@#1\endcsname
\fi
#2}}
\providecommand{\BIBdecl}{\relax}
\BIBdecl

\bibitem{olfati2007consensus}
R.~Olfati-Saber, J.~A. Fax, and R.~M. Murray, ``Consensus and cooperation in
  networked multi-agent systems,'' \emph{Proceedings of the IEEE}, vol.~95,
  no.~1, pp. 215--233, 2007.

\bibitem{oh2015survey}
K.-K. Oh, M.-C. Park, and H.-S. Ahn, ``A survey of multi-agent formation
  control,'' \emph{Automatica}, vol.~53, pp. 424--440, 2015.

\bibitem{choi2018detecting}
H.~Choi, W.-C. Lee, Y.~Aafer, F.~Fei, Z.~Tu, X.~Zhang, D.~Xu, and X.~Xinyan,
  ``Detecting attacks against robotic vehicles: A control invariant approach,''
  in \emph{Proceedings of the 2018 ACM SIGSAC Conference on Computer and
  Communications Security}.\hskip 1em plus 0.5em minus 0.4em\relax ACM, 2018,
  pp. 801--816.

\bibitem{hubaux2004security}
J.-P. Hubaux, S.~Capkun, and J.~Luo, ``The security and privacy of smart
  vehicles,'' \emph{IEEE Security \& Privacy}, vol.~2, no.~3, pp. 49--55, 2004.

\bibitem{ni2019toward}
Y.~Ni, L.~Cai, J.~He, A.~Vinel, Y.~Li, H.~Mosavat-Jahromi, and J.~Pan, ``Toward
  reliable and scalable internet of vehicles: Performance analysis and resource
  management,'' \emph{Proceedings of the IEEE}, vol. 108, no.~2, pp. 324--340,
  2019.

\bibitem{cardenas2008secure}
A.~A. Cardenas, S.~Amin, and S.~Sastry, ``Secure control: Towards survivable
  cyber-physical systems,'' in \emph{The 28th International Conference on
  Distributed Computing Systems Workshops}.\hskip 1em plus 0.5em minus
  0.4em\relax IEEE, 2008, pp. 495--500.

\bibitem{van2014encyclopedia}
H.~C. Van~Tilborg and S.~Jajodia, \emph{Encyclopedia of cryptography and
  security}.\hskip 1em plus 0.5em minus 0.4em\relax Springer Science \&
  Business Media, 2014.

\bibitem{shannon1949communication}
C.~E. Shannon, ``Communication theory of secrecy systems,'' \emph{The Bell
  System Technical Journal}, vol.~28, no.~4, pp. 656--715, 1949.

\bibitem{shokri2011quantifying}
R.~Shokri, G.~Theodorakopoulos, J.-Y. Le~Boudec, and J.-P. Hubaux,
  ``Quantifying location privacy,'' in \emph{2011 IEEE Symposium on Security
  and Privacy}.\hskip 1em plus 0.5em minus 0.4em\relax IEEE, 2011, pp.
  247--262.

\bibitem{li2020evolvegraph}
J.~Li, F.~Yang, M.~Tomizuka, and C.~Choi, ``Evolvegraph: Multi-agent trajectory
  prediction with dynamic relational reasoning,'' \emph{Advances in Neural
  Information Processing Systems}, vol.~33, pp. 19\,783--19\,794, 2020.

\bibitem{vasquez2018network}
B.~L.~M. V{\'a}squez and J.~Carlo~Barca, ``Network topology inference in swarm
  robotics,'' in \emph{IEEE International Conference on Robotics and Automation
  (ICRA)}.\hskip 1em plus 0.5em minus 0.4em\relax IEEE, 2018, pp. 7660--7666.

\bibitem{li2021topology}
Y.~Li, J.~He, and L.~Cai, ``Topology inference on partially observable mobile
  robotic networks under formation control,'' in \emph{2021 {{European Control
  Conference}} ({{ECC}})}, 2021, pp. 497--502.

\bibitem{zhang2014analysis}
D.~Zhang and J.~P. Sterbenz, ``Analysis of critical node attacks in mobile ad
  hoc networks,'' in \emph{2014 6th International Workshop on Reliable Networks
  Design and Modeling}, 2014, pp. 171--178.

\bibitem{deisenroth2015gaussian}
M.~P. Deisenroth, D.~Fox, and C.~E. Rasmussen, ``Gaussian processes for
  data-efficient learning in robotics and control,'' \emph{IEEE Transactions on
  Pattern Analysis and Machine Intelligence}, vol.~37, no.~2, pp. 408--423,
  2015.

\bibitem{tolstaya2020learning}
E.~Tolstaya, F.~Gama, J.~Paulos, G.~Pappas, V.~Kumar, and A.~Ribeiro,
  ``Learning decentralized controllers for robot swarms with graph neural
  networks,'' in \emph{Conference on Robot Learning}.\hskip 1em plus 0.5em
  minus 0.4em\relax PMLR, 2020, pp. 671--682.

\bibitem{quinonez2020savior}
R.~Quinonez, J.~Giraldo, L.~Salazar, E.~Bauman, A.~Cardenas, and Z.~Lin,
  ``{SAVIOR}: Securing autonomous vehicles with robust physical invariants,''
  in \emph{29th USENIX Security Symposium (USENIX Security 20)}, 2020, pp.
  895--912.

\bibitem{li2019learning}
Y.~Li, J.~He, C.~Chen, and X.~Guan, ``Learning-based intelligent attack against
  formation control with obstacle-avoidance,'' in \emph{2019 American Control
  Conference (ACC)}.\hskip 1em plus 0.5em minus 0.4em\relax IEEE, 2019, pp.
  2690--2695.

\bibitem{licitra2019single}
R.~A. Licitra, Z.~I. Bell, and W.~E. Dixon, ``Single-agent indirect herding of
  multiple targets with uncertain dynamics,'' \emph{IEEE Transactions on
  Robotics}, vol.~35, no.~4, pp. 847--860, 2019.

\bibitem{giraldo2017security}
J.~Giraldo, E.~Sarkar, A.~A. Cardenas, M.~Maniatakos, and M.~Kantarcioglu,
  ``Security and privacy in cyber-physical systems: A survey of surveys,''
  \emph{IEEE Design and Test}, vol.~34, no.~4, pp. 7--17, 2017.

\bibitem{sandberg2022secure}
H.~Sandberg, V.~Gupta, and K.~H. Johansson, ``Secure networked control
  systems,'' \emph{Annual Review of Control, Robotics, and Autonomous Systems},
  vol.~5, no.~1, pp. 072\,921--075\,953, 2022.

\bibitem{teixeira2015secure}
A.~Teixeira, I.~Shames, H.~Sandberg, and K.~H. Johansson, ``A secure control
  framework for resource-limited adversaries,'' \emph{Automatica}, vol.~51, pp.
  135--148, 2015.

\bibitem{zhang2015optimal}
H.~Zhang, P.~Cheng, L.~Shi, and J.~Chen, ``Optimal {DoS} attack scheduling in
  wireless networked control system,'' \emph{IEEE Transactions on Control
  Systems Technology}, vol.~24, no.~3, pp. 843--852, 2015.

\bibitem{conti2016survey}
M.~Conti, N.~Dragoni, and V.~Lesyk, ``A survey of man in the middle attacks,''
  \emph{IEEE Communications Surveys and Tutorials}, vol.~18, no.~3, pp.
  2027--2051, 2016.

\bibitem{anguluri2019probabilistic}
R.~Anguluri, V.~Katewa, and F.~Pasqualetti, ``A probabilistic approach to
  design switching attacks against interconnected systems,'' in \emph{American
  Control Conference}.\hskip 1em plus 0.5em minus 0.4em\relax IEEE, 2019, pp.
  4430--4435.

\bibitem{sanchez2019bibliographical}
H.~S. S{\'a}nchez, D.~Rotondo, T.~Escobet, V.~Puig, and J.~Quevedo,
  ``Bibliographical review on cyber attacks from a control oriented
  perspective,'' \emph{Annual Reviews in Control}, vol.~48, pp. 103--128, 2019.

\bibitem{pasqualetti2013attack}
F.~Pasqualetti, F.~D{\"o}rfler, and F.~Bullo, ``Attack detection and
  identification in cyber-physical systems,'' \emph{IEEE Transactions on
  Automatic Control}, vol.~58, no.~11, pp. 2715--2729, 2013.

\bibitem{bai2017data}
C.-Z. Bai, F.~Pasqualetti, and V.~Gupta, ``Data-injection attacks in stochastic
  control systems: Detectability and performance tradeoffs,''
  \emph{Automatica}, vol.~82, pp. 251--260, 2017.

\bibitem{guan2017distributed}
Y.~Guan and X.~Ge, ``Distributed attack detection and secure estimation of
  networked cyber-physical systems against false data injection attacks and
  jamming attacks,'' \emph{IEEE Transactions on Signal and Information
  Processing over Networks}, vol.~4, no.~1, pp. 48--59, 2017.

\bibitem{giraldo2018survey}
J.~Giraldo, D.~Urbina, A.~Cardenas, J.~Valente, M.~Faisal, J.~Ruths, N.~O.
  Tippenhauer, H.~Sandberg, and R.~Candell, ``A survey of physics-based attack
  detection in cyber-physical systems,'' \emph{ACM Computing Surveys}, vol.~51,
  no.~4, pp. 1--36, 2018.

\bibitem{anguluri2020centralized}
R.~Anguluri, V.~Katewa, and F.~Pasqualetti, ``Centralized {{versus
  decentralized detection}} of {{attacks}} in {{stochastic interconnected
  systems}},'' \emph{IEEE Transactions on Automatic Control}, vol.~65, no.~9,
  pp. 3903--3910, 2020.

\bibitem{fawzi2014secure}
H.~Fawzi, P.~Tabuada, and S.~Diggavi, ``Secure estimation and control for
  cyber-physical systems under adversarial attacks,'' \emph{IEEE Transactions
  on Automatic Control}, vol.~59, no.~6, pp. 1454--1467, 2014.

\bibitem{mo2015physical}
Y.~Mo, S.~Weerakkody, and B.~Sinopoli, ``Physical authentication of control
  systems: Designing watermarked control inputs to detect counterfeit sensor
  outputs,'' \emph{IEEE Control Systems Magazine}, vol.~35, no.~1, pp. 93--109,
  2015.

\bibitem{pajic2016attack}
M.~Pajic, I.~Lee, and G.~J. Pappas, ``Attack-resilient state estimation for
  noisy dynamical systems,'' \emph{IEEE Transactions on Control of Network
  Systems}, vol.~4, no.~1, pp. 82--92, 2016.

\bibitem{barboni2020towards}
A.~Barboni and T.~Parisini, ``Towards distributed accommodation of covert
  attacks in interconnected systems,'' in \emph{2020 59th IEEE Conference on
  Decision and Control}.\hskip 1em plus 0.5em minus 0.4em\relax IEEE, 2020, pp.
  5731--5736.

\bibitem{ding2020secure}
D.~Ding, Q.-L. Han, X.~Ge, and J.~Wang, ``Secure state estimation and control
  of cyber-physical systems: A survey,'' \emph{IEEE Transactions on Systems,
  Man, and Cybernetics: Systems}, vol.~51, no.~1, pp. 176--190, 2020.

\bibitem{teixeira2015securea}
A.~Teixeira, I.~Shames, H.~Sandberg, and K.~H. Johansson, ``A secure control
  framework for resource-limited adversaries,'' \emph{Automatica}, vol.~51, pp.
  135--148, 2015.

\bibitem{han2018privacy}
S.~Han and G.~J. Pappas, ``Privacy in control and dynamical systems,''
  \emph{Annual Review of Control, Robotics, and Autonomous Systems}, vol.~1,
  pp. 309--332, 2018.

\bibitem{mo2016privacy}
Y.~Mo and R.~M. Murray, ``Privacy preserving average consensus,'' \emph{IEEE
  Transactions on Automatic Control}, vol.~62, no.~2, pp. 753--765, 2016.

\bibitem{he2018privacy}
J.~He, L.~Cai, C.~Zhao, P.~Cheng, and X.~Guan, ``Privacy-preserving average
  consensus: Privacy analysis and algorithm design,'' \emph{IEEE Transactions
  on Signal and Information Processing over Networks}, vol.~5, no.~1, pp.
  127--138, 2018.

\bibitem{wang2019privacy}
Y.~Wang, ``Privacy-preserving average consensus via state decomposition,''
  \emph{IEEE Transactions on Automatic Control}, vol.~64, no.~11, pp.
  4711--4716, 2019.

\bibitem{ruan2019secure}
M.~Ruan, H.~Gao, and Y.~Wang, ``Secure and {{privacy-preserving consensus}},''
  \emph{IEEE Transactions on Automatic Control}, vol.~64, no.~10, pp.
  4035--4049, 2019.

\bibitem{dibaji2019systems}
S.~M. Dibaji, M.~Pirani, D.~B. Flamholz, A.~M. Annaswamy, K.~H. Johansson, and
  A.~Chakrabortty, ``A systems and control perspective of {{CPS}} security,''
  \emph{Annual Reviews in Control}, vol.~47, pp. 394--411, 2019.

\bibitem{fang2020stealthy}
C.~Fang, Y.~Qi, J.~Chen, R.~Tan, and W.~X. Zheng, ``Stealthy {{actuator signal
  attacks}} in {{stochastic control systems}}: {{Performance}} and
  {{limitations}},'' \emph{IEEE Transactions on Automatic Control}, vol.~65,
  no.~9, pp. 3927--3934, 2020.

\bibitem{sui2021vulnerability}
T.~Sui, Y.~Mo, D.~Marelli, X.~Sun, and M.~Fu, ``The {{vulnerability}} of
  {{cyber-physical system under stealthy attacks}},'' \emph{IEEE Transactions
  on Automatic Control}, vol.~66, no.~2, pp. 637--650, 2021.

\bibitem{zhou2021unified}
C.~Zhou, B.~Hu, Y.~Shi, Y.-C. Tian, X.~Li, and Y.~Zhao, ``A unified
  architectural approach for cyberattack-resilient industrial control
  systems,'' \emph{Proceedings of the IEEE}, vol. 109, no.~4, pp. 517--541,
  2021.

\bibitem{chong2015observability}
M.~S. Chong, M.~Wakaiki, and J.~P. Hespanha, ``Observability of linear systems
  under adversarial attacks,'' in \emph{2015 {{American Control Conference}}
  ({{ACC}})}.\hskip 1em plus 0.5em minus 0.4em\relax {IEEE}, 2015, pp.
  2439--2444.

\bibitem{guan2018distributed}
Y.~Guan and X.~Ge, ``Distributed {{attack detection}} and {{secure estimation}}
  of {{networked cyber-physical systems against false data injection attacks}}
  and {{jamming attacks}},'' \emph{IEEE Transactions on Signal and Information
  Processing over Networks}, vol.~4, no.~1, pp. 48--59, 2018.

\bibitem{mao2022computational}
Y.~Mao, A.~Mitra, S.~Sundaram, and P.~Tabuada, ``On the computational
  complexity of the secure state-reconstruction problem,'' \emph{Automatica},
  vol. 136, p. 110083, 2022.

\bibitem{ren2020secure}
X.~Ren, Y.~Mo, J.~Chen, and K.~H. Johansson, ``Secure state estimation with
  {Byzantine} sensors: A probabilistic approach,'' \emph{IEEE Transactions on
  Automatic Control}, vol.~65, no.~9, pp. 3742--3757, 2020.

\bibitem{weerakkody2017graphtheoretic}
S.~Weerakkody, X.~Liu, S.~H. Son, and B.~Sinopoli, ``A {{graph-theoretic
  characterization}} of {{perfect attackability}} for {{secure design}} of
  {{distributed control systems}},'' \emph{IEEE Transactions on Control of
  Network Systems}, vol.~4, no.~1, pp. 60--70, 2017.

\bibitem{lu2020privacy}
Y.~Lu and M.~Zhu, ``On privacy preserving data release of linear dynamic
  networks,'' \emph{Automatica}, vol. 115, p. 108839, 2020.

\bibitem{katewa2021security}
V.~Katewa, R.~Anguluri, and F.~Pasqualetti, ``On a security vs privacy
  trade-off in interconnected dynamical systems,'' \emph{Automatica}, vol. 125,
  p. 109426, 2021.

\bibitem{mo2017privacy}
Y.~Mo and R.~M. Murray, ``Privacy {{preserving average consensus}},''
  \emph{IEEE Transactions on Automatic Control}, vol.~62, no.~2, pp. 753--765,
  2017.

\bibitem{he2019consensus}
J.~He, L.~Cai, P.~Cheng, J.~Pan, and L.~Shi, ``Consensus-based data-privacy
  preserving data aggregation,'' \emph{IEEE Transactions on Automatic Control},
  vol.~64, no.~12, pp. 5222--5229, 2019.

\bibitem{yemini2021characterizing}
M.~Yemini, A.~Nedic, A.~J. Goldsmith, and S.~Gil, ``Characterizing {{trust}}
  and {{resilience}} in {{distributed consensus}} for {{cyberphysical
  systems}},'' \emph{IEEE Transactions on Robotics}, pp. 1--21, 2021.

\bibitem{cao2010sampled}
Y.~Cao and W.~Ren, ``Sampled-data discrete-time coordination algorithms for
  double-integrator dynamics under dynamic directed interaction,''
  \emph{International Journal of Control}, vol.~83, no.~3, pp. 506--515, 2010.

\bibitem{wainwright2019high}
M.~J. Wainwright, \emph{High-dimensional statistics: A non-asymptotic
  viewpoint}.\hskip 1em plus 0.5em minus 0.4em\relax Cambridge University
  Press, 2019, vol.~48.

\bibitem{candes2006stable}
E.~J. Candes, J.~K. Romberg, and T.~Tao, ``Stable signal recovery from
  incomplete and inaccurate measurements,'' \emph{Communications on Pure and
  Applied Mathematics: A Journal Issued by the Courant Institute of
  Mathematical Sciences}, vol.~59, no.~8, pp. 1207--1223, 2006.

\bibitem{hayden2016sparse}
D.~Hayden, Y.~H. Chang, J.~Goncalves, and C.~J. Tomlin, ``Sparse network
  identifiability via compressed sensing,'' \emph{Automatica}, vol.~68, pp.
  9--17, 2016.

\bibitem{dobbe2019blind}
R.~Dobbe, S.~Liu, Y.~Yuan, and C.~Tomlin, ``Blind identification of fully
  observed linear time-varying systems via sparse recovery,''
  \emph{Automatica}, vol. 100, pp. 330--335, 2019.

\bibitem{rezazadeh2018privacya}
N.~Rezazadeh and S.~S. Kia, ``Privacy preservation in a continuous-time static
  average consensus algorithm over directed graphs,'' in \emph{2018 American
  Control Conference}, 2018, pp. 5890--5895.

\bibitem{altafini2020systemtheoretic}
C.~Altafini, ``A system-theoretic framework for privacy preservation in
  continuous-time multiagent dynamics,'' \emph{Automatica}, vol. 122, p.
  109253, 2020.

\bibitem{liu2019dynamic}
C.~Liu, J.~He, S.~Zhu, and C.~Chen, ``Dynamic topology inference via external
  observation for multi-robot formation control,'' in \emph{2019 IEEE Pacific
  Rim Conference on Communications, Computers and Signal Processing
  (PACRIM)}.\hskip 1em plus 0.5em minus 0.4em\relax IEEE, 2019, pp. 1--6.

\bibitem{ahmed2009recovering}
A.~Ahmed and E.~P. Xing, ``Recovering time-varying networks of dependencies in
  social and biological studies,'' \emph{Proceedings of the National Academy of
  Sciences}, vol. 106, no.~29, pp. 11\,878--11\,883, 2009.

\bibitem{monti2014estimating}
R.~P. Monti, P.~Hellyer, D.~Sharp, R.~Leech, C.~Anagnostopoulos, and
  G.~Montana, ``Estimating time-varying brain connectivity networks from
  functional {MRI} time series,'' \emph{NeuroImage}, vol. 103, pp. 427--443,
  2014.

\bibitem{granger1969investigating}
C.~W. Granger, ``Investigating causal relations by econometric models and
  cross-spectral methods,'' \emph{Econometrica: Journal of the Econometric
  Society}, pp. 424--438, 1969.

\bibitem{brovelli2004beta}
A.~Brovelli, M.~Ding, A.~Ledberg, Y.~Chen, R.~Nakamura, and S.~L. Bressler,
  ``Beta oscillations in a large-scale sensorimotor cortical network:
  Directional influences revealed by granger causality,'' \emph{Proceedings of
  the National Academy of Sciences}, vol. 101, no.~26, pp. 9849--9854, 2004.

\bibitem{santos2019local}
A.~Santos, V.~Matta, and A.~H. Sayed, ``Local tomography of large networks
  under the low-observability regime,'' \emph{IEEE Transactions on Information
  Theory}, vol.~66, no.~1, pp. 587--613, 2020.

\bibitem{segarra2017network}
S.~Segarra, A.~G. Marques, G.~Mateos, and A.~Ribeiro, ``Network topology
  inference from spectral templates,'' \emph{IEEE Transactions on Signal and
  Information Processing over Networks}, vol.~3, no.~3, pp. 467--483, 2017.

\bibitem{schaub2019spectral}
M.~T. Schaub, S.~Segarra, and H.-T. Wai, ``Spectral partitioning of
  time-varying networks with unobserved edges,'' in \emph{2019 IEEE
  International Conference on Acoustics, Speech and Signal Processing
  (ICASSP)}.\hskip 1em plus 0.5em minus 0.4em\relax IEEE, 2019, pp. 4938--4942.

\bibitem{zhu2020network}
Y.~Zhu, M.~T. Schaub, A.~Jadbabaie, and S.~Segarra, ``Network inference from
  consensus dynamics with unknown parameters,'' \emph{IEEE Transactions on
  Signal and Information Processing over Networks}, vol.~6, pp. 300--315, 2020.

\bibitem{karanikolas2016multi}
G.~Karanikolas, G.~B. Giannakis, K.~Slavakis, and R.~M. Leahy, ``Multi-kernel
  based nonlinear models for connectivity identification of brain networks,''
  in \emph{2016 IEEE International Conference on Acoustics, Speech and Signal
  Processing (ICASSP)}.\hskip 1em plus 0.5em minus 0.4em\relax IEEE, 2016, pp.
  6315--6319.

\bibitem{karanikolas2017multi}
G.~V. Karanikolas, O.~Sporns, and G.~B. Giannakis, ``Multi-kernel change
  detection for dynamic functional connectivity graphs,'' in \emph{2017 51st
  Asilomar Conference on Signals, Systems, and Computers}.\hskip 1em plus 0.5em
  minus 0.4em\relax IEEE, 2017, pp. 1555--1559.

\bibitem{wang2018inferring}
S.~Wang, E.~D. Herzog, I.~Z. Kiss, W.~J. Schwartz, G.~Bloch, M.~Sebek,
  D.~Granados-Fuentes, L.~Wang, and J.-S. Li, ``Inferring dynamic topology for
  decoding spatiotemporal structures in complex heterogeneous networks,''
  \emph{Proceedings of the National Academy of Sciences}, vol. 115, no.~37, pp.
  9300--9305, 2018.

\bibitem{shahid2016fast}
N.~Shahid, N.~Perraudin, V.~Kalofolias, G.~Puy, and P.~Vandergheynst, ``Fast
  robust {PCA} on graphs,'' \emph{IEEE Journal of Selected Topics in Signal
  Processing}, vol.~10, no.~4, pp. 740--756, 2016.

\bibitem{onuki2016graph}
M.~Onuki, S.~Ono, M.~Yamagishi, and Y.~Tanaka, ``Graph signal denoising via
  trilateral filter on graph spectral domain,'' \emph{IEEE Transactions on
  Signal and Information Processing over Networks}, vol.~2, no.~2, pp.
  137--148, 2016.

\bibitem{giannakis2018topology}
G.~B. Giannakis, Y.~Shen, and G.~V. Karanikolas, ``Topology identification and
  learning over graphs: Accounting for nonlinearities and dynamics,''
  \emph{Proceedings of the IEEE}, vol. 106, no.~5, pp. 787--807, 2018.

\bibitem{brugere2018network}
I.~Brugere, B.~Gallagher, and T.~Y. Berger-Wolf, ``Network structure inference,
  a survey: Motivations, methods, and applications,'' \emph{ACM Computing
  Surveys}, vol.~51, no.~2, pp. 1--39, 2018.

\bibitem{lys2021cdc}
Y.~Li and J.~He, ``Topology inference for networked dynamical systems: A
  causality and correlation perspective,'' in \emph{2021 60th IEEE Conference
  on Decision and Control (CDC)}.\hskip 1em plus 0.5em minus 0.4em\relax IEEE,
  2021, pp. 1218--1223.

\bibitem{li2021topoJourna}
Y.~Li, J.~He, C.~Chen, and X.~Guan, ``On topology inference for networked
  dynamical systems: Principles and performances,'' \emph{arXiv preprint
  arXiv:2106.01031}, 2021.

\bibitem{matta2018consistent}
V.~Matta and A.~H. Sayed, ``Consistent tomography under partial observations
  over adaptive networks,'' \emph{IEEE Transactions on Information Theory},
  vol.~65, no.~1, pp. 622--646, 2019.

\bibitem{cirillo2021learninga}
M.~Cirillo, V.~Matta, and A.~H. Sayed, ``Learning {{Bollob\'as-Riordan graphs
  under partial observability}},'' in \emph{2021 IEEE International Conference
  on Acoustics, Speech and Signal Processing ({ICASSP})}, 2021, pp. 5360--5364.

\bibitem{6213241}
A.~Simpkins, ``System identification: Theory for the user,'' \emph{IEEE
  Robotics and Automation Magazine}, vol.~19, no.~2, pp. 95--96, 2012.

\bibitem{oymak2019non}
S.~Oymak and N.~Ozay, ``Non-asymptotic identification of {LTI} systems from a
  single trajectory,'' in \emph{2019 American control conference (ACC)}.\hskip
  1em plus 0.5em minus 0.4em\relax IEEE, 2019, pp. 5655--5661.

\bibitem{simchowitz2019learning}
M.~Simchowitz, R.~Boczar, and B.~Recht, ``Learning linear dynamical systems
  with semi-parametric least squares,'' in \emph{Conference on Learning
  Theory}.\hskip 1em plus 0.5em minus 0.4em\relax PMLR, 2019, pp. 2714--2802.

\bibitem{zheng2020non}
Y.~Zheng and N.~Li, ``Non-asymptotic identification of linear dynamical systems
  using multiple trajectories,'' \emph{IEEE Control Systems Letters}, vol.~5,
  no.~5, pp. 1693--1698, 2020.

\bibitem{ho1966effective}
B.~HO and R.~E. K{\'a}lm{\'a}n, ``Effective construction of linear
  state-variable models from input/output functions,''
  \emph{at-Automatisierungstechnik}, vol.~14, no. 1-12, pp. 545--548, 1966.

\bibitem{acar2019survey}
A.~Acar, H.~Aksu, A.~S. Uluagac, and M.~Conti, ``A survey on homomorphic
  encryption schemes: Theory and implementation,'' \emph{ACM Computing
  Surveys}, vol.~51, no.~4, pp. 1--35, 2019.

\bibitem{kogiso2015cybersecurity}
K.~Kogiso and T.~Fujita, ``Cyber-security enhancement of networked control
  systems using homomorphic encryption,'' in \emph{2015 54th {{IEEE
  Conference}} on {{Decision}} and {{Control}}}, 2015, pp. 6836--6843.

\bibitem{farokhi2017secure}
F.~Farokhi, I.~Shames, and N.~Batterham, ``Secure and private control using
  semi-homomorphic encryption,'' \emph{Control Engineering Practice}, vol.~67,
  pp. 13--20, 2017.

\bibitem{schulzedarup2019encrypted}
M.~Schulze~Darup, A.~Redder, and D.~E. Quevedo, ``Encrypted cooperative control
  based on structured feedback,'' \emph{IEEE Control Systems Letters}, vol.~3,
  no.~1, pp. 37--42, 2019.

\bibitem{lu2018privacy}
Y.~Lu and M.~Zhu, ``Privacy preserving distributed optimization using
  homomorphic encryption,'' \emph{Automatica}, vol.~96, pp. 314--325, 2018.

\bibitem{zhang2019admm}
C.~Zhang, M.~Ahmad, and Y.~Wang, ``{ADMM} based privacy-preserving
  decentralized optimization,'' \emph{IEEE Transactions on Information
  Forensics and Security}, vol.~14, no.~3, pp. 565--580, 2019.

\bibitem{geng2015optimal}
Q.~Geng and P.~Viswanath, ``The optimal noise-adding mechanism in differential
  privacy,'' \emph{IEEE Transactions on Information Theory}, vol.~62, no.~2,
  pp. 925--951, 2015.

\bibitem{dwork2006differential}
C.~Dwork, ``Differential privacy,'' in \emph{International Colloquium on
  Automata, Languages, and Programming}.\hskip 1em plus 0.5em minus 0.4em\relax
  Springer, 2006, pp. 1--12.

\bibitem{cortes2016differential}
J.~Cort{\'e}s, G.~E. Dullerud, S.~Han, J.~Le~Ny, S.~Mitra, and G.~J. Pappas,
  ``Differential privacy in control and network systems,'' in \emph{2016 55th
  IEEE Conference on Decision and Control (CDC)}.\hskip 1em plus 0.5em minus
  0.4em\relax IEEE, 2016, pp. 4252--4272.

\bibitem{he2020differential}
J.~He, L.~Cai, and X.~Guan, ``Differential private noise adding mechanism and
  its application on consensus algorithm,'' \emph{IEEE Transactions on Signal
  Processing}, vol.~68, pp. 4069--4082, 2020.

\bibitem{katewa2018privacy}
V.~Katewa, F.~Pasqualetti, and V.~Gupta, ``On privacy vs. cooperation in
  multi-agent systems,'' \emph{International Journal of Control}, vol.~91,
  no.~7, pp. 1693--1707, 2018.

\bibitem{he2018preservingTIT}
J.~He, L.~Cai, and X.~Guan, ``Preserving data-privacy with added noises:
  Optimal estimation and privacy analysis,'' \emph{IEEE Transactions on
  Information Theory}, vol.~64, no.~8, pp. 5677--5690, 2018.

\bibitem{li2020unpredictable}
J.~Li, J.~He, Y.~Li, and X.~Guan, ``Unpredictable trajectory design for mobile
  agents,'' in \emph{2020 American Control Conference (ACC)}.\hskip 1em plus
  0.5em minus 0.4em\relax IEEE, 2020, pp. 1471--1476.

\bibitem{ztw2022acc}
Z.~Wang, Y.~Li, C.~Fang, and J.~He, ``Distributed topology-preserving
  collaboration algorithm against inference attack,'' in \emph{2022 {{American
  Control Conference}} ({{ACC}})}.\hskip 1em plus 0.5em minus 0.4em\relax
  {IEEE}, to be published.

\bibitem{yuan2019data}
Y.~Yuan, X.~Tang, W.~Zhou, W.~Pan, X.~Li, H.-T. Zhang, H.~Ding, and
  J.~Goncalves, ``Data driven discovery of cyber physical systems,''
  \emph{Nature Communications}, vol.~10, no.~1, pp. 1--9, 2019.

\bibitem{anderson2007optimal}
B.~D. Anderson and J.~B. Moore, \emph{Optimal control: Linear quadratic
  methods}.\hskip 1em plus 0.5em minus 0.4em\relax Courier Corporation, 2007.

\bibitem{montgomery2021introduction}
D.~C. Montgomery, E.~A. Peck, and G.~G. Vining, \emph{Introduction to linear
  regression analysis}.\hskip 1em plus 0.5em minus 0.4em\relax John Wiley \&
  Sons, 2021.

\bibitem{jiao2021topology}
Q.~Jiao, Y.~Li, and J.~He, ``Topology inference for consensus-based cooperation
  under time-invariant latent input,'' in \emph{2021 {{IEEE}} 94th {{Vehicular
  Technology Conference}} ({{VTC2021-Fall}})}, 2021, pp. 1--5.

\bibitem{zheng2021equivalence}
Y.~Zheng, L.~Furieri, A.~Papachristodoulou, N.~Li, and M.~Kamgarpour, ``On the
  equivalence of {{Youla}}, system-level, and input-output parameterizations,''
  \emph{IEEE Transactions on Automatic Control}, vol.~66, no.~1, pp. 413--420,
  2021.

\bibitem{xymdec2022acc}
X.~Mao and J.~He, ``Decentralized system identification method for large-scale
  networks,'' in \emph{2022 American Control Conference (ACC)}.\hskip 1em plus
  0.5em minus 0.4em\relax {IEEE}, to be published.

\bibitem{xyminput2022acc}
X.~Mao, J.~He, and C.~Zhao, ``An improved subspace identification method with
  variance minimization and input design,'' in \emph{2022 American Control
  Conference (ACC)}.\hskip 1em plus 0.5em minus 0.4em\relax {IEEE}, to be
  published.

\bibitem{kawano2020design}
Y.~Kawano and M.~Cao, ``Design of privacy-preserving dynamic controllers,''
  \emph{IEEE Transactions on Automatic Control}, vol.~65, no.~9, pp.
  3863--3878, 2020.

\bibitem{kawano2021modular}
Y.~Kawano, K.~Kashima, and M.~Cao, ``Modular control under privacy protection:
  {{Fundamental}} trade-offs,'' \emph{Automatica}, vol. 127, p. 109518, 2021.

\bibitem{nekouei2022randomized}
E.~Nekouei, M.~Pirani, H.~Sandberg, and K.~H. Johansson, ``A randomized
  filtering strategy against inference attacks on active steering control
  systems,'' \emph{IEEE Transactions on Information Forensics and Security},
  vol.~17, pp. 16--27, 2022.

\bibitem{yazdani2022differentially}
K.~Yazdani, A.~Jones, K.~Leahy, and M.~Hale, ``Differentially private {LQ}
  control,'' \emph{IEEE Transactions on Automatic Control}, 2022.

\bibitem{hawkins2020differentially}
C.~Hawkins and M.~Hale, ``Differentially private formation control,'' in
  \emph{2020 59th IEEE Conference on Decision and Control (CDC)}.\hskip 1em
  plus 0.5em minus 0.4em\relax IEEE, 2020, pp. 6260--6265.

\bibitem{levine2018reinforcement}
S.~Levine, ``Reinforcement learning and control as probabilistic inference:
  Tutorial and review,'' \emph{arXiv preprint arXiv:1805.00909}, 2018.

\bibitem{de2020formulas}
C.~De~Persis and P.~Tesi, ``Formulas for data-driven control: Stabilization,
  optimality, and robustness,'' \emph{IEEE Transactions on Automatic Control},
  vol.~65, no.~3, pp. 909--924, 2020.

\bibitem{hewing2020learning}
L.~Hewing, K.~P. Wabersich, M.~Menner, and M.~N. Zeilinger, ``Learning-based
  model predictive control: Toward safe learning in control,'' \emph{Annual
  Review of Control, Robotics, and Autonomous Systems}, vol.~3, pp. 269--296,
  2020.

\bibitem{chen2019reinforcement}
C.~Chen, H.~Modares, K.~Xie, F.~L. Lewis, Y.~Wan, and S.~Xie, ``Reinforcement
  learning-based adaptive optimal exponential tracking control of linear
  systems with unknown dynamics,'' \emph{IEEE Transactions on Automatic
  Control}, vol.~64, no.~11, pp. 4423--4438, 2019.

\bibitem{westenbroek2020learning}
T.~Westenbroek, F.~Casta{\~n}eda, A.~Agrawal, S.~S. Sastry, and K.~Sreenath,
  ``Learning min-norm stabilizing control laws for systems with unknown
  dynamics,'' in \emph{2020 59th IEEE Conference on Decision and Control
  (CDC)}.\hskip 1em plus 0.5em minus 0.4em\relax IEEE, 2020, pp. 737--744.

\bibitem{westenbroek2021combining}
T.~Westenbroek, A.~Agrawal, F.~Casta{\~n}eda, S.~S. Sastry, and K.~Sreenath,
  ``Combining model-based design and model-free policy optimization to learn
  safe, stabilizing controllers,'' \emph{IFAC-PapersOnLine}, vol.~54, no.~5,
  pp. 19--24, 2021.

\bibitem{he2022model}
Z.~He, S.~Bolognani, J.~He, F.~D{\"o}rfler, and X.~Guan, ``Model-free nonlinear
  feedback optimization,'' \emph{arXiv preprint arXiv:2201.02395}, 2022.

\bibitem{pukelsheim1994three}
F.~Pukelsheim, ``The three sigma rule,'' \emph{The American Statistician},
  vol.~48, no.~2, pp. 88--91, 1994.

\end{thebibliography}

\begin{IEEEbiographynophoto}{Jianping He} 
(SM'19) is currently an associate professor in the Department of Automation at Shanghai Jiao Tong University. He received the Ph.D. degree in control science and engineering from Zhejiang University, Hangzhou, China, in 2013, and had been a research fellow in the Department of Electrical and Computer Engineering at University of Victoria, Canada, from Dec. 2013 to Mar. 2017. His research interests mainly include the distributed learning, control and optimization, security and privacy in network systems.

Dr. He serves as an Associate Editor for IEEE Trans. Control of Network Systems, IEEE Open Journal of Vehicular Technology, and KSII Trans. Internet and Information Systems. He was also a Guest Editor of IEEE TAC, International Journal of Robust and Nonlinear Control, etc. He was the winner of Outstanding Thesis Award, Chinese Association of Automation, 2015. He received the best paper award from IEEE WCSP'17, the best conference paper award from IEEE PESGM'17, and was a finalist for the best student paper award from IEEE ICCA'17, and the finalist best conference paper award from IEEE VTC'20-FALL.
\end{IEEEbiographynophoto}

\begin{IEEEbiographynophoto}{Yushan Li}
(S'19) received the B.E. degree in School of Artificial Intelligence and Automation from Huazhong University of Science and Technology, Wuhan, China, in 2018. 
He is currently working toward the Ph.D. degree with the Department of Automation, Shanghai Jiaotong University, Shanghai, China. 
He is a member of Intelligent of Wireless Networking and Cooperative Control group. 
His research interests include robotics, security of cyber-physical system, and distributed computation and optimization in multi-agent networks. 
\end{IEEEbiographynophoto}

\begin{IEEEbiographynophoto}{Lin Cai}
(F'20) received her M.A.Sc. and Ph. D. degrees (awarded Outstanding Achievement in Graduate Studies) in electrical and computer engineering from the University of Waterloo, Waterloo, Canada, in 2002 and 2005, respectively. Since 2005, she has been with the Department of Electrical and Computer Engineering at the University of Victoria, and she is currently a Professor. She is an NSERC E.W.R. Steacie Memorial Fellow, an Engineering Institute of Canada (EIC) Fellow, and an IEEE Fellow. In 2020, she was elected as a Member of the Royal Society of Canada's College of New Scholars, Artists and Scientists. She was also elected as a 2020 ``Star in Computer Networking and Communications" by N2Women. Her research interests span several areas in communications and networking, with a focus on network protocol and architecture design supporting emerging multimedia traffic and the Internet of Things. 

She was a recipient of the NSERC Discovery Accelerator Supplement (DAS) Grants in 2010 and 2015, respectively, and the Best Paper Awards of IEEE ICC 2008 and IEEE WCNC 2011. She has co-founded and chaired the IEEE Victoria Section Vehicular Technology and Communications Joint Societies Chapter. She has been elected to serve the IEEE Vehicular Technology Society Board of Governors, 2019 - 2024. She has served as an area editor for IEEE Transactions on Vehicular Technology, a member of the Steering Committee of the IEEE Transactions on Big Data (TBD) and IEEE Transactions on Cloud Computing (TCC), an Associate Editor of the IEEE Internet of Things Journal, IEEE Transactions on Wireless Communications, IEEE Transactions on Vehicular Technology, IEEE Transactions on Communications, EURASIP Journal on Wireless Communications and Networking, International Journal of Sensor Networks, and Journal of Communications and Networks (JCN), and as the Distinguished Lecturer of the IEEE VTS Society and the IEEE ComSoc Society. She has served as a TPC co-chair for IEEE VTC2020-Fall, and a TPC symposium co-chair for IEEE Globecom'10 and Globecom'13. She is a registered professional engineer in British Columbia, Canada.
\end{IEEEbiographynophoto}

\begin{IEEEbiographynophoto}{Xinping Guan} (F’18) received the B.S. degree in Mathematics from Harbin Normal University, Harbin, China, in 1986, and the Ph.D. degree in Control Science and Engineering from Harbin Institute of Technology, Harbin, China, in 1999. He is currently a Chair Professor with Shanghai Jiao Tong University, Shanghai, China, where he is the Dean of School of Electronic, Information and Electrical Engineering, and the Director of the Key Laboratory of Systems Control and Information Processing, Ministry of Education of China. Before that, he was the Professor and Dean of Electrical Engineering, Yanshan University, Qinhuangdao, China.

Dr. Guan’s current research interests include industrial cyber-physical systems, wireless networking and applications in smart factory, and underwater networks. He has authored and/or coauthored 5 research monographs, more than 270 papers in IEEE Transactions and other peer-reviewed journals, and numerous conference papers. As a Principal Investigator, he has finished/been working on many national key projects. He is the leader of the prestigious Innovative Research Team of the National Natural Science Foundation of China (NSFC). Dr. Guan received the First Prize of Natural Science Award from the Ministry of Education of China in both 2006 and 2016, and the Second Prize of the National Natural Science Award of China in both 2008 and 2018. He was a recipient of IEEE Transactions on Fuzzy Systems Outstanding Paper Award in 2008. He is a National Outstanding Youth honored by NSF of China, Changjiang Scholar by the Ministry of Education of China and State-level Scholar of New Century Bai Qianwan Talent Program of China.
\end{IEEEbiographynophoto}
\vspace{-20pt}

\end{document}